\documentclass[prd,aps,showpacs,tightenlines,notitlepage,nofootinbib,superscriptaddress]{revtex4-1}
\usepackage{mathrsfs}
\usepackage{amsmath}
\usepackage{amsfonts}
\usepackage{array}
\usepackage{verbatim}
\usepackage{epsfig}
\usepackage{graphicx} 
\usepackage[usenames, dvipsnames]{color}
\usepackage{marginnote}
\usepackage{slashed}
\usepackage{bm}

\begin{document}
\title{The role of the chiral anomaly in polarized deeply inelastic scattering III:\\ Wess-Zumino-Witten contributions and chiral Ward identities for finite quark mass}
\author{Andrey Tarasov}
\affiliation{Department of Physics, North Carolina State University, Raleigh, NC 27695, USA}
\author{Raju Venugopalan}
\affiliation{Physics Department, Brookhaven National Laboratory, Upton, NY 11973, USA}
\affiliation{CFNS, Department of Physics and Astronomy, Stony Brook University, Stony Brook, NY 11794, USA}

\begin{abstract}  
We extend our prior results on the worldline computation of the axial vector-vector-vector (AVV) triangle anomaly in polarized deeply inelastic scattering (DIS) to the finite mass case by computing in addition the pseudoscalar-vector-vector (PVV) triangle graph. For the well-studied QED case, we show explicitly how the off-forward AVV pole exactly cancels  an identical PVV pole. We then demonstrate the dramatic difference in QCD due to the chiral condensate, which qualitatively modifies anomalous Ward identities. As in the massless case, the anomaly pole in QCD is canceled by the dynamics of a primordial isosinglet pseudoscalar $\bar \eta$-meson, whose Wess-Zumino-Witten coupling to the topological charge density shifts the pole to the physical $\eta^\prime$ mass, with the finite quark mass contribution differing by $O(10\%)$ from the Witten-Veneziano formula. We obtain a compact analytic expression for the finite mass corrections to Shore and Veneziano's result that the proton's net quark helicity $\Delta \Sigma\propto \sqrt{\chi_{\rm QCD}' |_{m=0}(0)}$, the forward slope of the topological susceptibility in the chiral limit, and show they are of the order of a few percent. Our prior prediction that the polarized DIS structure function $g_1$ is quenched by sphaleron-like topological transitions at small $x$ is unaffected by quark mass effects. Our results illustrate how worldline computations of anomalous processes, in synergy with lattice computations and nonet chiral perturbation theory, can uncover novel nonperturbative features of QCD at the Electron-Ion collider.  
\end{abstract}

\maketitle


\section{Introduction}
In two previous papers~\cite{Tarasov:2020cwl,Tarasov:2021yll}, we addressed the longstanding problem of the role of the chiral anomaly in polarized deeply inelastic scattering (DIS). In the first paper (henceforth Paper I)~\cite{Tarasov:2020cwl}, we employed the worldline formalism~\cite{Strassler:1992zr,Schubert:2001he} in quantum field theory to compute the box diagram in polarized DIS in exact off-forward kinematics. We showed that this computation of the box diagram resulted in a pole that had the structure $\frac{l^\mu}{l^2}F\tilde F$, where $l^\mu = {P'}^\mu - P^\mu$ is the four-momentum transfer between the outgoing and incoming polarized proton, $F^{\mu\nu}$ is the QCD field strength tensor and ${\tilde F}_{\mu\nu} = \frac{1}{2}\epsilon_{\mu\nu\rho\sigma}F^{\rho\sigma}$, is its dual. In particular, we showed that this pole is present in both Bjorken  asymptotics ($Q^2=-q^2 \rightarrow \infty$, $s\rightarrow \infty$, with $x\sim Q^2/s={\rm fixed}$) and Regge asymptotics 
($Q^2= \rm {fixed}\gg \Lambda_{\rm QCD}^2$, with $x\rightarrow 0$ for $s\rightarrow \infty$), where 
$q^2 = -Q^2$ represents the photon virtuality and $s$ the squared center-of-mass energy in the DIS process. Our results were subsequently reproduced in a Feynman diagram analysis of the box diagram~\cite{Bhattacharya:2022xxw,Bhattacharya:2023wvy}.
To relate this quantity to the inclusive cross-section, one has to take the limit $l\rightarrow 0$ which, on the surface, poses a problem. Clearly the pole is spurious but how it is regulated is extremely subtle, touching on several fundamental aspects of quantum field theory. 

The resolution of the apparent problem, and its deeper implications, were addressed at length in our second paper (Paper II)~\cite{Tarasov:2021yll}. Fundamentally, it has to do with the physics of the chiral anomaly and the topology of the QCD vacuum. In other words, as first noted by Veneziano, it follows from the fact that the QCD ``spin puzzle" is related to the so-called $U_A(1)$ puzzle~\cite{Veneziano:1989ei}. The latter is to be anticipated since since the first moment of the isosinglet piece of the polarized inclusive structure function $g_1(x)$, is proportional to the net quark helicity
\begin{equation}
\label{eq:quark-helicity}
    S^\mu \Delta \Sigma = \frac{1}{2M_N}\langle P',S |J_5^\mu|P,S\rangle_{ l\rightarrow 0 } \,,
\end{equation}
where $M_N$ is the nucleon mass and $J_\mu^5$ is the isosinglet axial-vector current, defined as the flavor sum $\sum_f J_{\mu,f}^5$. As is well-known, this current is not conserved due to the chiral anomaly, with $J_{\mu,f}^5$ satisfying the relation
\begin{equation}
\label{eq:anomaly-mass}
    \partial^\mu J_{\mu,f}^5 = 2 im_f \,{\bar q}_f\gamma_5 q_f + \frac{\alpha_S}{2\pi} {\rm Tr}\left(F_{\mu\nu} {\tilde F}^{\mu\nu}\right)\,.
\end{equation}
Here $m_f$ denotes quark mass of a given flavor, $q_f$ (${\bar q}_f$) are the respective quark (antiquark) spinors, and $\alpha_S = \frac{g^2}{4\pi}$, where $g$ is the QCD coupling constant.

In \cite{Tarasov:2021yll}, working in the chiral limit (with $m_f=0$, causing the first term in Eq.~(\ref{eq:anomaly-mass}) to vanish), we recovered the result first obtained by Shore and Veneziano~\cite{Shore:1990zu,Shore:1991dv,Shore:1997tq},\footnote{
In Paper II, we used instead $\Sigma = 2\,\Delta \Sigma$. In other words, our definition here equates $\Delta\Sigma=G_A(0)$, the isosinglet axial vector charge, while in Paper II, $\Sigma=2\,G_A(0)$. We will discuss $G_A$ further later in the paper.}
\begin{equation}
\label{eq:chiral-DeltaSigma}
\Delta \Sigma(Q^2) = \sqrt{\frac{2}{3}}\,\frac{2 N_f}{2M_N}\, g_{\eta_0 NN}\,\sqrt{\chi'(0)}\,.
\end{equation}
In this expression, $M_N$ is the proton's mass, $g_{\eta_0 NN}$ the coupling of a primordial $\eta_0$ meson to the nucleon\footnote{Here $\eta_0$ is the primordial $N_f=2$ projection ($\langle 0|u\bar u+d\bar d|0\rangle$) of the $N_f=3$ ($\langle 0|u\bar u+d\bar d+s\bar s|0\rangle$) primordial $\bar\eta$, which can be understood as the ninth Goldstone corresponding to the coset of the broken chiral group $U(N_f)_{L+R}$ in the $N_c\rightarrow \infty$ limit. Its relation to the physical $\eta^\prime$ and its coupling to the nucleon has been discussed extensively elsewhere~\cite{Shore:1997tq,Narison:1998aq,Narison:2021kny}. Since the nucleon contains only up and down valence quarks, a non-perturbative violation of the Okubo-Zwieg-Iizuka (OZI) rule (corresponding to soft gluon exchanges),  must occur, transforming $\eta_0\rightarrow \bar \eta$. One can alternatively write, as we will do later, $\Delta \Sigma(Q^2) = \frac{2 N_f}{2M_N}\, g_{{\bar\eta} NN}\,\sqrt{\chi'(0)}$, with the $\sqrt{\frac{2}{3}}$ factor absorbed into the coupling of $\bar \eta$ to the nucleon.   For a discussion of the corresponding OZI expansion of vacuum-vacuum connected diagrams, see \cite{Kaiser:2000gs,HerreraSiklody:1996pm,Leutwyler:1996sa} for its implementation in chiral perturbation theory.}, and $\chi'(0) = \frac{\partial}{\partial l^2}\chi$, evaluated in the forward limit $l^2\rightarrow 0$, where $\chi$ is the topological susceptibility of the QCD vacuum. As this striking result shows, the proton's net quark helicity is directly related to the topology of the QCD vacuum\footnote{$\Delta \Sigma$ can also be computed within the framework of instanton models that incorporate this physics~\cite{Forte:1989qq,Forte:1990xb,Zahed:2022wae}.}. Note that the quantity on the r.h.s, while having the right dimensions, is obtained by taking the Fourier transform of a dimension-eight operator, and further, the square root of its slope, which cannot be intuited from the QCD parton model. The Coleman-Witten theorem however tells us that this is unsurprising since the triangle operator is intrinsically non-analytic in the forward limit and therefore not subject to the twist expansion~\cite{Coleman:1980mx}. An alternative intuition is however provided by the Atiyah-Singer index theorem that relates the topological charge to the difference in the helicities of left and right handed quarks~\cite{Bilal:2008qx}.

Shore and Veneziano derived Eq.~(\ref{eq:chiral-DeltaSigma}) by starting from the 
Wess-Zumino action~\cite{Wess:1971yu} for QCD and constructing the simplest effective action consistent with anomalous Ward identities and renormalization group constraints imposed on the composite operators appearing in the effective action. Specifically, one adds to the QCD action, external scalar, pseudoscalar, vector and axial vector sources that couple to the corresponding low energy degrees of freedom of the spontaneously broken theory. Taking functional derivatives with respect to the external sources, one obtains the functional Ward identities that generate $n$-point correlation functions for these low energy degrees of freedom. In other words, Eq.~(\ref{eq:anomaly-mass}) corresponds to only the first in a hierarchy of equations that must be satisfied by correlators of the axial vector current with the low energy scalar and pseudoscalar fields.  

In the worldline approach, one recovers key features of these results by working with the worldline effective action. Specifically, when the Dirac operator couples to scalar, pseudoscalar and axial-vector fields,  besides the usual vector gauge fields, its determinant acquires an imaginary part (the phase of the determinant), that fully captures the nonperturbative physics of the chiral anomaly. A convenient feature of the worldline approach is that it captures both the high $Q^2$, $l^2$ dynamics of spinning partons in background gauge fields and the low $Q^2$, $l^2$ dynamics of broken chiral symmetry in QCD, represented by the additional composite scalar, pseudoscalar and axial-vector fields. Because QCD is an anomaly free theory (both in the chiral limit and for finite quark masses) poles generated by anomaly sensitive operators in perturbative computations must be canceled exactly by identical poles generated by the dynamics of the low energy degrees of freedom~\cite{tHooft:1979rat}. We note that this is fundamentally different from QED where the anomaly pole vanishes for finite lepton mass but is unregulated in the limit of zero mass. The presence of this unregulated pole confirms that massless QED is not a well-defined theory. 

In Paper II, we showed explicitly in the chiral limit that the $\frac{l^\mu}{l^2} F\tilde F $ pole of the anomaly in QCD is canceled\footnote{Without this pole cancellation, QCD in the chiral limit would not be a well-defined theory, just as in the case of massless QED. As a corollary, the absence of such a cancellation for 1-flavor QCD (no $\eta$ !) explains why $N_f=1$ QCD in the chiral limit is anomalous. We thank Mike Creutz for a discussion on this point.} by the pole corresponding to $\bar \eta$-exchange in the $t$ channel, where $t = -l^2 < 0$. Further, we derived from the worldline action corresponding to the imaginary part of the worldline determinant, the Wess-Zumino-Witten term,
\begin{equation}
\label{eq:S-WZW}
    S_{\rm WZW}^{\bar\eta}= -i\frac{\sqrt{2N_f}}{F_{\bar \eta}}\int d^4 x\, {\bar \eta}\,\Omega\,,
\end{equation}
where $\Omega = \frac{\alpha_s}{4\pi}{\rm Tr}\left(F_{\mu\nu} {\tilde F}^{\mu\nu}\right)$ represents the composite field corresponding to the topological charge density, and $F_{\bar \eta}\propto \sqrt{N_c}$ is the ${\bar \eta}$ decay constant. Since this term couples the primordial ${\bar \eta}$ to the topological charge density, the  propagation of the former proceeds via loops of time ordered correlators of the {\it Yang-Mills} topological charge density. These correlators correspond to the Yang-Mills topological susceptibility,  
\begin{equation}
    \chi_{\rm YM}(l^2) = i\int d^4 x\, e^{il\cdot x}\,\langle 0|T\Omega(x)\Omega(0)|0\rangle_{\rm YM}\,.
\end{equation}
The net effect of this loop resummation, to all orders in the expansion parameter $N_f/N_c$ (which is understood to be kept fixed but $\ll 1$), shifts the anomaly pole $l^2\rightarrow l^2-{\tilde m}_{\eta^\prime}^2$, where we obtain the result ${\tilde m}_{\eta^\prime}^2= -\frac{2 N_f}{F_{\bar \eta}^2}\,\chi_{\rm YM}(0)$, which is nothing but the Witten-Veneziano formula~\cite{Witten:1979vv,Veneziano:1979ec} for the $\eta^\prime$ mass in the chiral limit. The latter is of course a remarkable manifestation of the role of topology in hadron physics, especially considering that the $\eta^\prime$ is heavier than the proton! 

The final key step in obtaining Eq.~(\ref{eq:chiral-DeltaSigma}) is the Goldberger-Treiman (G-T) identity that relates the isosinglet axial-vector form factor to the isosinglet pseudoscalar form factor. In the language of generalized parton distributions, these correspond to the expressions $\tilde H$ and $\tilde E$, respectively. We stress that though they correspond to different tensorial structures, the G-T identity allows one to express one in terms of the other~\cite{CSSMQCDSFUKQCD:2021lkf}. As clearly explained in \cite{Shore:2007yn}, G-T identities, and the related Dashen-Gell-Mann-Oakes-Renner (DGMOR) relations, can be understood as direct consequence of the zero-momentum Ward identities in the OZI expansion we alluded to previously. 

We further argued in Paper II that the cancellations that occur for the first moment ($\Delta \Sigma$) of the structure function $g_1$ must also hold for the isosinglet piece of the structure function itself since it too exhibits an identical $\frac{l^\mu}{l^2} F\tilde F$ structure. The same conclusion was also arrived at subsequently in an independent Feynman diagram analysis~\cite{Bhattacharya:2024geo}. Our argument relied on applying simplicity and generality to our worldline results for the box diagram. The $\bar\eta$ pole is independent of kinematics as is the mass of the $\eta^\prime$. The coupling of the $\eta_0$ to the nucleon will however change, and we wrote down an effective action that generalizes $g_{\eta_0 NN}$ to a distribution at small $x_{\rm Bj}$. Further, for the short lived small $x$ configurations in the polarized nucleon, the axion-like propagation of the $\bar \eta$ is no longer in the ground state corresponding to the vacuum topological susceptibility but is influenced by the presence of the large density of color charges characterized by a saturation scale $Q_S$~\cite{Gelis:2010nm}. We showed in \cite{Tarasov:2021yll} that this leads to a drag effect on the propagation of the $\bar \eta$ proportional to the transition rate of sphaleron-like topological transitions. An important consequence is the rapid quenching of $g_1$ at small $x$, which can be measured at the future Electron-Ion Collider (EIC)~\cite{Accardi:2012qut,Aschenauer:2017jsk}.

All of our results were obtained in the chiral limit $m_f=0$. Away from it, as noted, the first term in Eq.~(\ref{eq:anomaly-mass}) will contribute.  How does turning on quark masses modify the results? As lattice QCD studies have shown, the theory (for $N_f\geq 2$) is robust in the chiral limit, albeit with a different phase structure from QCD with physical quark masses~\cite{Brown:1990ev}: one expects all poles to cancel in physical quantities in this limit. 
In the context of $\Delta \Sigma$, the question was addressed previously in the effective action framework in \cite{Narison:1998aq} and it was found that the G-T relation is valid for finite quark masses. While the effect of quark masses is small, quantifying their contribution is nontrivial since quark masses couple the isosinglet ($U_A(1)$) and isooctet sectors ($SU(3)_{L+R}$) of broken chiral symmetry. Our focus in this paper is to address the problem in the worldline framework. In particular, we would like to understand the finite mass effects on the triangle anomaly due to the WZW terms arising from the imaginary part of the worldline effective action. As we will discuss, this leads to nontrivial conceptual generalizations of our prior work; however, as anticipated by \cite{Narison:1998aq}, their quantitative impact on $\Delta \Sigma$ is small. 

We will also address the recent claim in \cite{Castelli:2024eza} that there is no anomaly pole in the forward limit, with the argument being that the pole in the anomaly term in Eq.~(\ref{eq:anomaly-mass}) is exactly canceled by the pole appearing in the mass term\footnote{This discussion was prefigured in early debates more than thirty years ago regarding the so-called ``spin crisis" ~\cite{Carlitz:1988ab,Altarelli:1988nr,Veneziano:1989ei}, nicely summarized in \cite{Jaffe:1989jz}. As pointed out there, the Chern-Simons (CS) current in several papers was rewritten simply, employing perturbation theory, as proportional to the gluon helicity $\Delta G$. However neither the CS current, nor its forward matrix element, are  gauge invariant; further, the perturbative framework does not take into account the constraint that the CS number is integer valued, corresponding to winding numbers that  characterize the topology of the QCD vacuum.}. We will show explicitly that while this result is {\it correct}\footnote{It was first obtained in Adler and Bardeen's seminal work on the anomaly~\cite{Adler:1969er}. Indeed, the very fact that massless QED is anomalous was used by Adler and collaborators to simplify computations demonstrating that the anomaly is robust under radiative corrections~\cite{Adler:1971uvk}. For an important clarification regarding misinterpretations of the Adler-Bardeen theorem, see page 12 of Adler's survey of the topic~\cite{Adler:2004qt}. } in QED, it is 
{\it not correct} in QCD. As we noted previously, massless QED is not a well-defined theory (in the 't Hooft sense~\cite{tHooft:1979rat}), while QCD in the chiral limit is perfectly well-defined. In particular, we will show  in the worldline formalism that this follows from the presence of a chiral condensate whether one is in the chiral limit or not. The result is of course more general: independently of the worldline formalism, it can be deduced in the zero momentum limit of the anomalous Ward identitities~\cite{Shore:2007yn}. The latter, by the Banks-Casher relation~\cite{Banks:1979yr}, is proportion to the density of zero modes of the  Dirac operator; the role of topologically nontrivial configurations in arriving at this relation is discussed in \cite{Leutwyler:1992yt}.

The paper is organized as follows. In Section~\ref{sec:DIS}, we will briefly review the formalism relating inclusive measurements in polarized DIS to the diagram representing the anomalous Axial Vector-Vector-Vector (AVV) triangle operator. The worldline computation of the triangle anomaly is discussed in Section~\ref{sec:worldline-imaginary}. We will rederive the result obtained in Paper I in a generalization of the worldline formalism that also includes the coupling of the Dirac operator to background scalar and pseudoscalar fields, in addition to the axial vector background fields discussed in Paper I. The finite mass result for the AVV anomaly is obtained in Section~\ref{sec:Anomaly-quark-mass}. We first perform the worldline computation for the AVV triangle graph in Section~\ref{sec:AVV-finite-mass}, and then in Section~\ref{sec:PVV-finite-mass}, the analogous computation of the pseudoscalar-Vector-Vector (PVV) triangle diagram. The resulting cancellations between the AVV and PVV triangle in QED, and lessons for QCD, are discussed further in Section~\ref{sec:Anomaly-cancellation}. In Section~\ref{sec:Ward}, we will relate our worldline results to the discussion in the aforementioned papers by Narison, Shore and Veneziano, summarized in \cite{Shore:2007yn} in the language of the Wess-Zumino effective action. Specifically, in Section~\ref{sec:chiral-Ward}, we will derive the chiral Ward identities for the two-point correlators of the axial vector current with isosinglet pseudoscalar fields and with the topological charge density. We will then show in Section~\ref{sec:polology-topology} how ``polology" and topology enable us to extract fundamental nonperturbative information to the $\eta^\prime$ mass, both in and beyond the chiral limit, and likewise obtain results for $\Delta \Sigma$. In Section~\ref{sec:large-N}, we will discuss some of the phenomenology anticipated from large $N_c$ arguments, and connect our results to lattice and sum rule computations of the topological susceptibility in pure Yang-Mills theory and in QCD. In Section~\ref{sec:Summary}, we will provide a summary of our key results, outline some of the consequences, and point to future directions to pursue. 
Some details of the worldline computation are provided in Appendix A. In Appendix B, we provide a short derivation of a result employed in Section~\ref{sec:polology-topology}.

\section{Brief review of polarized DIS and its sensitivity to the chiral anomaly}
\label{sec:DIS}

The $g_1(x_B,Q^2)$ structure function, proportional to the longitudinal spin-spin asymmetry in polarized DIS, can be extracted most generally from the antisymmetric piece of the hadron tensor~\cite{Anselmino:1994gn,Leader:2001gr}, which can be expressed as 
\begin{eqnarray}
\tilde{W}_{\mu\nu}(q, P, S)= \frac{2M_N}{P\cdot q}\epsilon_{\mu\nu\alpha\beta}\, q^\alpha\Big\{ S^\beta g_1(x_B, Q^2) + \Big[S^\beta - \frac{(S\cdot q)P^\beta}{P\cdot q}\Big]g_2(x_B, Q^2)\Big\}\,.
\label{WA}
\end{eqnarray}
Here $M_N$ denotes the proton mass and the totally antisymmetric Levi-Civita tensor $\epsilon_{\mu\nu\alpha\beta}$  is defined with $\epsilon_{0123} = -1$. For a  longitudinally polarized target, $S^\mu(\lambda) \simeq \frac{2{\tilde \lambda}_P}{M_N}P^\mu$, with ${\tilde \lambda}_P = \pm \frac{1}{2}$ representing the proton's helicity; in this case, the $g_2$ structure function does not contribute. 

The hadron tensor itself can be expressed as the imaginary part of the expectation value of the polarization tensor: 
\begin{eqnarray}
W^{\mu\nu}(q, P, S) = \frac{1}{\pi e^2}{\rm Im}\  \int d^4x~ e^{iqx}\langle P,S| \frac{\delta^2 \Gamma[a, A]}{\delta a_\mu(\frac{x}{2})\delta a_\nu(-\frac{x}{2})} |P,S\rangle\,,
\label{TprodEffex}
\end{eqnarray}
 where $\Gamma[a,A]$ is the QED+QCD worldline effective action,  $a_\mu(x)$ denotes the QED electromagnetic field and $A$ is the four-vector denoting the QCD gauge field. Its antisymmetric piece, which appears on the l.h.s of Eq.\,(\ref{WA}), can be written as 
 \begin{eqnarray}
i\tilde{W}^{\mu\nu}(q, P, S) = \frac{ 1 }{ 2 \pi e^2}\,{\rm Im}\ \int d^4x \,e^{-iqx} 
 \int \frac{d^4k_1}{(2\pi)^4} \int \frac{d^4k_4}{(2\pi)^4} e^{-ik_1 \frac{x}{2}} e^{ik_4 \frac{x}{2} }\langle P,S| {\tilde{\Gamma}}^{\mu\nu}_A[k_1, k_4] |P,S\rangle\,.
\label{WModFu}
\end{eqnarray}
Here ${\tilde{\Gamma}}^{\mu\nu}_A[k_1, k_4] \equiv {\tilde{\Gamma}}^{\mu\nu}[k_1, k_4] - (\mu\leftrightarrow\nu)$, with 
\begin{eqnarray}
{\tilde{\Gamma}}^{\mu\nu}[k_1, k_4] \equiv \int d^4z_1 d^4z_4 \frac{\delta^2 \Gamma[a, A]}{\delta a_\mu(z_1)\delta a_\nu(z_4)}|_{a=0} \,e^{ik_1 z_1} e^{ik_4 z_4}\,,
\label{secder}
\end{eqnarray}
where $k_1$ and $k_4$ denote the incoming photon four-momenta. Because the r.h.s of Eq.\,(\ref{WModFu}) corresponds to the same in-out ground state of the proton,  $k_1=-k_4=-q$ in the forward limit. To extract the infrared pole of the anomaly, it is important to work in exact kinematics and keep the incoming photon momenta distinct in computing the off-forward matrix element $\langle P'|\dots|P\rangle$ in Eq.\,(\ref{TprodEffex}), with $P'-P\equiv l$ and $t=l^2$, and then subsequently take $t\to0$ in the final expression~\cite{Tarasov:2020cwl}. 

 \begin{figure}[htb]
 \begin{center}
\includegraphics[width=160mm]{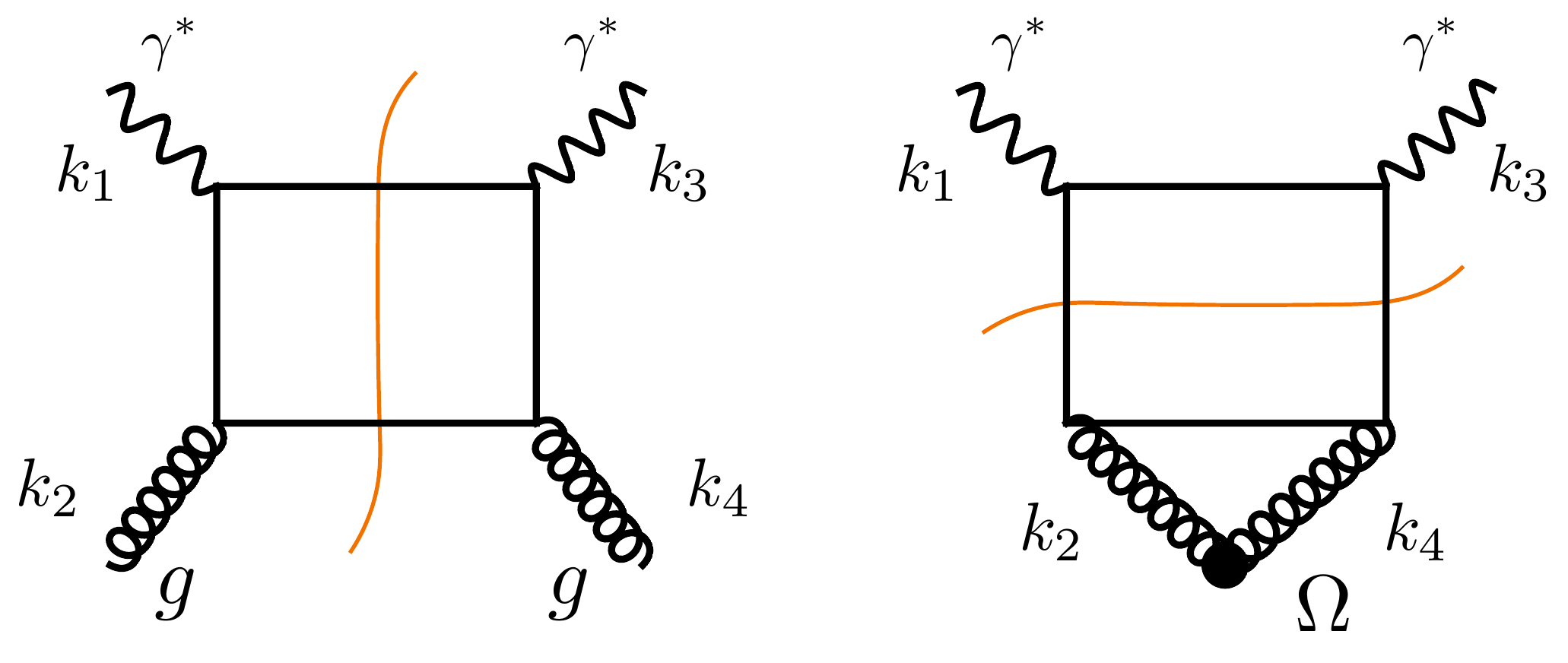}
 \end{center}
\caption{The box diagram in polarized DIS. Left: The figure depicts the imaginary part of the box, corresponding to a contribution to $g_1(x_B,Q^2)$, which can be understood as the virtual photon scattering off a polarized gluon. Right: The contribution to the imaginary part of box from the horizontal cut. This graph gives the leading ``anomaly" contribution $\propto \lim_{l\rightarrow 0}\frac{l^\mu}{l^2}F\tilde F$~\cite{Tarasov:2020cwl,Tarasov:2021yll,Bhattacharya:2022xxw,Bhattacharya:2023wvy,Bhattacharya:2024geo}, with the background gluon fields fusing into the topological charge density $\Omega=F\tilde F$. Since the large four-momentum $k_1$ from the incoming virtual photon is balanced by the four-momentum $k_4$ of the outgoing photon, the contribution of this graph is large for arbitrary $Q^2$. }
\label{fig:box-diagram}
 \end{figure}

In the worldline formalism, to one loop accuracy, the  QED+QCD effective action is given by~\cite{Schubert:2001he}
\begin{eqnarray}
&&\Gamma [a, A] = -\frac{1}{2} \int^T_0 \frac{dT}{T} {\rm Tr_c} \int \mathcal{D}x \int \mathcal{D} \psi \exp\Big\{-\int^T_0 d\tau \Big(\frac{1}{4} \dot{x}^2 + \frac{1}{2}\psi_\mu\dot{\psi}^\mu + ig\dot{x}^\mu (A_\mu + a_\mu) - ig \psi^\mu \psi^\nu F_{\mu\nu}(A+a)\Big)\Big\}\,,
\nonumber\\
\label{MLag}
\end{eqnarray}
where $x^\mu(\tau)$  and $\psi^\mu(\tau)$ are, respectively, 0+1-dimensional scalar coordinate and Grassmann variables coupled to the background electromagnetic ($a_\mu$) and gluon ($A_\mu$) fields. Note that the scalar functional integral has periodic (P) boundary conditions while the Grassmannian functional integral has anti-periodic (AP) boundary conditions. Further, we can express ${\tilde{\Gamma}}^{\mu\nu}_A[k_1, k_4]$ on 
the r.h.s of Eq.\,(\ref{WModFu}) as 
\begin{eqnarray}
&&{\tilde \Gamma}^{\mu\nu}_A[k_1, k_4] = \int \frac{d^4k_2}{(2\pi)^4} \int \frac{d^4k_4}{(2\pi)^4}~
\Gamma^{\mu\nu\alpha\beta}_A[k_1, k_4, k_2, k_4]~ {\rm Tr_c}({\tilde A}_\alpha(k_2) {\tilde A}_\beta(k_4))\,,
\label{Hadgamma}
\end{eqnarray}
with $ {\tilde A}$ denoting Fourier transforms of the background gauge fields (the trace is over their color degrees of freedom) and $\Gamma^{\mu\nu\alpha\beta}_A[k_1, k_4, k_2, k_4]$ 
is the box diagram of polarized DIS we discussed at length in \cite{Tarasov:2020cwl,Tarasov:2021yll}. 

The box diagram is illustrated in Fig.~\ref{fig:box-diagram}. The left figure shows the ``usual" imaginary part with the vertical line depicting the cut, and corresponds to an interpretation in terms of photon-(polarized) gluon fusion. However there is another less familiar ``horizontal" cut in polarized DIS, depicted in the right figure\footnote{For a nice discussion, see \cite{Vasquez-Mozo}.}. Here the two gluons can be understood to fuse into a composite field, the topological charge density $\Omega$, we will discuss at length in this paper. 
This cut corresponds to the triangle anomaly, and it is the nontrivial coupling of $\Omega$ to the proton that gives the leading contribution to the polarized DIS cross-section in the forward limit. 

The worldline representation of the box diagram provides useful intuition by mapping the ordering of momentum labels of the four vertices to corresponding proper times. The AVV triangle limit of the box diagram in Bjorken asymptotics can be understood as the ``pinching" $\tau_1\rightarrow \tau_4$ when $k_1\rightarrow k_4$ (with corrections of order $1/Q$), and likewise in Regge asymptotics as 
$\tau_2\rightarrow \tau_4$ when $k_2\rightarrow k_4$ of the background gauge fields, with corrections of order $O(\frac{x_B}{x})$, as $x_B\rightarrow 0$ for  gluon momenta carrying a finite but small fraction $x$ of the hadron's large ``+" momentum. The latter corresponds to an inverted AVV triangle which too is sensitive to the anomaly. 

An explicit computation of the box diagram in the chiral limit for exact $k_1,\cdots, k_4$ kinematics in \cite{Tarasov:2020cwl} employing the worldline formalism,  subsequent insertion of  the result in Eqs.~(\ref{Hadgamma}), (\ref{WModFu}), and comparison of this result to Eq.~(\ref{WA}), demonstrated that the leading contribution to $g_1$ in both Bjorken and Regge limits is given by 
\begin{eqnarray}
\label{eq:g1-Bj-Regge}
 S^\mu g_1(x_B, Q^2) = \sum_f e_f^2 \frac{\alpha_s}{i\pi M_N}
 \int^1_{x_B} \frac{dx}{x} 
~ \Big( 1  - \frac{x_B}{x} \Big) \int \frac{d\xi}{2\pi} e^{-i\xi x }\lim_{l_\mu \to 0} \frac{ l^\mu }{l^2  }\langle P^\prime,S| {\rm Tr_c} F_{\alpha\beta}(\xi n) \tilde{F}^{\alpha\beta}(0) |P,S\rangle\,.
\end{eqnarray}
Here $\alpha_s$ is the QCD coupling, the sum is over $N_f$  quark  flavors\footnote{We will assume these to be massless and $N_f=3$ for our discussion.} with electric charge $e_f$.  The structure of the r.h.s is dominated by a nonlocal generalization of the AVV triangle anomaly governed by the topological charge density $\Omega=\frac{\alpha_s}{4\pi} {\rm Tr}\left(F {\tilde F}\right)$, with $\tilde{F}_{\mu\nu} = \frac{1}{2}\,\epsilon_{\mu\nu\rho\sigma}F^{\rho\sigma}$.  Indeed, the first moment of the above equation gives a contribution to the net quark helicity 
\begin{eqnarray}
   S^\mu \Delta \Sigma \propto S^\mu \int dx_{B}\, g_1(x_B,Q^2)\equiv \sum_f e_f^2 \frac{\alpha_s}{2i\pi M_N} 
    \lim_{l_\mu \to 0} \frac{ l^\mu }{l^2  }\langle P^\prime,S| {\rm Tr} (F {\tilde F}) |P,S\rangle\,.
\end{eqnarray}
This expression on the surface presents a puzzle since on the l.h.s one has $S^\mu$ while on the r.h.s one has 
$l^\mu$, and further, in the forward limit
\begin{eqnarray}
    \frac{1}{2M_N}\langle P,S|J_5^\mu|P,S\rangle= S^\mu \Delta \Sigma \equiv S^\mu \,G_A(0)\,,
\end{eqnarray}
where the first equality was noted in Eq.~(\ref{eq:quark-helicity}), and $G_A$ in the second equality is the isosinglet axial vector form factor of the proton.
As we alluded to earlier, the resolution of this puzzle is that from the anomaly equation and the Dirac equation, one obtains a Goldberger-Treiman (G-T) relation\footnote{In the language of generalized parton distributions~\cite{CSSMQCDSFUKQCD:2021lkf}, this relates the functions $\tilde H$ and $\tilde E$ in the forward limit.}
\begin{eqnarray}
\label{eq:G-T}
    G_A(0) = \frac{\sqrt{2 N_f}}{2 M_N}\,F_{\bar \eta}\, g_{\bar \eta NN}\,,
\end{eqnarray}
 relating the coefficients of $S^\mu$ and $l^\mu$ in the axial vector and pseudoscalar sectors, respectively. On the r.h.s of this expression, $F_{\bar \eta}$ is the decay constant of a primordial $\bar \eta$ which, as discussed earlier, can be understood as the ninth ``prodigal" Goldstone boson of broken $U(3)_{L+R}$ in the large $N_c$ limit.  As also noted earlier, this primordial $\bar \eta$ is dressed by the QCD topological susceptibility to become the physical $\eta^\prime$, with a mass greater than that of the proton. Thus Eq.~(\ref{eq:G-T}) fundamentally relates the physics of the proton's net quark helicity (for arbitrarily large $Q^2$ !) to that $U_A(1)$ and chiral symmetry breaking, and to the topology of the QCD vacuum. 

 \begin{figure}[htb]
 \begin{center}
\includegraphics[width=80mm]{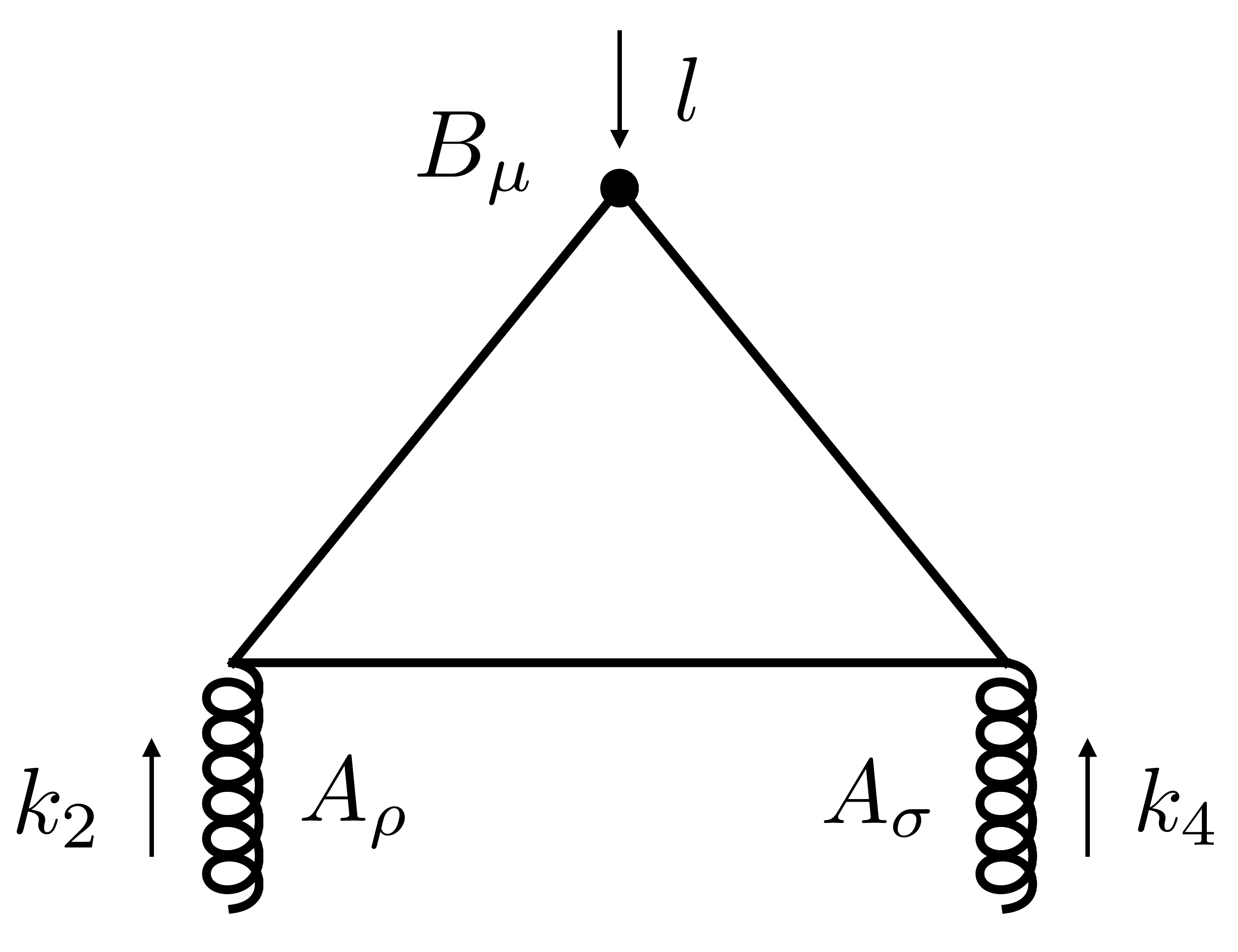}
 \end{center}
\caption{Depiction of the triangle AVV vertex function $\Gamma^{\kappa\alpha\beta}_5[l, k_2, k_4]$ in 
Eq.~(\ref{trianglFour}), with $l=k_2+k_4$, obtained by taking the functional derivative of the worldline effective action with respect the external axial vector field $B_\mu$ and the 
vector gauge field $A_\mu$.}
\label{fig:AVV-triangle}
 \end{figure}

As we stated in the introduction, our purpose here is to articulate what happens to this picture when one goes away from the chiral limit. For this purpose, we will focus first on the AVV triangle operator which, as we argued above, provides the leading contribution to the box diagram in Bjorken and Regge asymptotics. 
In order to obtain a full description of the triangle anomaly, we will replace the worldline efffective action   $\Gamma[A,a]$ that appears in Eq.~(\ref{secder}) with $\Gamma\rightarrow \cal W$, where the latter is the effective action in the presence of additional axial vector, scalar and pseudoscalar fields. As we will discuss shortly,  $\cal W$ in the presence of axial vector and pseudoscalar fields acquires an imaginary piece $\cal W_I$. Since $J_5^\mu$ couples directly to this axial vector background field, which we will denote henceforth as $B^\mu$, its expectation value in the proton can be obtained by taking  the functional derivative of ${\cal W}_I[A, B]$ with respect to $B$ and then setting the latter equal to zero: 
\begin{eqnarray}
\label{eq:functional-AVV}
\langle P^\prime,S| J^\kappa_5 |P,S\rangle = \int d^4y\, \frac{\partial {\cal W}_I[A, B]}{\partial B_{\kappa}(y)}\Big|_{B_\kappa=0} e^{ily} \equiv \Gamma^\kappa_5[l]\,,
\end{eqnarray}
where  
\begin{eqnarray}
&&\Gamma^\kappa_5[l] = \int \frac{d^4k_2}{(2\pi)^4} \int \frac{d^4k_4}{(2\pi)^4} ~\Gamma^{\kappa\alpha\beta}_5[l,k_2,k_4]~{\rm Tr_c} A_\alpha(k_2)  A_\beta(k_4) \,.
\label{trianglFour}
\end{eqnarray}
The function $\Gamma^{\kappa\alpha\beta}_5[l,k_2,k_4]$ in this expression is the AVV vertex function (colloquially known as the ``anomaly") shown in Fig. \ref{fig:AVV-triangle}. 

An explicit derivation of this quantity in the chiral limit can be found in Appendix C of Paper I (see also \cite{Schubert:2001he}), to obtain the well-known result,
\begin{eqnarray}
\label{eq:VVA1}
&&\Gamma^{\kappa\alpha\beta}_5[l, k_2, k_4] =  \frac{ 1 }{2\pi^2}\frac{k^\kappa_{2} + k^\kappa_{4}}{ (k_2 + k_4)^2 } ~ \epsilon^{\alpha\sigma\beta\lambda} k_{2\sigma} k_{4\lambda}  (2\pi)^4 \delta^4(l + k_2 + k_4)\,.
\end{eqnarray}
Substituting the anomaly back into Eq.\,(\ref{trianglFour}), we obtain in the chiral limit, 
\begin{eqnarray}
&&\Gamma^\kappa_5[l] = \frac{ 1 }{4\pi^2} \frac{l^\kappa}{ l^2 } \int \frac{d^4k_2}{(2\pi)^4} \int \frac{d^4k_4}{(2\pi)^4} ~ {\rm Tr_c} F_{\alpha\beta}(k_2)  \tilde{F}^{\alpha\beta}(k_4)~(2\pi)^4 \delta^4(l + k_2 + k_4)\,,
\end{eqnarray}
the operator structure of which, up to kinematic factors, can be generalized to arrive at the principal result of Paper I--given here in Eq.\,(\ref{eq:g1-Bj-Regge}). 

In the next section, we will discuss at length the imaginary part of the worldline effective action in the presence of additional external scalar and pseudoscalar background fields. This formalism, in subsequent sections, will be used to derive the AVV triangle for finite mass, as well as the PVV triangle that only appears for finite quark mass, and to discuss their interplay, first with QED in mind, and subsequently, the qualitative and quantitative differences that appear in QCD.

\section{Imaginary part of the worldline effective action}
\label{sec:worldline-imaginary}
\begin{figure}[htb]
 \begin{center}
\includegraphics[width=90mm]{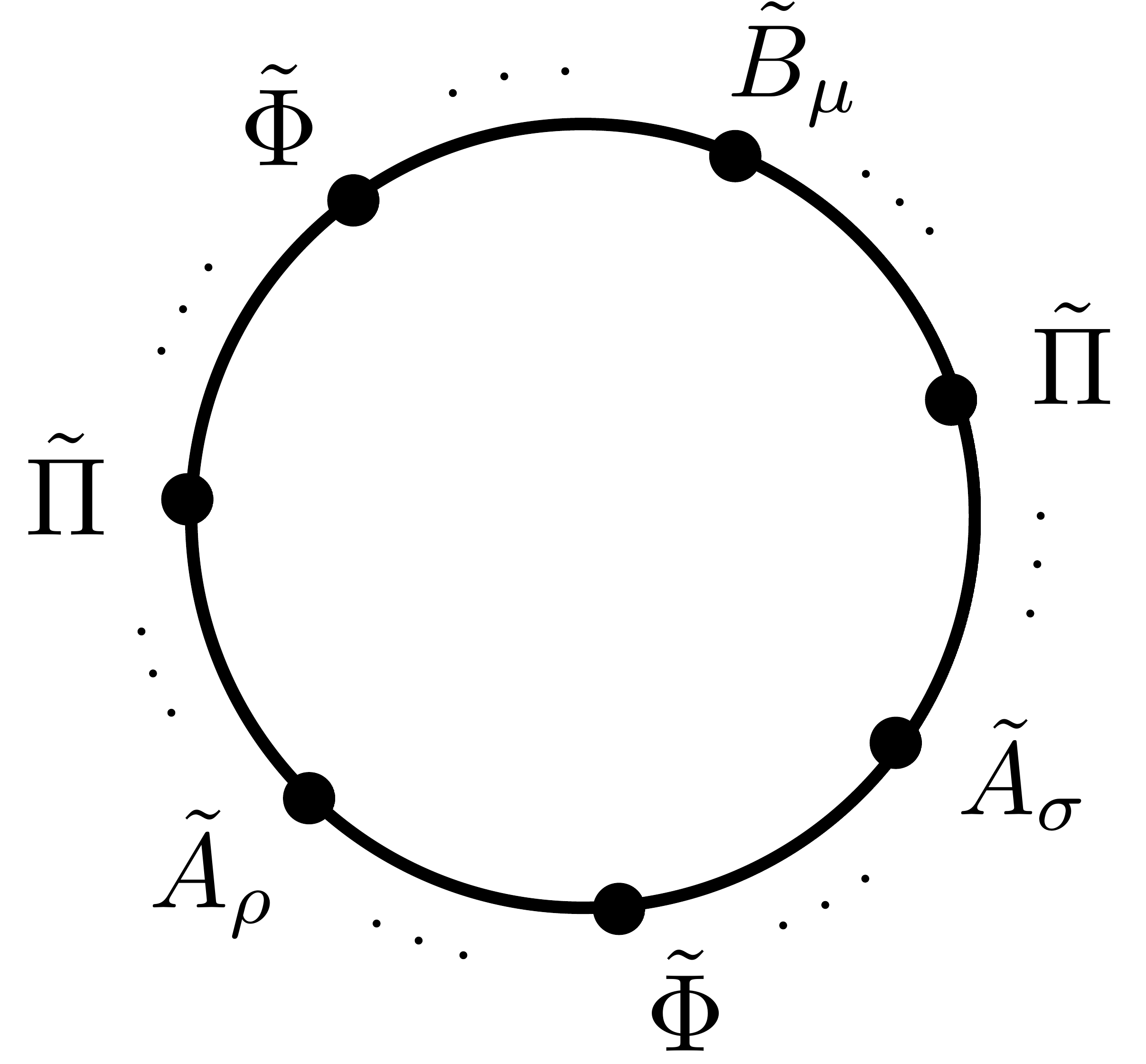}
 \end{center}
\caption{The contribution of the fermion determinant to the worldline effective action containing arbitrary numbers of insertions of background scalar, pseudoscalar, vector and axial vector fields.}
\label{fig:effective-action}
 \end{figure}

The effective action $W$ for a flavor multiplet of Dirac fermions coupled to matrix valued  scalar, pseudoscalar, vector and axial-vector background fields, can be expressed as
\begin{eqnarray}
&&e^{iW[\Phi,\Pi,A,B]} = \int \mathcal{D} \bar{\Psi} \mathcal{D}\Psi e^{iS_{\rm fermion}[{\bar \Psi},\Phi,\Pi,A,B,\Psi]}\,,
\label{eq:WDirac}
\end{eqnarray}
where the action for the theory is given by
\begin{eqnarray}
\label{eq:Dirac}
S_{\rm fermion}[{\bar \Psi},\Phi,\Pi,A,B,\Psi] = \int d^4 x\,{\bar \Psi}^I\left[i\slashed{\partial}-\Phi+i\gamma^5 \Pi + \slashed{A} + \gamma^5 \slashed{B} \right]^{IJ}\Psi^J\,.
\end{eqnarray}
Here $\Phi$, $\Pi$, $A$ and $B$ are, respectively, the scalar, pseudoscalar, vector and axial-vector fields, whose couplings to the higher frequency fermion fields ${\bar \Psi}$, $\Psi$, are absorbed into field definitions. The superscripts $I$ and $J$ denote the internal quantum numbers of the fermion multiplet as well as those of the matrix valued ``source" fields. The perturbative expression for this effective action corresponding to a quark loop in the background fields is 
\begin{eqnarray}
{\cal W}[\Phi,\Pi,A,B] &=& \sum_{n=1}^\infty \frac{1}{n} \int \frac{d^4 k_1}{(2\pi)^4}\cdots \frac{d^4 k_n}{(2\pi)^4}\, \delta^{(4)}(k_1+\cdots +k_n)\int \frac{d^4 q}{(2\pi)^4}\, \frac{\slashed{q}+ i m}{q^2+m^2}\nonumber \\
&\times&\left(i{\tilde \Phi}_1 + \gamma_5\, {\tilde \Pi}_1+ {\tilde {\slashed{A}}}_1 + \gamma_5 {\tilde {\slashed{B}}}_1\right)\cdots \frac{\slashed{q}-\slashed{k_1}-\cdots\slashed{k_n}+im}{\left(\left(q-k_1\cdots k_n\right)^2 + m^2\right)} \left(i{\tilde \Phi}_n + \gamma_5\, {\tilde \Pi}_n+ {\tilde {\slashed{A}}}_n+ \gamma_5{\tilde {\slashed{B}}}_n \right)\,.
\end{eqnarray}
This expression, illustrated in Fig.~\ref{fig:effective-action}, exhibits the useful mnemonic that even numbers of insertions of pseudoscalar and axial vector fields contribute to the real part of ${\cal W}$ while odd numbers contribute to the imaginary part. The map between the worldline and Feynman diagram computations in the presence of these external fields was discussed previously\footnote{This perturbative expansion sets up the treatment of the so-called consistent anomaly in the worldline/heat kernel framework. For a detailed discussion of its relation to the covariant anomaly of Bardeen and Zumino~\cite{Bardeen:1984pm}, see \cite{Alvarez-Gaume:1984zlq}. } in \cite{DHoker:1995uyv,Mondragon:1995ab,Schubert:2001he}, and in the DIS context in \cite{Tarasov:2019rfp,Tarasov:2020cwl,Tarasov:2021yll}. 

To discuss the heat kernel regularized effective action, we need to consider the Euclidean  fermion effective action in the presence of the external sources, 
\begin{equation}
    -{\cal W}[A,B,\Phi,\Pi] = {\rm Ln}\,{\rm Det }\,\left[{\cal D}\right]\,,
\end{equation}
with the Dirac operator, 
\begin{equation}
    {\cal D} = \slashed{p} -i\Phi(x) - \gamma_5\,\Pi\,-\slashed{A}-\gamma_5 \slashed{B}\,.
    \label{eq:Dirac-operator}
\end{equation}
This effective action can be split into real and 
imaginary parts, with~\cite{DHoker:1995aat} 
\begin{eqnarray}
    {\cal W}_R = - \frac{1}{2} {\rm Ln} \left({\cal D}^\dagger {\cal D}\right)\,\,\,;\,\,\,
    {\cal W}_I = \frac{1}{2}{\rm Arg}\, {\rm Det}\left({\cal D}^2\right)\,.
    \label{eq:Real-Imaginary-action}
\end{eqnarray}
Since the anomaly is sensitive to the imaginary part of the effective action, we will primarily focus on the latter in the rest of this paper\footnote{The derivation of the real part of the effective action in  the presence of sources is discussed at length in \cite{DHoker:1995aat,DHoker:1995uyv}.} but will return to the real part in Section~\ref{sec:Ward}.

Using a clever trick invented in \cite{DHoker:1995aat,DHoker:1995uyv}, one can also reexpress the imaginary part of the effective action, corresponding to {\it the phase of the Dirac determinant},  as a worldline path integral\footnote{In the ``doubling" framework of \cite{DHoker:1995aat,DHoker:1995uyv}, the $4\times 4$ gamma matrices are replaced with the $8\times 8$ matrices,
\begin{eqnarray}
\Gamma_\mu=  \begin{pmatrix} 0 & \gamma_\mu\\ \gamma_\mu & 0
  \end{pmatrix}\,\,\,;\,\,\, \Gamma_5 = \begin{pmatrix} 0 & \gamma_5\\ \gamma_5 & 0
  \end{pmatrix}
  \,\,\,;\,\,\,
 \Gamma_6= \begin{pmatrix} 0 & i I\\ -i I & 0
  \end{pmatrix}\,,
\end{eqnarray}
where $I$ is the $4\times 4$ unit matrix and the six Hermitean $\Gamma$-matrices satisfy 
$\left\{\Gamma_A,\Gamma_B\right\}= 2\delta_{AB} \,I_{8\times 8}$. This alternative formulation does not modify the results. Its primary purpose is to obtain a manifestly positive definite heat kernel operator that can be employed to construct worldline 0+1-dimensional path integrals for bosonic and Grassmannian worldlines.}
\begin{eqnarray}
\label{eq:Imag-worldline}
W_\mathcal{I} = - \frac{i\mathcal{E}}{64}\int^1_{-1}d\alpha \int^\infty_0 dT \mathcal{N} \int_P \mathcal{D}x \mathcal{D}\psi~ {\rm tr} ~\chi \bar{\omega}(0) \exp\Big[-\int^T_0 d\tau \mathcal{L}_{(\alpha)}(\tau)\Big]\,,
\end{eqnarray}
where $\mathcal{N}$ is a $T$-dependent normalization factor, the Grassmanian path integral measure\footnote{The field $\psi_6$ corresponding to $\Gamma_6$ in the $8\times 8$  representation decouples since it does not correspond to a dynamical field in the worldline effective action.} $\mathcal{D}\psi = \mathcal{D}\psi_\mu \mathcal{D}\psi_5$ (with $\mu=1,\cdots,4$). The corresponding worldline Lagrangian is 
\begin{eqnarray}
\mathcal{L}_{(\alpha)}(\tau) = \mathcal{L}(\tau)\Big|_{\Phi\to \alpha\Phi, B\to \alpha B}\,,
\end{eqnarray}
where the {\it usual} worldline Lagrangian corresponding to the real part of the effective action is
\begin{eqnarray}
\mathcal{L}(\tau) = \frac{\dot{x}^2}{2\mathcal{E}} + \frac{1}{2}\psi\dot{\psi} - i \dot{x}^\mu \mathcal{A}_\mu + + \frac{i\mathcal{E}}{2} \psi^\mu \psi^\nu \mathcal{F}_{\mu\nu}+\frac{\mathcal{E}}{2}\mathcal{H}^2 + i\mathcal{E}\psi^\mu\psi_5 \mathcal{D}_\mu \mathcal{H} \,.
\end{eqnarray}
In this expression, $x(\tau)$ and $\psi(\tau)$ are the boson and fermion worldlines parametrized by the proper time $\tau$, and the first two terms in the action correspond to their kinetic terms (with $\mathcal{E}$, the vierbein). The next two terms are the ``Wilson line"  and ``spin precession" phases representing the phase rotation and precession of the worldlines in the background gauge field $\cal A$, and last two terms are their counterparts in the background field $\cal H$, with these fields being defined as 
\begin{eqnarray}
\label{eq:combo-gauge-scalar}
\mathcal{A}_\mu \equiv \begin{pmatrix} A^L_\mu & 0 \\ 0 & A^R_\mu \end{pmatrix} = \begin{pmatrix} A_\mu + B_\mu & 0 \\ 0 & A_\mu - B_\mu \end{pmatrix}\qquad;\qquad
\mathcal{H} \equiv \begin{pmatrix} 0 & iH \\ -iH^\dag & 0 \end{pmatrix} = \begin{pmatrix} 0 & i \Phi + \Pi \\ -i \Phi + \Pi & 0 \end{pmatrix}\,.
\end{eqnarray}
It is remarkable that the worldline Lagrangian can be expressed in this representation as a chiral doublet of 
vector and axial vector background gauge fields. The corresponding phase of the Dirac determinant in 
Eq.~(\ref{eq:Imag-worldline}) is manifestly chirally symmetric for $\alpha=\pm 1$, but breaks chiral symmetry for 
$-1<\alpha < 1$. This reflects the well-known fact that it is not possible to achieve a robust heat kernel representation of the phase and maintain chiral symmetry simultaneously -- in other words, the phase generates the chiral anomaly\footnote{This ``$\alpha$-representation" of the phase of the fermion determinant is strongly reminiscent of the so-called $\eta$-invariant parametrization of the 5-D Dirac operator in chiral gauge theories~\cite{Alvarez-Gaume:1985jck,Ball:1985tya}.}.

Further, in Eq.~(\ref{eq:Imag-worldline}), we define 
$\chi \equiv \begin{pmatrix} I_{N_f} & 0 \\ 0  & -I_{N_f} \end{pmatrix}$, which only acts on the internal flavor degrees of freedom, and
\begin{eqnarray}
&&\bar{\omega} = 2\psi^\mu\Big( -\frac{2i\dot{x}_\mu}{\mathcal{E}}(\mathcal{H} - \mathcal{H}^c) - \{(\mathcal{A}_\mu - \mathcal{A}^c_\mu), (\alpha_+\mathcal{H} + \alpha_-\mathcal{H}^c)\}\Big)
\nonumber\\
&&+ 2\psi_5 \Big(-i\partial^\mu(\mathcal{A}_\mu - \mathcal{A}^c_\mu) + [\mathcal{A}^\mu, \mathcal{A}^c_\mu] + [\mathcal{H}, \mathcal{H}^c]\Big) + 2\psi^\mu\psi^\nu\psi_5 \frac{4i\dot{x}_\mu}{\mathcal{E}}(\mathcal{A}_\nu - \mathcal{A}^c_\nu)\,,
\end{eqnarray}
where $\alpha_\pm \equiv \frac{1\pm\alpha}{2}$. The conjugate fields ${\cal A}_\mu^c$ and ${\cal H}^c$ are obtained by replacing $A_\mu^{L,R}\leftrightarrow A_\mu^{R,L}$, and likewise, $iH\leftrightarrow -iH^\dagger$, in 
${\cal A}_\mu$ and ${\cal H}$, respectively.

We will be taking products of the ``Jacobian" $\chi{\bar \omega}$ and terms in the expansion of the Lagrangian in Eq.~(\ref{eq:Imag-worldline}), so it will be useful to write these, respectively, in their $2\times2$ chiral doublet matrix form\footnote{In writing these expressions, we are focusing on the flavor singlet $U_A(1)$ sector alone and not considering the other flavor-octet internal degrees of freedom. As discussed in \cite{DHoker:1995uyv}, these can be straightforwardly accommodated by introducing additional Grassmann fields, which we have exploited in other contexts~\cite{Mueller:2019gjj,Mueller:2019qqj}. While the full $U(3)_{L+R}$ flavor structure will be important for the description of the relevant low energy dynamics, it is sufficient for our purpose here to focus on the flavor-singlet structure of the worldline Lagrangian. }: 
\begin{eqnarray}
\label{eq:matrix-Jacobian}
&&\chi\bar{\omega}
= - 4 \psi_5 \Big( i \partial^\mu B_\mu + [A^\mu, B_\mu] - i\{\Phi, \Pi\} - \frac{4 i }{\mathcal{E}} \psi^\mu\psi^\nu \dot{x}_\mu B_\nu\Big)\begin{pmatrix} 1 & 0 \\  0 & 1 \end{pmatrix}
+ 4 \psi^\mu ( \frac{ 2 }{\mathcal{E}} \dot{x}_\mu \Phi - [ B_\mu, \Pi] ) \begin{pmatrix} 0 & 1 \\ 1 & 0 \end{pmatrix}\nonumber\\
&&- 4 i\alpha \psi^\mu [ B_\mu, \Phi ] \begin{pmatrix} 0 & 1 \\  -1 & 0 \end{pmatrix}\,,
\end{eqnarray}
and 
\begin{eqnarray}
\label{eq:matrix-Lagrangian}
&&\frac{\delta \mathcal{L}_{(\alpha)}(\tau_1)}{\delta B_\mu(x)} \Big|^{\rm singlet} =
 \alpha \Big( - i \dot{x}^\mu_1 + \frac{i\mathcal{E}}{2} 2 \psi^\nu_1 \psi^\mu_1  \partial^1_\nu \Big) \delta(x_1 - x)\begin{pmatrix} 1 & 0 \\ 0 & - 1 \end{pmatrix} 
\nonumber\\
&&+ \alpha \mathcal{E}\psi^\mu_1 \psi^5_1 d_{000} \Pi_1 \delta(x_1 - x) \begin{pmatrix} 0 & 1 \\ -1 & 0 \end{pmatrix} 
+ i \alpha^2 \mathcal{E}\psi^\mu_1 \psi^5_1 d_{000} \Phi_1 \delta(x_1 - x) \begin{pmatrix} 0 & 1 \\ 1 & 0 \end{pmatrix}\,.
\end{eqnarray}
The ``1" subscripts here refer to the proper time variable-for example, $x_1 \equiv x(\tau_1)$.  We will work extensively with these expressions in the following sections.

 \section{Computation of the triangle anomaly for finite quark mass}
 \label{sec:Anomaly-quark-mass}

We will now use the formalism in the previous section to compute the AVV triangle for the case of finite quark masses as well as the PVV triangle which of course only appears away from the chiral limit. 
Our analysis is simplified by focusing on the flavor singlet sector. In this case, we can remove all commutators of the internal flavor degrees of freedom\footnote{Since the low energy degrees of freedom are color singlets, its only the color singlet projection of the Dirac operator that couples to these degrees of freedom and to the two virtual photons.}. Eq.~(\ref{eq:matrix-Jacobian}) then simplifies to 
\begin{eqnarray}
\label{eq:matrix-Jacobian-singlet}
&&\chi\bar{\omega}(0)\Big|^{\rm singlet}
= - 4 \psi^5_0 \Big( i \partial^\mu_0 B_{0\mu} - \frac{4 i }{\mathcal{E}} \psi^\mu_0\psi^\nu_0 \dot{x}_{0\mu} B_{0\nu} - i\{\Phi_0, \Pi_0\} \Big)\begin{pmatrix} 1 & 0 \\  0 & 1 \end{pmatrix}
+ 4 \psi^\mu_0 \frac{ 2 }{\mathcal{E}} \dot{x}_{0\mu} \Phi_0 \begin{pmatrix} 0 & 1 \\ 1 & 0 \end{pmatrix}\,.
\end{eqnarray}
Here the ``0" in parenthesis on the l.h.s and in subscripts on the r.h.s denotes the proper time at which this  Jacobian is evaluated. Likewise, the Lagrangian in Eq.~(\ref{eq:matrix-Lagrangian}) simplifies to  
\begin{eqnarray}
\label{eq:matrix-Lagrangian-singlet}
&&\mathcal{L}_{(\alpha)}(\tau)\Big|^{\rm singlet} = \Big( \frac{\dot{x}^2}{2\mathcal{E}} + \frac{1}{2}\psi\dot{\psi} - i \dot{x}^\mu A_\mu + \frac{i\mathcal{E}}{2}\psi^\mu \psi^\nu (\partial_\mu A_\nu - \partial_\nu A_\mu ) +  \frac{\mathcal{E}}{2} \alpha^2 \Phi^2 + \frac{\mathcal{E}}{2} \Pi^2 \Big) \begin{pmatrix} 1 & 0 \\  0 & 1 \end{pmatrix}
\nonumber\\
&&+ \Big( - i \alpha \dot{x}^\mu B_\mu + \alpha \frac{i\mathcal{E}}{2}\psi^\mu \psi^\nu  ( \partial_\mu B_\nu -  \partial_\nu B_\mu) \Big) \begin{pmatrix} 1 & 0 \\ 0 & - 1 \end{pmatrix} 
\nonumber\\
&&- \alpha \mathcal{E}\psi^\mu \psi_5 \Big(  \partial_\mu \Phi - \{B_\mu, \Pi\} \Big)\begin{pmatrix} 0 & 1 \\ -1 & 0 \end{pmatrix} 
+ i\mathcal{E}\psi^\mu \psi_5 \Big(\partial_\mu \Pi + \alpha^2 \{B_\mu, \Phi\}\Big) \begin{pmatrix} 0 & 1 \\ 1 & 0 \end{pmatrix}\,.
\end{eqnarray}

\subsection{The AVV triangle at finite quark mass}
\label{sec:AVV-finite-mass}
Recalling Eq.~(\ref{eq:functional-AVV}), we need to take the functional derivative of the imaginary part of the effective action with respect to $B^\mu$, which gives 
\begin{eqnarray}
&&\frac{\delta W_\mathcal{I}}{\delta B_\mu(x)} = - \frac{i\mathcal{E}}{64}\int^1_{-1}d\alpha \int^\infty_0 dT \mathcal{N} \int_P \mathcal{D}x \mathcal{D}\psi~ \exp\Big[-\int^T_0 d\tau \mathcal{L}_{(\alpha)}(\tau)\Big]
\Big[ {\rm tr} ~\chi \frac{\delta \bar{\omega}(0) }{\delta B_\mu(x)} -\int^T_0 d\tau_1 ~ {\rm tr} ~\chi \bar{\omega}(0) \frac{\delta \mathcal{L}_{(\alpha)}(\tau_1) }{\delta B_\mu(x)} \Big]\nonumber \,.\\
\end{eqnarray}

Using Eqs.~(\ref{eq:matrix-Jacobian-singlet}) and (\ref{eq:matrix-Lagrangian-singlet}), we can then straightforwardly compute
\begin{eqnarray}
&&\chi\frac{\delta \bar{\omega}(0)}{\delta B_\mu(x)}\Big|^{\rm singlet}
= - 4 \psi^5_0 \Big( i \partial^\mu_0 - \frac{4 i }{\mathcal{E}} \psi^\nu_0\psi^\mu_0 \dot{x}_{0\nu} \Big) \delta(x_0 - x) \begin{pmatrix} 1 & 0 \\  0 & 1 \end{pmatrix}\,,
\end{eqnarray}
and
\begin{eqnarray}
&&\frac{\delta \mathcal{L}_{(\alpha)}(\tau_1)}{\delta B_\mu(x)} \Big|^{\rm singlet} =
 \alpha \Big( - i \dot{x}^\mu_1 + \frac{i\mathcal{E}}{2} 2 \psi^\nu_1 \psi^\mu_1  \partial^1_\nu \Big) \delta(x_1 - x)\begin{pmatrix} 1 & 0 \\ 0 & - 1 \end{pmatrix} 
\nonumber\\
&&+ \alpha \mathcal{E}\psi^\mu_1 \psi^5_1 d_{000} \Pi_1 \delta(x_1 - x) \begin{pmatrix} 0 & 1 \\ -1 & 0 \end{pmatrix} 
+ i \alpha^2 \mathcal{E}\psi^\mu_1 \psi^5_1 d_{000} \Phi_1 \delta(x_1 - x) \begin{pmatrix} 0 & 1 \\ 1 & 0 \end{pmatrix}\,.
\end{eqnarray}
Here $d_{000}$ denotes the symmetric structure constant of $U(3)$ flavor corresponding to the singlet sector- in general $d_{abc}$ has non-zero labels that span $a,b,c=0,3,8$.

The scalar field $\Phi$ appearing in Eq.~(\ref{eq:Dirac}) has the structure of a mass term - we can write it as $\Phi(x)= m-\mathcal{S}(x)$, where $\mathcal{S}$ sources the quark condensate. For the moment, we will not consider the existence of the quark condensate (as in the case of QED) and set $\Phi=m=\rm{constant}$. This allows us to simplify the derivative
\begin{eqnarray}
&&\frac{\delta W_\mathcal{I}}{\delta B_\mu(x)}\Big|^{\rm singlet}_{B_\mu=0;\Pi=0;\Phi=m} = - \frac{i\mathcal{E}}{64}\int^1_{-1}d\alpha \int^\infty_0 dT \mathcal{N} \int_P \mathcal{D}x \mathcal{D}\psi~ 
\nonumber\\
&&\times \exp\Big[-\int^T_0 d\tau \Big( \frac{\dot{x}^2}{2\mathcal{E}} + \frac{1}{2}\psi\dot{\psi} - i \dot{x}^\mu A_\mu + \frac{i\mathcal{E}}{2}\psi^\mu \psi^\nu (\partial_\mu A_\nu - \partial_\nu A_\mu ) +  \frac{\mathcal{E}}{4}d_{000} \alpha^2 m^2 \Big) \Big]
\nonumber\\
&&\times \Big[ - 8 \psi^5_0 \Big( i \partial^\mu_0 - \frac{4 i }{\mathcal{E}} \psi^\nu_0\psi^\mu_0 \dot{x}_{0\nu} \Big) \delta(x_0 - x)
- 8 i \alpha^2  \frac{ 2 }{\mathcal{E}} \psi^\nu_0 \dot{x}_{0\nu} m \int^T_0 d\tau_1 \Big\{ \mathcal{E}\psi^\mu_1 \psi^5_1 d_{000} m\delta(x_1 - x) \Big\} \Big]\,.
\end{eqnarray}
Rewriting this equation as
\begin{eqnarray}
&&\frac{\delta W_\mathcal{I}}{\delta B_\mu(x)}\Big|^{\rm singlet}_{B_\mu=0;\Pi=0;\Phi=m} = \frac{\mathcal{E}}{8}\int^1_{-1}d\alpha \int^\infty_0 dT \exp\Big[- T \frac{\mathcal{E}d_{000} \alpha^2 m^2}{4} \Big] \mathcal{N} \int_P \mathcal{D}x \mathcal{D}\psi~ \exp\Big[-\int^T_0 d\tau \Big\{ \frac{\dot{x}^2}{2\mathcal{E}} + \frac{1}{2}\psi\dot{\psi} \Big\} \Big]
\nonumber\\
&&\times \exp\Big[ i \int^T_0 d\tau \Big\{ \dot{x}^\rho + \mathcal{E} \psi^\rho \psi^\alpha \partial_\alpha  \Big\} A_\rho \Big]
 \Big[ - \psi^5_0 \Big( \partial^\mu_0 - \frac{4 }{\mathcal{E}} \psi^\nu_0\psi^\mu_0 \dot{x}_{0\nu} \Big) \delta(x_0 - x)
- 2 d_{000} m^2 \alpha^2 \psi^\nu_0 \dot{x}_{0\nu} \int^T_0 d\tau_1 \Big\{ \psi^\mu_1 \psi^5_1 \delta(x_1 - x) \Big\} \Big]\,,\nonumber\\
\end{eqnarray}
replacing the $\delta$-functions by their Fourier representation, and expanding to quadratic order in $A$, gives 
\begin{eqnarray}
&&\int d^4x \,e^{ilx} \frac{\delta W_\mathcal{I}}{\delta B_\mu(x)}\Big|^{\rm singlet}_{B_\mu=0;\Pi=0;\Phi=m} = \frac{\mathcal{E}}{8}\int^1_{-1}d\alpha \int^\infty_0 dT \exp\Big[- T \frac{\mathcal{E}d_{000} \alpha^2 m^2}{4} \Big] \mathcal{N} \int_P \mathcal{D}x \mathcal{D}\psi~ \exp\Big[-\int^T_0 d\tau \Big\{ \frac{\dot{x}^2}{2\mathcal{E}} + \frac{1}{2}\psi\dot{\psi} \Big\} \Big]
\nonumber\\
&&\times \Big[ i \int^T_0 d\tau_2 \Big\{ \dot{x}^\rho_2 + \mathcal{E} \psi^\rho_2 \psi^\alpha_2 \partial_\alpha  \Big\} A_\rho(x_2) \Big]\Big[ i \int^T_0 d\tau_4 \Big\{ \dot{x}^\sigma_4 + \mathcal{E} \psi^\sigma_4 \psi^\beta_4 \partial_\beta  \Big\} A_\sigma(x_4) \Big]
 \Big[ - \psi^5_0 \Big( ip^\mu - \frac{4 }{\mathcal{E}} \psi^\nu_0\psi^\mu_0 \dot{x}_{0\nu} \Big) e^{ipx_0}
\nonumber\\
&&- 2 d_{000} m^2 \alpha^2 \psi^\nu_0 \dot{x}_{0\nu} \Big( \int^T_0 d\tau_1  \psi^\mu_1 \psi^5_1 e^{ilx_1} \Big) \Big]\,.
\end{eqnarray}
Further replacing the background (photon) fields by their Fourier transforms 
\begin{eqnarray}
&&A_\mu(x_i) = \int \frac{d^4p_i}{(2\pi)^4} A_\mu(p_i) e^{ip_i x_i}\,,
\end{eqnarray}
the expression can be simplified to read, 
\begin{eqnarray}
&&\int d^4x\, e^{ilx} \frac{\delta W_\mathcal{I}}{\delta B_\mu(x)}\Big|^{\rm singlet}_{B_\mu=0;\Pi=0;\Phi=m} 
\nonumber\\
&&= - \frac{\mathcal{E}}{8} \int \frac{d^4k_2}{(2\pi)^4} A_\rho(k_2) \int \frac{d^4k_4}{(2\pi)^4} A_\sigma(k_4) \int^1_{-1}d\alpha \int^\infty_0 dT \exp\Big[- T \frac{\mathcal{E}d_{000} \alpha^2 m^2}{4} \Big] \int^T_0 d\tau_2 \int^T_0 d\tau_4
\nonumber\\
&&\times \mathcal{N} \int_P \mathcal{D}x \mathcal{D}\psi~ \exp\Big[-\int^T_0 d\tau \Big\{ \frac{\dot{x}^2}{2\mathcal{E}} + \frac{1}{2}\psi\dot{\psi} \Big\} \Big] \Big[ \dot{x}^\rho_2 \dot{x}^\sigma_4 + i \mathcal{E} k_{4\beta} \dot{x}^\rho_2 \psi^\sigma_4 \psi^\beta_4 + i \mathcal{E} k_{2\alpha} \dot{x}^\sigma_4  \psi^\rho_2 \psi^\alpha_2  - \mathcal{E}^2 k_{2\alpha} k_{4\beta} \psi^\rho_2 \psi^\alpha_2 \psi^\sigma_4 \psi^\beta_4 \Big] 
\nonumber\\
&&\times \Big[ - \psi^5_0 \Big( il^\mu - \frac{4 }{\mathcal{E}} \psi^\nu_0\psi^\mu_0 \dot{x}_{0\nu} \Big) e^{ilx_0} - 2 d_{000} m^2 \alpha^2 \psi^\nu_0 \dot{x}_{0\nu} \Big( \int^T_0 d\tau_1 \psi^\mu_1 \psi^5_1 e^{ilx_1} \Big) \Big] e^{ik_2 x_2} e^{ik_4 x_4}\,.
\label{eq:wl-diag-init}
\end{eqnarray}

To obtain the anomaly equation, one has to take the partial derivative of the l.h.s of the above expression. Doing so, and after a considerable amount of ``worldline algebra" in performing the bosonic and Grassmann path integrals along the lines discussed in \cite{Tarasov:2020cwl,Tarasov:2021yll}, we obtain\footnote{Some of the details of the derivation are given in Appendix A. } 
\begin{eqnarray}
\label{eq:AVV-worldline-final}
&&\int d^4x e^{ilx} \partial_\mu \frac{\delta W_\mathcal{I}}{\delta B_\mu(x)}\Big|^{\rm singlet}_{B_\mu=0;\Pi=0;\Phi=m} 
\\
&&= \frac{1}{2\pi^{2}} \int \frac{d^4k_2}{(2\pi)^4} A_\rho(k_2) \int \frac{d^4k_4}{(2\pi)^4} A_\sigma(k_4) \epsilon^{\rho\alpha\sigma\beta} k_{2\alpha} k_{4\beta} (2\pi)^4\delta^4(l + k_2 + k_4)
\nonumber\\
&&- 2 m^2 \frac{1}{4\pi^{2}} \int \frac{d^4k_2}{(2\pi)^4} A_\rho(k_2) \int \frac{d^4k_4}{(2\pi)^4} A_\sigma(k_4) \epsilon^{\rho\alpha\sigma\beta} k_{2\alpha} k_{4\beta} (2\pi)^4\delta^4(l + k_2 + k_4) \int^1_0 du_2 \int^1_0 du_4 \frac{1}{\mathcal{G}(u_0, u_2, u_4) + m^2}\,. 
\nonumber
\end{eqnarray}
Here, we used the compact notation
\begin{eqnarray}
\label{eq:Green-function-combo}
&&\mathcal{G}(u_0, u_2, u_4) \equiv k_2 \cdot k_4 \Big( G_B(u_0, u_2) + G_B(u_0, u_4) -  G_B(u_2, u_4)\Big)\,,
\end{eqnarray}
 where $G_B$ is the bosonic worldline Green function~\cite{Schubert:2001he}
 \begin{eqnarray}
G_B(u_1, u_2) = |u_1 - u_2| - 2(u_1-u_2)^2\,,
\end{eqnarray}
fixing the einbein $\mathcal{E}=2$. In the chiral limit $m=0$, the above expression reduces to 
Eq.~(\ref{eq:VVA1}). As may be anticipated, the second term can be expressed as the second term in Eq.~(\ref{eq:anomaly-mass}). To see this in the worldline formalism, we will turn now to discuss the PVV triangle. 

\subsection{The PVV triangle at finite quark mass}
\label{sec:PVV-finite-mass}

\begin{figure}[htb]
 \begin{center}
\includegraphics[width=110mm]{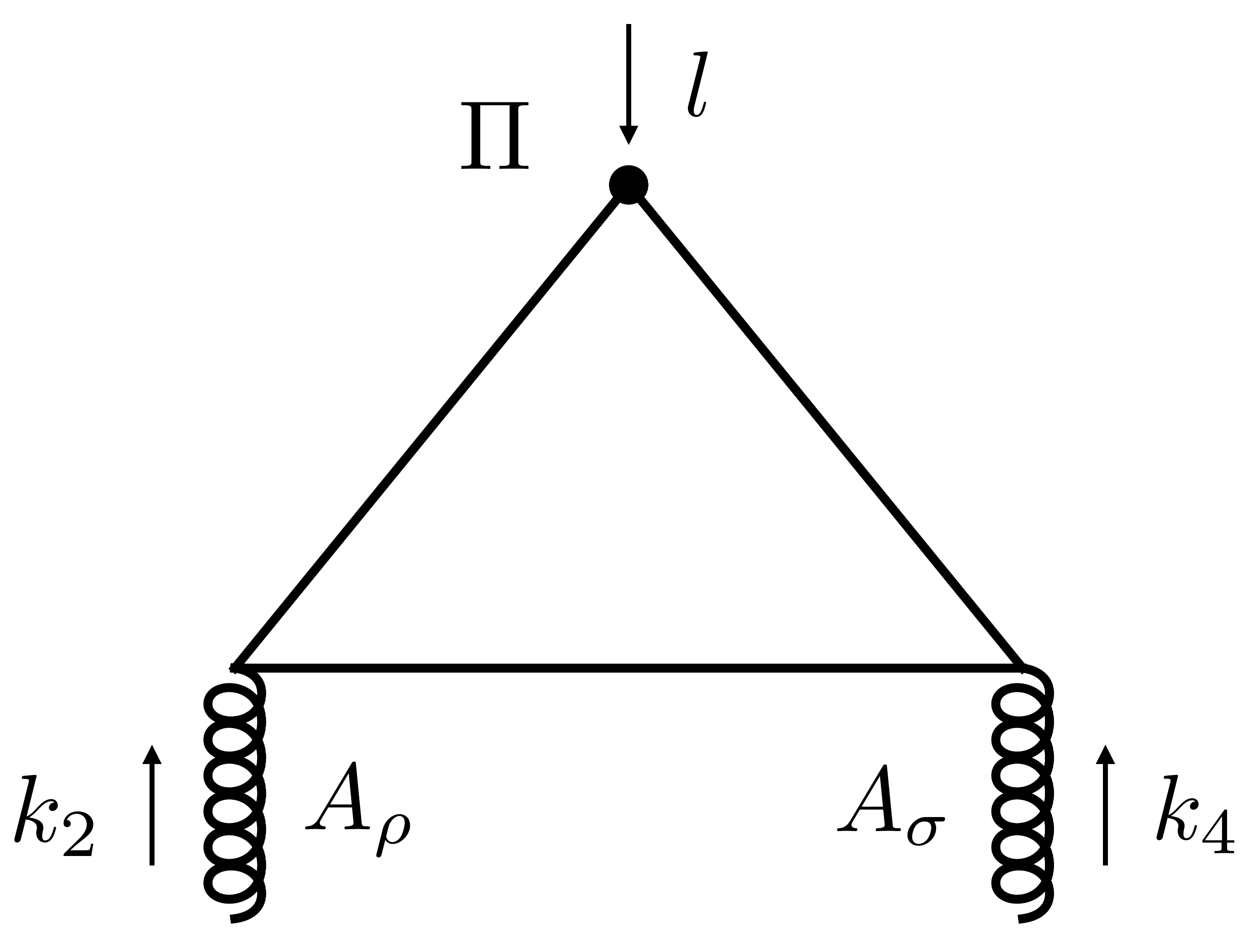}
 \end{center}
\caption{The PVV triangle obtained by taking the functional derivative of the effective action, first with respect to the pseudoscalar source $\Pi$, and then with respect to the vector gauge field $A_\mu$. }
\label{fig:PVV-triangle}
 \end{figure}

The PVV triangle is illustrated in Fig.~\ref{fig:PVV-triangle}. As its structure indicates, it can be obtained by 
taking the functional derivative of Eq.~(\ref{eq:Imag-worldline}) with respect to the pseudoscalar source field $\Pi$, which is given by 
\begin{eqnarray}
\label{eq:action-Pi-derivative}
\frac{\delta W_\mathcal{I}}{\delta \Pi(x)} = - \frac{i\mathcal{E}}{64}\int^1_{-1}d\alpha \int^\infty_0 dT \mathcal{N} \int_P \mathcal{D}x \mathcal{D}\psi~ \exp\Big[-\int^T_0 d\tau \mathcal{L}_{(\alpha)}(\tau)\Big]
 \Big[ {\rm tr} ~\chi \frac{\delta \bar{\omega}(0) }{\delta \Pi(x)} -\int^T_0 d\tau_1 ~ {\rm tr} ~\chi \bar{\omega}(0) \frac{\delta \mathcal{L}_{(\alpha)}(\tau_1) }{\delta \Pi(x)} \Big]\,.
\end{eqnarray}
The subsequent steps follow as in the previous subsection.

Since we are focusing on the flavor singlet sector, we can put the external sources $B_\mu$ and $\Pi$ to zero (after taking appropriate functional derivatives thereof), and $\Phi=m$ (as discussed previously). The expressions simplify tremendously. 
We obtain for the Jacobian and its functional derivative,
\begin{eqnarray}
\chi\bar{\omega}(0)\Big|^{\rm singlet}_{B_\mu=0;\Pi=0} = 4 \psi^\mu_0 \frac{ 2 }{\mathcal{E}} \dot{x}_{0\mu} \Phi_0 \begin{pmatrix} 0 & 1 \\ 1 & 0 \end{pmatrix};\qquad
\chi \frac{\delta \bar{\omega}(0)}{\delta \Pi(x)}\Big|^{\rm singlet}_{B_\mu=0;\Pi=0}
= 4 i \psi^5_0\, d_{000} m \,\delta(x_0 - x) \begin{pmatrix} 1 & 0 \\  0 & 1 \end{pmatrix}\,.
\end{eqnarray}
Likewise, the worldline Lagrangian reduces to  
\begin{eqnarray}
&&\mathcal{L}_{(\alpha)}(\tau)\Big|^{\rm singlet}_{B_\mu=0;\Pi=0;\Phi=m} = \Big( \frac{\dot{x}^2}{2\mathcal{E}} + \frac{1}{2}\psi\dot{\psi} - i \dot{x}^\mu A_\mu + \frac{i\mathcal{E}}{2}\psi^\mu \psi^\nu (\partial_\mu A_\nu - \partial_\nu A_\mu ) +  \frac{\mathcal{E}}{4}d_{000} \alpha^2 m^2 \Big) \begin{pmatrix} 1 & 0 \\  0 & 1 \end{pmatrix}\,,
\end{eqnarray}
and its functional derivative is 
\begin{eqnarray}
&&\frac{\delta\mathcal{L}_{(\alpha)}(\tau_1)}{\delta \Pi(x)}\Big|^{\rm singlet}_{B_\mu=0;\Pi=0} = i\mathcal{E}\psi^\mu_1 \psi^5_1 \partial^1_\mu \delta(x_1-x) \begin{pmatrix} 0 & 1 \\ 1 & 0 \end{pmatrix}\,.
\end{eqnarray}
Plugging these expressions into Eq.~(\ref{eq:action-Pi-derivative}), we obtain 
\begin{eqnarray}
&&\frac{\delta W_\mathcal{I}}{\delta \Pi(x)}\Big|^{\rm singlet}_{B_\mu=0;\Pi=0;\Phi=m} = - \frac{i\mathcal{E}}{64}\int^1_{-1}d\alpha \int^\infty_0 dT \mathcal{N} \int_P \mathcal{D}x \mathcal{D}\psi~ 
\nonumber\\
&&\times \exp\Big[-\int^T_0 d\tau \Big( \frac{\dot{x}^2}{2\mathcal{E}} + \frac{1}{2}\psi\dot{\psi} - i \dot{x}^\mu A_\mu + \frac{i\mathcal{E}}{2}\psi^\mu \psi^\nu (\partial_\mu A_\nu - \partial_\nu A_\mu ) \Big) \Big] \exp\Big[- T \frac{\mathcal{E}}{4}d_{000} \alpha^2 m^2 \Big]
\nonumber\\
&&\times \Big[  8 i \psi^5_0 d_{000} m \delta(x_0 - x)
- 8 i \frac{ 2 }{\mathcal{E}} \psi^\nu_0 \dot{x}_{0\nu} m  \int^T_0 d\tau_1 \Big\{ \mathcal{E} \psi^\mu_1 \psi^5_1 \partial^1_\mu \delta(x_1-x) \Big\} \Big]\,.
\end{eqnarray}

As in the previous subsection, we now replace the $\delta$-functions by their Fourier transforms, expand the exponential to second order in the gauge fields, and replace them with their respective Fourier transforms. 
As a result, we get
\begin{eqnarray}
&&\int d^4x\, e^{ilx} \frac{\delta W_\mathcal{I}}{\delta \Pi(x)}\Big|^{\rm singlet}_{B_\mu=0;\Pi=0;\Phi=m}
= - \frac{\mathcal{E}m}{8} \int \frac{d^4k_2}{(2\pi)^4} \int \frac{d^4k_4}{(2\pi)^4} \int^1_{-1}d\alpha \int^\infty_0 dT \exp\Big[- T \frac{\mathcal{E} d_{000} \alpha^2 m^2 }{4}  \Big] \int^T_0 d\tau_2 \int^T_0 d\tau_4
\nonumber\\
&&\times \mathcal{N} \int_P \mathcal{D}x \mathcal{D}\psi~ \exp\Big[-\int^T_0 d\tau \Big\{ \frac{\dot{x}^2}{2\mathcal{E}} + \frac{1}{2}\psi\dot{\psi} \Big\} \Big] \Big[ \dot{x}^\rho_2 + i \mathcal{E}\psi^\rho_2 \psi^\alpha_2 k_{2\alpha} \Big] A_\rho(k_2) e^{ik_2 x_2}  \Big[ \dot{x}^\sigma_4 + i\mathcal{E}\psi^\sigma_4 \psi^\beta_4 k_{4\beta} \Big] A_\sigma(k_4) e^{ik_4 x_4}
\nonumber\\
&&\times 
 \Big[ \psi^5_0 d_{000} e^{ipx_0}
- 2 i \psi^\mu_0 \dot{x}_{0\mu} \int^T_0 d\tau_1 \Big\{ \psi^\nu_1 \psi^5_1 l_\nu e^{ilx_1} \Big\} \Big]\,.
\end{eqnarray}

We can rewrite this equation as
\begin{eqnarray}
&&\int d^4x\, e^{ilx} \frac{\delta W_\mathcal{I}}{\delta \Pi(x)}\Big|^{\rm singlet}_{B_\mu=0;\Pi=0;\Phi=m}
\nonumber\\
&&= - \frac{\mathcal{E}\Phi}{8} \int \frac{d^4k_2}{(2\pi)^4} A_\rho(k_2) \int \frac{d^4k_4}{(2\pi)^4} A_\sigma(k_4) \int^1_{-1}d\alpha \int^\infty_0 dT \exp\Big[- T \frac{\mathcal{E} d_{000} \alpha^2 \Phi^2 }{4}  \Big] \int^T_0 d\tau_2 \int^T_0 d\tau_4
\nonumber\\
&&\times \mathcal{N} \int_P \mathcal{D}x \mathcal{D}\psi~ \psi_5 \exp\Big[-\int^T_0 d\tau \Big\{ \frac{\dot{x}^2}{2\mathcal{E}} + \frac{1}{2}\psi\dot{\psi} \Big\} \Big] \Big[ \dot{x}^\rho_2 \dot{x}^\sigma_4 + i\mathcal{E} k_{4\beta} \dot{x}^\rho_2 \psi^\sigma_4 \psi^\beta_4 + i \mathcal{E} k_{2\alpha} \dot{x}^\sigma_4 \psi^\rho_2 \psi^\alpha_2   - \mathcal{E}^2 k_{2\alpha} k_{4\beta} \psi^\rho_2 \psi^\alpha_2 \psi^\sigma_4 \psi^\beta_4 \Big]
\nonumber\\
&&\times 
 \Big[ d_{000} e^{ilx_0}
- 2 i \psi^\mu_0 \dot{x}_{0\mu} \int^T_0 d\tau_1 \Big\{ \psi^\nu_1 l_\nu e^{ilx_1} \Big\} \Big] e^{ik_2 x_2} e^{ik_4 x_4}\,.
\end{eqnarray}
As in the case of the AVV triangle, one then performs the bosonic and Grassmann path integrals, the details of which are outlined in Appendix A. We obtain finally 
\begin{eqnarray}
&&\int d^4x\, e^{ilx} \frac{\delta W_\mathcal{I}}{\delta \Pi(x)}\Big|^{\rm singlet}_{B_\mu=0;\Pi=0;\Phi=m} = - 2 m\frac{1}{(4\pi )^2} \int \frac{d^4k_2}{(2\pi)^4} A_\rho(k_2) \int \frac{d^4k_4}{(2\pi)^4} A_\sigma(k_4) \epsilon^{\rho\alpha\sigma\beta} k_{2\alpha} k_{4\beta} (2\pi)^4\delta^4(l + k_2 + k_4)
\nonumber\\
&&\times  \int^1_0 du_2 \int^1_0 du_4 \int^1_{-1}d\alpha \frac{ \alpha^2 m^2 - \mathcal{G}(u_0, u_2, u_4)}{\Big\{ \mathcal{G}(u_0, u_2, u_4) + \alpha^2 m^2\Big\}^2}\,,
\end{eqnarray}
where $\mathcal{G}(u_0, u_2, u_4)$ was defined previously in Eq.~(\ref{eq:Green-function-combo}). 
Integrating this expression over $\alpha$ gives  
\begin{eqnarray}
\label{eq:PVV-worldline-1}
&&\int d^4x\, e^{ilx} \frac{\delta W_\mathcal{I}}{\delta \Pi(x)}\Big|^{\rm singlet}_{B_\mu=0;\Pi=0;\Phi=m}
\nonumber\\
&&= \frac{m}{4\pi^2} \int \frac{d^4k_2}{(2\pi)^4} A_\rho(k_2) \int \frac{d^4k_4}{(2\pi)^4} A_\sigma(k_4) \epsilon^{\rho\alpha\sigma\beta} k_{2\alpha} k_{4\beta} (2\pi)^4\delta^4(l + k_2 + k_4) \int^1_0 du_2 \int^1_0 du_4 \frac{1}{\mathcal{G}(u_0, u_2, u_4) + m^2}\,.
\end{eqnarray}

Substituting this PVV expression into Eq.~(\ref{eq:AVV-worldline-final}), we see immediately that it can be written as 
\begin{eqnarray}
\label{eq:AVV-final}
\int d^4x \,e^{ilx} \partial_\mu \frac{\delta W_\mathcal{I}}{\delta B_\mu(x)}\Big|^{\rm singlet}_{B_\mu=0;\Pi=0;\Phi=m} 
&&= -\frac{1}{4\pi^{2}} \int d^4x\, e^{i l x} F_{\alpha\rho}(x) \tilde{F}^{\alpha\rho}(x)
\nonumber\\
&&-  2 m \int d^4x\, e^{ilx} \frac{\delta W_\mathcal{I}}{\delta \Pi(x)}\Big|^{\rm singlet}_{B_\mu=0;\Pi=0;\Phi=m}\,.
\end{eqnarray}
This expression is nothing but the Fourier transform of the anomaly equation given in Eq.~(\ref{eq:anomaly-mass}), now expressed as the chiral Ward identity relating the 1-point functions $J_\mu^5$, $F\tilde F$ and $\bar \psi \gamma_5 \psi$.

 \subsection{Anomaly cancellation in QED and QCD in the forward limit}
 \label{sec:Anomaly-cancellation}

In the expression in Eq.~(\ref{eq:Green-function-combo}),
\begin{eqnarray}
&&\mathcal{G}(u_0, u_2, u_4) \equiv k_2 \cdot k_4 \Big( G_B(u_0, u_2) + G_B(u_0, u_4) -  G_B(u_2, u_4)\Big)\nonumber\,.
\end{eqnarray}
Since $k^2_2 = k^2_4$, we can write $k_2 \cdot k_4 = \frac{1}{2} (k_2 + k_4)^2 = \frac{1}{2} l^2$. Therefore, in the forward limit where $l^2 = 0$, we have
\begin{eqnarray}
&&\mathcal{G}(u_0, u_2, u_4)\Big|_{l^2 = 0} = 0\,.
\end{eqnarray}

In this limit,
\begin{eqnarray}
&&\int d^4x\, e^{ilx} \partial_\mu \frac{\delta W_\mathcal{I}}{\delta B_\mu(x)}\Big|^{\rm singlet}_{B_\mu=0;\Pi=0;\Phi=m} \Big|_{l^2 = 0}
\\
&&= \frac{1}{2\pi^{2}}  (2\pi)^4\delta^4(l + k_2 + k_4) \int \frac{d^4k_2}{(2\pi)^4} A_\rho(k_2) \int \frac{d^4k_4}{(2\pi)^4} A_\sigma(k_4) \epsilon^{\rho\alpha\sigma\beta} k_{2\alpha} k_{4\beta}
\nonumber\\
&&- 2 m^2 \frac{1}{4\pi^{2}}  (2\pi)^4\delta^4(l + k_2 + k_4) \int \frac{d^4k_2}{(2\pi)^4} A_\rho(k_2) \int \frac{d^4k_4}{(2\pi)^4} A_\sigma(k_4) \epsilon^{\rho\alpha\sigma\beta} k_{2\alpha} k_{4\beta} \int^1_0 du_2 \int^1_0 du_4 \frac{1}{m^2} \,.
\nonumber
\end{eqnarray}

We can now integrate over the proper-time variables in the second term on the r.h.s trivially to obtain 
\begin{eqnarray}
&&\int d^4x \,e^{ilx} \partial_\mu \frac{\delta W_\mathcal{I}}{\delta B_\mu(x)}\Big|^{\rm singlet}_{B_\mu=0;\Pi=0;\Phi=m} \Big|_{l^2 = 0}
\\
&&= \frac{1}{2\pi^{2}}  (2\pi)^4\delta^4(l + k_2 + k_4) \int \frac{d^4k_2}{(2\pi)^4} A_\rho(k_2) \int \frac{d^4k_4}{(2\pi)^4} A_\sigma(k_4) \epsilon^{\rho\alpha\sigma\beta} k_{2\alpha} k_{4\beta}
\nonumber\\
&&- \frac{1}{2\pi^{2}}  (2\pi)^4\delta^4(l + k_2 + k_4) \int \frac{d^4k_2}{(2\pi)^4} A_\rho(k_2) \int \frac{d^4k_4}{(2\pi)^4} A_\sigma(k_4) \epsilon^{\rho\alpha\sigma\beta} k_{2\alpha} k_{4\beta}\,,
\nonumber
\end{eqnarray}
resulting in 
\begin{eqnarray}
&&\int d^4x\, e^{ilx} \partial_\mu \frac{\delta W_\mathcal{I}}{\delta B_\mu(x)}\Big|^{\rm singlet}_{B_\mu=0;\Pi=0;\Phi=m} \Big|_{l^2 = 0}= 0 \,.
\end{eqnarray}
This shows explicitly how the ``anomaly" pole from the AVV triangle is exactly canceled by the PVV pole in the forward limit. Our worldline result agrees with the recent computation in \cite{Castelli:2024eza}, which, in turn, is a well-known result in QED~\cite{Adler:1969er,Adler:1969gk}, fundamental to the consistency of the theory.

Our result also tells us that such a cancellation would not occur for $m=0$, confirming that massless QED is not a well-defined theory\footnote{For an elegant discussion, see \cite{Weinberg:1965nx}, in particular footnote 3.}. The situation is dramatically different in QCD (for $N_f\geq 2$), which is a well-defined theory in the chiral limit. Since the above cancellation will not work, how does this occur? One may consider this question to be a statement of the famous $U_A(1)$ puzzle. 

\begin{figure}[htb]
 \begin{center}
\includegraphics[width=160mm]{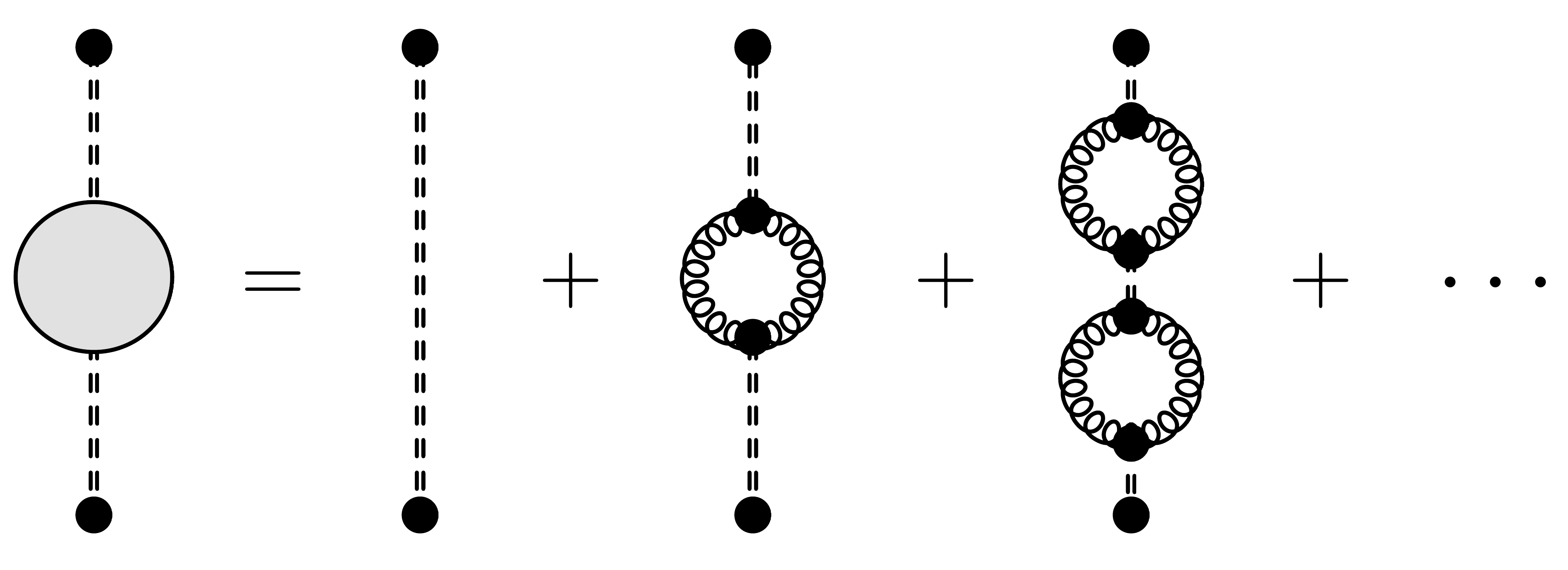}
 \end{center}
\caption{Topological generation of the $\eta'$ mass in the $N_f/N_c$ expansion in the chiral limit. The first term on the r.h.s denotes the propagation of the massless primordial $\bar \eta$, which is the ninth Goldstone boson of $U(3)_{L+R}$  in the chiral and large $N_c$ limit. Because of the WZW term coupling $\bar \eta$ with the topological charge density $\Omega$, one generates vacuum bubbles of the Yang-Mills topological susceptibility, as depicted, which can be iterated in the $N_f/N_c$ expansion, as shown on the r.h.s. The summation of these bubbles, in the chiral limit, shifts the bare $\bar \eta$ pole to one at ${\tilde m}_{\eta^
\prime}$, where ${\tilde m}_{\eta^
\prime}$ is given by the Witten-Veneziano formula for the $\eta^\prime$ mass.}
\label{fig:etapr-mass-generation}
 \end{figure}

The answer, as we discussed earlier, is of course well-known, being fundamentally tied to features of the topology of the QCD vacuum that gives the $\eta^\prime$ its mass~\cite{tHooft:1976snw,Witten:1979vv,Veneziano:1979ec};  their role in determining the proton's quark helicity was first discussed more than three decades ago~\cite{Veneziano:1989ei,Jaffe:1989jz}. It is nevertheless interesting to see how it arises in the worldline framework because it provides novel insight into some of the details of the dynamics and is a useful tool in understanding how this nonperturbative dynamics plays out in different kinematical regions at the high energies of interest. 
 
In the worldline framework, we see immediately that our derivation will change if $\Phi\neq m\equiv {constant}$ but instead if we write the scalar source density of the quark condensate as $\Phi=-\mathcal{S}(x)+m$. Firstly, one can't ignore derivatives with respect to $\mathcal{S}$ when taking further functional derivatives of the 
effective action ${\cal W}$ in deriving chiral Ward identities for two-point correlators of the axial current. We will discuss this key point further in the next section. Further, even if we ignore the spatial variation of $\Phi$, it gives a finite contribution to our results in the chiral limit $m=0$. 
For example, we derived a nontrivial Wess-Zumino-Witten (WZW)  (see Appendix A in Paper II) by expanding ${\cal W}_I$ and keeping terms to linear order in $\Pi$, $\Phi$ and quadratic order in $A$. In this case, going through a similar derivation as described in the previous subsections, we obtained 
\begin{eqnarray}
    {\cal W}_I[\Pi,A^2,\Phi] = \frac{-1}{8\pi^2}\frac{1}{\Phi} {\rm Tr}_c\int d^4x\, \Pi(x) F_{\mu\nu}{\tilde F}^{\mu\nu}\,.
\end{eqnarray}
Performing the field redefinition,
\begin{eqnarray}
    {\bar \eta}=-\sqrt{2N_f} \frac{\Pi}{\Phi}F_{\bar \eta}\,,
\end{eqnarray}
with the ${\bar \eta}$ decay constant defined as the forward limit $l^2\rightarrow 0$  of the decay constant in  the identity
\begin{eqnarray}
    \langle 0|J_\mu^5|{\bar \eta}\rangle=i\sqrt{2N_f}\, l^\mu F_{\tilde \eta}(l^2)\,,
\end{eqnarray}
we obtained Eq.~(\ref{eq:S-WZW}). The structure of this result is analogous to Eq.~(\ref{eq:AVV-worldline-final}) in AVV (and Eq.~(\ref{eq:PVV-worldline-1}) in PVV), where we now instead have the structure
\begin{eqnarray}
\int^1_0 du_2 \int^1_0 du_4 \frac{1}{1+ \frac{\mathcal{G}(u_0, u_2, u_4)}{\Phi^2}}\rightarrow \int^1_0 du_2 \int^1_0 du_4 \frac{1}{1+ \frac{l^2}{\Phi^2}f(u_2,u_4)}\,,
\end{eqnarray}
with $f(u_2,u_4)=\Big( G_B(0, u_2) + G_B(0, u_4) -  G_B(u_2, u_4)\Big)$. Taking the forward limit is equivalent to 
taking $\Phi\rightarrow \infty$, corresponding effectively to a large mass expansion\footnote{A similar expansion was employed in \cite{DHoker:1995aat} to derive the WZW term corresponding to $\pi^0\rightarrow 2\gamma$ in the worldline formalism.}. Such a Pauli-Villars type regularization can be used to derive all the WZW terms arising from ${\cal W}_I$, as discussed in \cite{Alvarez-Gaume:1985jck,Ball:1985tya}.

In Paper II, we showed that the existence of the ${\bar \eta}\,\Omega$ WZW term leads to the possibility that the AVV anomaly $\frac{l^\mu}{l^2} \Omega$ can couple to the hadron via the diagrams shown in Fig.~\ref{fig:etapr-mass-generation} corresponding to the propagation of the $\bar \eta$; the bare exchange can be iterated in a systematic $N_f/N_c$ expansion, as shown, corresponding to resumming bubbles of the Yang-Mills topological susceptibility. The end result is the 
shift of the anomaly pole from $l^2\rightarrow {\tilde m}_{\eta^\prime}^2$, where ${\tilde m}_{\eta^\prime}$ is given by the Witten-Veneziano formula in the chiral limit. Our result therefore explained why, unlike QED, chiral QCD is not anomalous. The effect of adding finite quark masses will be discussed at length in the following section. Specifically, we will discuss how chiral Ward identities are modified away from the chiral limit, and correspondingly, how the result for $\Delta \Sigma$ in Eq.~(\ref{eq:chiral-DeltaSigma}) is modified. 

 \section{Chiral Ward identities: from worldlines to Wess-Zumino effective action}
 \label{sec:Ward}
We will demonstrate in this section that the QED mechanism for regularizing the ``anomaly" pole in the AVV triangle via its exact cancellation by an identical pole of the finite quark mass PVV triangle, in the forward limit, is not realized in QCD. The physics governing pole cancellation in the latter is much richer and is valid both in the chiral limit and for finite quark masses. Unsurprisingly, the fundamental difference between QED and QCD in this context is due to the nontrivial structure of the QCD vacuum, which modifies the relation of matrix elements of vacuum operators to those sandwiched between hadron states. These matrix elements are intrinsically nonperturbative objects. Nevertheless, we can systematically study their properties employing the Wess-Zumino effective action, and the chiral Ward identities that result by taking functional derivatives of this effective action. Further progress can be made, as we will discuss,  with the assumptions of polology and expectations based on the smooth behavior of the topological susceptibility in pure Yang-Mills and in QCD in the forward limit. Finally, we will end this section with a discussion of large $N_c$ phenomenology, and relate our conclusions to experimental data, and to nonperturbative results from QCD sum rules and lattice QCD. 

\begin{figure}[htb]
 \begin{center}
\includegraphics[width=140mm]{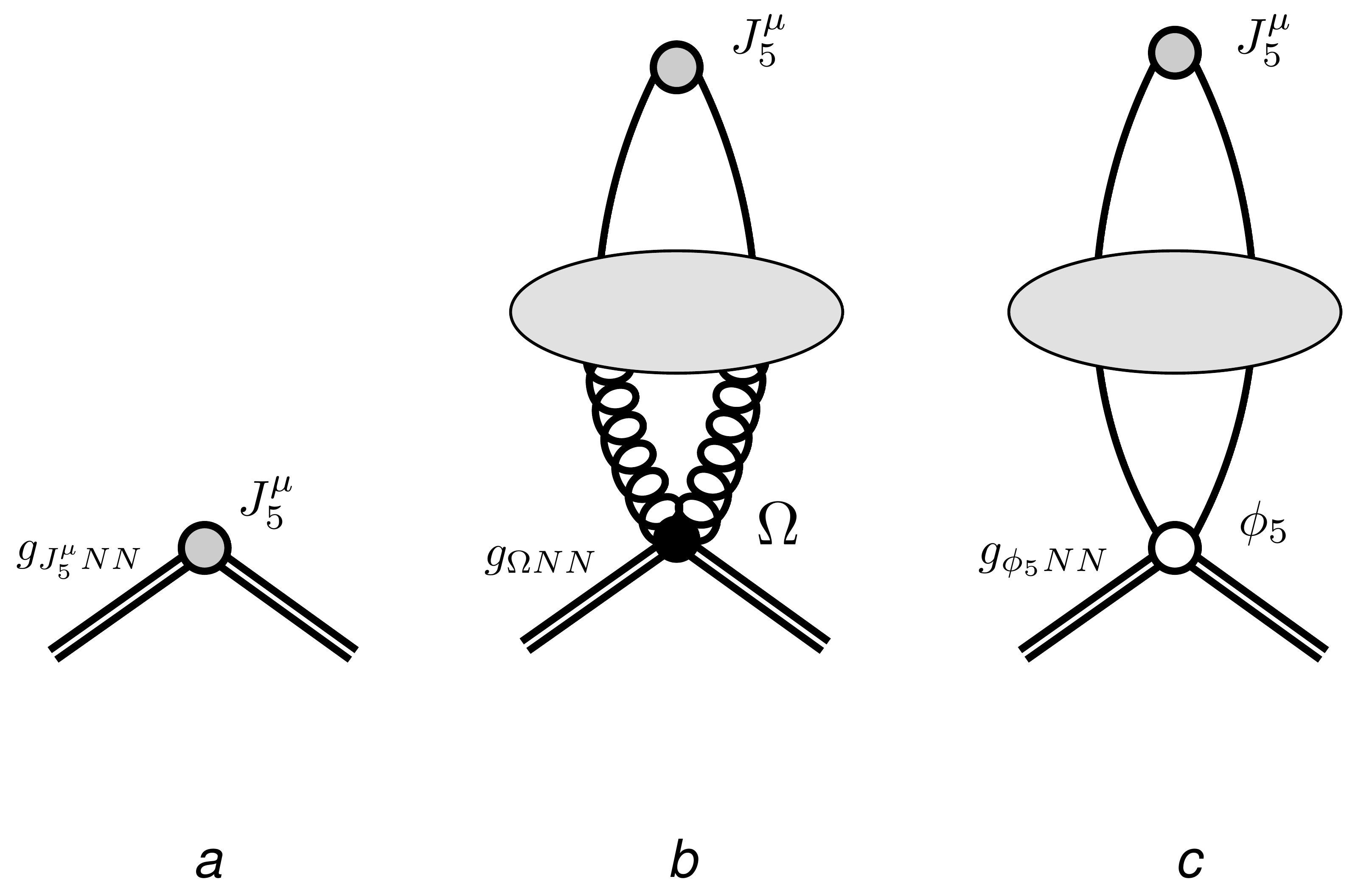}
 \end{center}
\caption{Contributions to the matrix element of the isosinglet axial-vector current $J_\mu^5$ in the polarized proton. 
The first term a) denotes the direct nonperturbative coupling of $J_\mu^5$ to the proton, whose strength, given by $g_{J_\mu^5 NN}$,  is simply related to its isosinglet axial charge. The other two terms, b) and c), respectively, represent indirect couplings of the composite fields $\Omega$ and $\phi_5$, corresponding to the topological charge density and an isosinglet pseudoscalar field $\phi_5$. The latter field is simply related to the physical $\eta^\prime$. Further discussion in text.}
\label{fig:j5-coupling-hadron}
 \end{figure}

\subsection{Deriving chiral Ward identities from the Wess-Zumino effective action}
\label{sec:chiral-Ward}
To understand the interplay between perturbative and nonperturbative effects in the proton's quark helicity, it is not sufficient to use the effective action for the Dirac fermions in Eq.~(\ref{eq:WDirac}). Instead, we need to start with the full QCD generating functional
\begin{eqnarray}
&&e^{iZ[B,\mathcal{S},\Pi,\Theta]} = \int \mathcal{D}A \int \mathcal{D} \bar{\Psi} \mathcal{D}\Psi e^{iS[A, {\bar \Psi}, \Psi, B,\mathcal{S},\Pi,\Theta]}\,,
\label{eq:WQCD}
\end{eqnarray}
where the action for the theory, in the presence of the external sources $B_\mu$, $\mathcal{S}$, and $\Pi$, we discussed previously, is given by
\begin{eqnarray}
\label{eq:SQCD}
S[A, {\bar \Psi}, \Psi, B,\mathcal{S},\Pi,\Theta] = \int d^4 x\, \Big(\mathcal{L}_{QCD} + B_\mu{\bar \Psi} \gamma^\mu \gamma_5 \Psi + \mathcal{S}{\bar \Psi} \Psi + \Pi {\bar \Psi} i\gamma_5 \Psi + \Theta \Omega\Big) \,.
\end{eqnarray}
Further, $\Theta$ is the external source for the topological charge density $\Omega$.

As shown in \cite{Shore:1991dv}, by taking functional derivatives of the effective action in Eq.~(\ref{eq:WQCD}) with respect to the sources, one can represent the matrix element of the axial-vector current at finite momentum transfer $l^\mu = P'^\mu - P^\mu$ as
\begin{eqnarray}
&&\langle P', S| J^\mu_5 |P, S\rangle = -i g_{J^\mu_{5}NN}(l^2)\, \bar{u}(P',S)\gamma^\mu \gamma_5 u(P,S)
\nonumber\\
&&-i \frac{\delta^2 Z}{\delta B_\mu \delta\Theta}(l^2)\, g_{\Omega NN}(l^2)\,\bar{u}(P',S) \gamma_5 u(P,S) -i \frac{\delta^2 Z}{\delta B_\mu \delta\Pi}(l^2)\, g_{\phi_5 NN}(l^2)\,\bar{u}(P',S) \gamma_5 u(P,S)
\,,
\label{eq:jmu5-Wdecomp}
\end{eqnarray}
where $g_{J^\mu_{5}NN}$, $g_{\Omega NN}$, and $g_{\phi_5 NN}$ denote the 1-particle irreducible (1PI) couplings of the fields $J^\mu_{5}$, $\Omega$, and $\phi_5$ to the nucleon, and $u(P,S)$ ($\bar u (P',S)$)  represents an appropriately normalized spinor representation of the polarized nucleon (adjoint) wavefunction. 
The Fourier transforms of derivatives of the generating functional in Eq.~(\ref{eq:WQCD}) defines the two-point time-ordered vacuum Green functions:
\begin{eqnarray}
&&\frac{\delta^2 Z}{\delta B_\mu \delta\Theta}(l^2) = i \int d^4x \,e^{ilx}\, \langle 0|T J^\mu_5 (x) \Omega(0)|0\rangle\,,
\label{eq:prop-BTh}
\end{eqnarray}
 and
 \begin{eqnarray}
&&\frac{\delta^2 Z}{\delta B_\mu \delta\Pi}(l^2) = i \int d^4x \,e^{ilx} \,\langle 0|T J^\mu_5 (x) \phi_5(0)|0\rangle\,,
\label{eq:prop-BPi}
\end{eqnarray}
where $\phi_5 \equiv i\bar{\Psi}\gamma_5\Psi$. Fig. \ref{fig:j5-coupling-hadron} provides\footnote{Note that the diagrams in Figs. \ref{fig:j5-coupling-hadron}a and \ref{fig:j5-coupling-hadron}b correspond, respectively, to diagrams Figs. 2a and 2d in \cite{Tarasov:2021yll}. Likewise, Fig. \ref{fig:j5-coupling-hadron}c  was split into the two contributions in Figs. 2b and 2c of \cite{Tarasov:2021yll}.} a physical interpretation of the terms in the r.h.s. of Eq.~(\ref{eq:jmu5-Wdecomp}). The first term in Fig. \ref{fig:j5-coupling-hadron}a represents the direct coupling of the $J^\mu_5$ current to the polarized nucleon, while Figs. \ref{fig:j5-coupling-hadron}b and \ref{fig:j5-coupling-hadron}c correspond to the second and third terms on the r.h.s. depicting, respectively, the indirect coupling of $J^\mu_5$ to the target via the vacuum propagators in Eqs.~(\ref{eq:prop-BTh}) and (\ref{eq:prop-BPi}). 

It is instructive to compare Eq. (\ref{eq:jmu5-Wdecomp}) with the decomposition of the nucleon axial-vector current
\begin{eqnarray}
&&\langle P', S|J^\mu_5|P,S\rangle = \bar{u}(P', S)\Big[\gamma^\mu\gamma_5 G_A(l^2) + l^\mu\gamma_5G_P(l^2)\Big]u(P,S)\,,
\label{eq:jmu5-Wdecomp-ff}
\end{eqnarray}
conventionally represented in terms of the axial and pseudoscalar form factors $G_A$ and $G_P$. Comparing Eqs. (\ref{eq:jmu5-Wdecomp}) and (\ref{eq:jmu5-Wdecomp-ff}), they are equivalent for 
\begin{eqnarray}
&&G_A(l^2) = -i g_{J^\mu_{5}NN}(l^2)\,,
\label{eq:ff-rel1}
\end{eqnarray}
and 
\begin{eqnarray}
&& G_P(l^2) = -i \frac{l_\mu}{l^2} \Big[  \frac{\delta^2 Z}{\delta B_\mu \delta\Theta}(l^2)\, g_{\Omega NN}(l^2) + \frac{\delta^2 Z}{\delta B_\mu \delta\Pi}(l^2)\, g_{\phi_5 NN}(l^2) \Big]\,.
\label{eq:ff-rel2}
\end{eqnarray}

We emphasize that the propagators in Eq.~(\ref{eq:prop-BTh}) and (\ref{eq:prop-BPi}) in the decomposition of the matrix element in Eq.~(\ref{eq:jmu5-Wdecomp}) are intrinsically nonperturbative objects. However, as discussed at length in \cite{Veneziano:1989ei,Shore:1990zu,Shore:1991dv}, their properties are constrained by  chiral Ward identities for the two-point Green functions, as well as the dynamical assumptions (``polology"- see for example, Chapter 10 in \cite{Weinberg:1995mt}) corresponding to pole dominance of single-particle poles of the flavor nonet, the weak momentum dependence of the 1-particle irreducible couplings and decay constants, as well as the existence of smoothly behaved topologically nontrivial configurations represented by the topological susceptibilities of QCD and pure Yang-Mills theory~\cite{Shore:1997tq}. 

We will discuss here first the relations derived from the chiral Ward identities, and in the following subsection, the dynamical considerations following from polology. The first of these Ward identities, obtained  from the generating functional in Eq.~(\ref{eq:WQCD}), has the form~\cite{Wess:1971yu}
\begin{eqnarray}
&&\partial_\mu \frac{\delta Z}{\delta B_\mu} - 2N_f \frac{\delta Z}{\delta \Theta} - 2m \frac{\delta Z}{\delta \Pi} + 2\mathcal{S}\frac{\delta Z}{\delta \Pi} - 2\Pi\frac{\delta Z}{\delta \mathcal{S}} = 0\,.
\label{ChWid}
\end{eqnarray}
Indeed, if after performing the functional derivatives, we set the sources to zero, we recover nothing but the singlet form of Eq.~(\ref{eq:anomaly-mass}). The role of the source-dependent terms in this equation are however  crucial for our analysis of the chiral anomaly in QCD. Further functional differentiation with respect to these provide nontrivial contributions to the two-point functions in Eqs.~(\ref{eq:prop-BTh}) and (\ref{eq:prop-BPi}), and therefore contribute to the matrix element of the axial-vector current.

Specifically, differentiating Eq. (\ref{ChWid}) w.r.t $\Theta$ and $\Pi$, and then setting all sources to zero, we obtain, respectively, 
\begin{eqnarray}
&&il_\mu \frac{\delta^2 Z}{\delta B_\mu \delta\Theta}(l^2) - 2N_f \frac{\delta^2 Z}{\delta \Theta\delta\Theta}(l^2) - 2m \frac{\delta^2 Z}{\delta \Pi\delta\Theta}(l^2) = 0\,,
\label{eq:tpf-Om}
\end{eqnarray}
and
\begin{eqnarray}
&&il_\mu \frac{\delta^2 Z}{\delta B_\mu\delta\Pi}(l^2) - 2N_f \frac{\delta^2 Z}{\delta \Theta\delta\Pi}(l^2) - 2m \frac{\delta^2 Z}{\delta \Pi\delta\Pi}(l^2) - 2\langle\phi\rangle = 0\,.
\label{eq:tmp-Phy}
\end{eqnarray}
In the last equation,  $ \langle\phi\rangle \equiv \frac{\delta Z}{\delta S}|_{\mathcal{S}=0} = \langle \bar{\Psi}\Psi\rangle $ is the quark condensate responsible for spontaneous chiral symmetry breaking in QCD. The  appearance of this fundamental quantity in the chiral Ward identity follows directly from the source-dependent terms in Eq.~(\ref{ChWid}). This term explains why the QED resolution of the anomaly pole doesn't work in the case of QCD. 

Indeed, if that mechanism worked, the contribution of the anomaly (second) and mass (third) terms in Eq. (\ref{ChWid}) would cancel each other in the forward limit. Therefore, in the forward limit, we would have a cancellation between the anomaly and mass terms in Eq. (\ref{eq:tpf-Om}) as well as in Eq. (\ref{eq:tmp-Phy}). 
However this QED (or purely perturbative QCD) mechanism for the regularization of the infrared singularity in the forward limit doesn't regulate the pole of the quark condensate term $\frac{\delta W}{\delta \mathcal{S}}$ in Eq. (\ref{eq:tmp-Phy}).
Since the first term in this equation  contributes to the matrix element of the axial-vector current in Eq. (\ref{eq:jmu5-Wdecomp}), this would lead to an unregulated pole in the matrix element. 

We conclude therefore that the QED mechanism of the infrared pole regularization simply fails in QCD and a completely different scenario must take its place, as understood already in the seminal papers \cite{Veneziano:1989ei,Jaffe:1989jz} and discussed at length by us in  \cite{Tarasov:2020cwl,Tarasov:2021yll}. The regularization of the anomaly pole  is due to a ``primordial" $\bar{\eta}$ Goldstone boson exchange (for $N_c\rightarrow \infty$) and its WZW coupling to the topological charge density $\Omega$; as is well-known, this latter coupling topologically generates the large mass of the $\eta'$. Our discussion in \cite{Tarasov:2021yll} was in the chiral limit. As we will discussion in the next subsection, our conclusion is not modified qualitatively for massive quarks, though there will be interesting consequences that we will emphasize. 

Before we discuss this mechanism in detail, 
we will address further the consistency of the form of the matrix element in Eq.~(\ref{eq:jmu5-Wdecomp}) with the anomaly equation for the axial-vector current in Eq.~(\ref{eq:anomaly-mass}). Using Eqs. (\ref{eq:jmu5-Wdecomp}), (\ref{eq:tpf-Om}), and (\ref{eq:tmp-Phy}) to write the divergence of this matrix element as
\begin{eqnarray}
\label{eq:anomaly-mass-functional}
&&\langle P', S| \partial_\mu J^\mu_5 |P, S\rangle =  \Bigg[2M_N \, g_{J^\mu_{5}NN}(l^2)\, 
- 2\langle\phi\rangle \, i g_{\phi_5 NN}(l^2)
\nonumber\\
&&+ 2N_f \Big( -i  \frac{\delta^2 Z}{\delta \Theta\delta\Theta}(l^2) \, g_{\Omega NN}(l^2)\, -i \frac{\delta^2 Z}{\delta \Theta\delta\Pi}(l^2)\, g_{\phi_5 NN}(l^2)\, \Big)
\nonumber\\
&&+ 2m\Big( -i \frac{\delta^2 Z}{\delta \Pi\delta\Theta}(l^2)\, g_{\Omega NN}(l^2)\, -i \frac{\delta^2 Z}{\delta \Pi\delta\Pi}(l^2)\, g_{\phi_5 NN}(l^2)\ \Big) \Bigg]\bar{u}(P',S) \gamma_5 u(P,S)
\,,
\end{eqnarray}
where $M_N$ is the nucleon mass. 

Now following the same 1PI logic as previously, and rewriting the derivatives of the generating functional with respect to $\Theta$ and $\Pi$ as
\begin{eqnarray}
\langle P', S| \Omega |P, S\rangle = -i  \frac{\delta^2 Z}{\delta \Theta\delta\Theta}(l^2) \, g_{\Omega NN}(l^2)\,\bar{u}(P',S) \gamma_5 u(P,S) -i \frac{\delta^2 Z}{\delta \Theta\delta\Pi}(l^2)\, g_{\phi_5 NN}(l^2)\,\bar{u}(P',S) \gamma_5 u(P,S)\,,
\end{eqnarray}
as well as  
\begin{eqnarray}
\langle P', S| \phi_5 |P, S\rangle = -i \frac{\delta^2 Z}{\delta \Pi\delta\Theta}(l^2)\, g_{\Omega NN}(l^2)\,\bar{u}(P',S) \gamma_5 u(P,S) -i \frac{\delta^2 Z}{\delta \Pi\delta\Pi}(l^2)\, g_{\phi_5 NN}(l^2)\,\bar{u}(P',S) \gamma_5 u(P,S)\,,
\end{eqnarray}
we can substitute the r.h.s expressions, which both appear in Eq.~(\ref{eq:anomaly-mass-functional}), with their l.h.s counterparts. Comparing the result with Eq. (\ref{eq:anomaly-mass}),  we see that to satisfy the anomaly equation for the axial-vector current we have to require for consistency that 
\begin{eqnarray}
&&  2M_N \, g_{J^\mu_{5}NN}(l^2) =  2\langle\phi\rangle \, i g_{\phi_5 NN}(l^2)
\,.
\label{eq:cfuncs}
\end{eqnarray}

Using Eq.~(\ref{eq:ff-rel1}) we can rewrite the expression above as
\begin{eqnarray}
&&  2M_N \, G_A(l^2) =  2\langle\phi\rangle \, g_{\phi_5 NN}(l^2)\,.
\label{eq:cfuncs-ff}
\end{eqnarray}
This result is valid at all values of the momenta $l$ and will play an important role in our subsequent discussion. In particular, in the forward limit, it will yield us the Goldberger-Treiman relation for the isosinglet axial form factor $G_A$.

\subsection{The $\eta^\prime$ mass and $\Delta \Sigma$ from polology and topology}
\label{sec:polology-topology}
To make further progress in employing the chiral Ward identities and WZW terms, as mentioned, we will have to make further dynamical assumptions on the pole dominance of  nonperturbative correlation functions and the smooth behavior of the 1PI couplings as a function of $l^2$. As previously, for simplicity, we will restrict our discussion of the finite quark mass generalization of \cite{Tarasov:2021yll} to the flavor singlet sector corresponding to the dynamics of the $\eta^\prime$ meson. How one should extend our discussion of this dynamics to include mixing between the singlet and octet sectors is well-known~\cite{HerreraSiklody:1996pm,Leutwyler:1996sa}, and highly developed~\cite{Gan:2020aco}. However for simplicity, we will omit this sophisticated and important discussion for the purposes of this work. In the context of polarized DIS, this generalization is included in the work of Narison, Shore and Veneziano, summarized in \cite{Shore:2007yn}. As noted, our approach is complementary to theirs and can be generalized similarly. 

Our starting point is a systematic $N_f/N_c$ expansion around the $N_c\rightarrow \infty$ limit~\cite{DiVecchia:1980yfw,Leutwyler:1997yr}, where the triangle graph is suppressed and $U_A(1)$ is restored. This limit is characterized by a primordial $\eta^\prime$ pseudoscalar Goldstone boson labeled $\bar{\eta}$, whose  exchange is defined by the propagator
\begin{eqnarray}
&&\int d^4x\, e^{ilx}\, \langle 0|T \bar{\eta}(x) \bar{\eta}(0)|0\rangle
= \frac{i}{l^2}\,.
\label{eq:prop-eta-bar}
\end{eqnarray}
In the chiral limit, the $\bar \eta$ is a QCD axion that couples to the topological charge density $\Omega$ via the WZW term in Eq.~(\ref{eq:S-WZW}), with strength $O(1/\sqrt{N_c})$. As a consequence of the multiple interactions of the $\bar \eta$, as it propagates through the vacuum, we noted earlier in Sec.~\ref{sec:Anomaly-cancellation}  that its  pole is shifted to ${\tilde m}^2_{\eta'}$. Specifically, in the chiral limit, when $l^2\to 0$, and as illustrated in Fig. \ref{fig:etapr-mass-generation}, this corresponds to iteration of ``bubbles" of the Yang-Mills topological susceptibility, resulting in the modified $\eta^\prime$ propagator 
\begin{eqnarray}
\int d^4x\, e^{ilx}\, \langle 0|T \eta^\prime(x) \eta^\prime(0)|0\rangle|_{\rm m=0} &&\simeq \frac{i}{l^2} + \frac{i}{l^2} \cdot \Big[-i\frac{\sqrt{2N_f}}{F_{\bar \eta}}\Big] \cdot \Big[-i \chi_{\rm YM}(0)\Big]\cdot \Big[-i\frac{\sqrt{2N_f}}{F_{\bar \eta}}\Big] \cdot \frac{i}{l^2} + \dots
\nonumber\\
&&= \frac{i}{l^2- {\tilde m}_{\eta'}^2}\,,
\label{eq:prop-eta-pr}
\end{eqnarray}
with 
\begin{eqnarray}
\label{eq:YM-top-susceptibility}
    \chi_{\rm YM}(l^2)=\int d^4x\, e^{ilx}\, \langle 0|T \Omega(x) \Omega(0)|0\rangle_{\rm YM}\,,
\end{eqnarray}
where $\chi_{\rm YM}(l^2)$ is the Yang-Mills topological susceptibility, and
\begin{eqnarray}
&&{\tilde m}_{\eta'}^2 = -\frac{2N_f}{F^2_{\bar \eta}} \chi_{\rm YM}(0)\,.
\label{eq:WV-etapr-mass}
\end{eqnarray}
The latter is of course the well-known Witten-Veneziano formula for the mass $\tilde m_{\eta'}$ of the $\eta'$ meson in the chiral limit~\cite{Witten:1979vv,Veneziano:1979ec}. 
For finite quark masses,  this must be replaced by the physical $\eta^\prime$ propagator 
\begin{eqnarray}
\int d^4x\, e^{ilx}\, \langle 0|T \eta^\prime(x) \eta^\prime(0)|0\rangle = \frac{i}{l^2- m^2_{\eta'}}
\label{eq:prop-eta-pr-mass}
\end{eqnarray}
where $m^2_{\eta'}$ is the measured mass of the $\eta^\prime$ meson.

To obtain a consistent extension of the results for $\Delta \Sigma$ in Paper II  to finite quark mass, Eq.~(\ref{eq:anomaly-mass-functional}) indicates that we need to know the Green functions corresponding to 
$\frac{\delta^2 Z}{\delta \Theta\delta\Theta}(l^2)$, $\frac{\delta^2 Z}{\delta \Theta\delta\Pi}(l^2)$ and 
$\frac{\delta^2 Z}{\delta \pi\delta\Pi}(l^2)$. We will consider each of these individually, beginning with the topological susceptibility. 

\subsubsection{Topological susceptibility for finite quark mass}

\begin{figure}[htb]
 \begin{center}
\includegraphics[width=170mm]{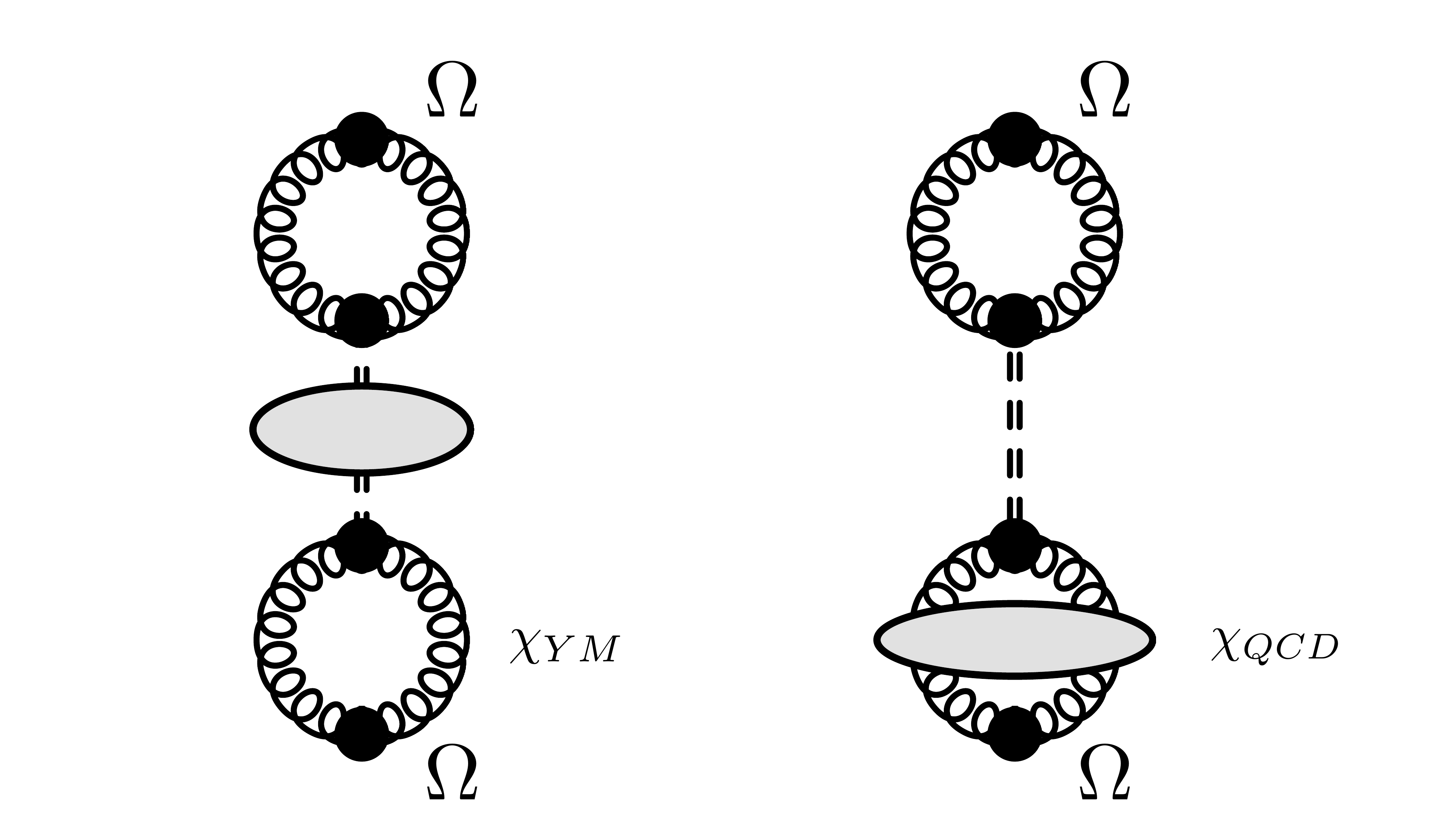}
 \end{center}
\caption{The two-point Green function $\frac{\delta^2 W}{\delta \Theta \delta\Theta}$ representing the QCD topological susceptibility $\chi_{\rm QCD}(l^2)$ for finite quark mass. Left: The WZW term generates iterations of Yang-Mills bubbles, that for finite quark mass, are resummed into the grey blob representing the dressing of the $\bar \eta$ propagator to generate the physical $\eta^\prime$ propagator with mass $m_{\eta'}$. This diagram is expressed as the second term in Eq.(\ref{eq:top-suspt1}). Right:  The QCD topological susceptibility alternately depicted as a YM topological bubble interacting with the $\bar \eta$ via the WZW vertex. The $\bar \eta$ propagating here has mass 
$m_{\bar \eta}$. It further interacts via another WZW vertex with the full QCD topological susceptibility represented by the bubble with the grey blob at bottom. The diagram is expressed as the second term in Eq.~(\ref{eq:top-suspt2}).}
\label{fig:ThTh-gree-func}
 \end{figure}

 We begin with the formal expression for the QCD topological susceptibility,
\begin{eqnarray}
\frac{\delta^2 Z}{\delta \Theta\delta\Theta}(l^2)\Big|_{\theta=0}\equiv \chi_{\rm QCD}(l^2)\,.
\end{eqnarray}
Employing the WZW term coupling the topological charge density to the $\bar \eta$, we can express the r.h.s either as 
\begin{eqnarray}
&& \chi_{\rm QCD}(l^2) = \chi_{\rm YM}(l^2) + i \Big[ -i\chi_{\rm YM}(l^2) \Big] \, \Big[-i\frac{\sqrt{2N_f}}{F_{\bar{\eta}}}b(l^2)\Big] \, \frac{i}{l^2 - m^2_{\eta'}} \, \Big[-i\frac{\sqrt{2N_f}}{F_{\bar{\eta}}}b(l^2)\Big] \, \Big[ -i\chi_{\rm YM}(l^2) \Big]\,,
\label{eq:top-suspt1}
\end{eqnarray}
or 
\begin{eqnarray}
&&\chi_{\rm QCD}(l^2) = \chi_{\rm YM}(l^2) + i \Big[ -i\chi_{\rm YM}(l^2) \Big] \, \Big[-i\frac{\sqrt{2N_f}}{F_{\bar{\eta}}}b(l^2)\Big] \, \frac{i}{l^2 - m^2_{\bar{\eta}}} \, \Big[-i\frac{\sqrt{2N_f}}{F_{\bar{\eta}}}b(l^2)\Big] \, \Big[ -i\chi_{\rm QCD}(l^2) \Big]\,.
\label{eq:top-suspt2}
\end{eqnarray}
These two expressions, which are equivalent ways of decomposing the QCD topological susceptibility, require some discussion. The second term in Eq.~(\ref{eq:top-suspt1}) is shown in Fig.~\ref{fig:ThTh-gree-func} (left). The bubble at the top is the Yang-Mills topological susceptibility. Its coupling to the $\bar \eta$ is given by the  usual WZW coefficient but we have now multiplied it by a factor $b(l^2)$ to reflect the possible $l^2$ dependence of this term. As we shall see, its impact, while quantitatively very small, will be important for the consistency of our framework. The pseudoscalar that's exchanged is now the physical $\eta^\prime$, depicted by the dashed lines with the grey blob insertion. This quantity therefore includes all the WZW iterations as well as finite mass corrections. This $\eta^\prime$ meson then connects to the YM bubble at the bottom with the vertex again represented by the generalized WZW coefficient. 

However the QCD topological susceptibility can be equally well decomposed in the manner shown in Fig.~\ref{fig:ThTh-gree-func} (right) corresponding to the second term in Eq.~(\ref{eq:top-suspt2}). In this case, the exchange is that of the bare $\bar \eta$ with mass $m_{\bar \eta}$, denoted by the grey blob on the dashed lines. This quantity again couples via the generalized WZW term, but the quantity at the bottom (represented by the gluon bubble dressed with the grey blob) is the full QCD topological susceptibility. 
In the chiral limit, $m_{\bar\eta}\rightarrow 0$ and $m_{\eta'}\rightarrow {\tilde m}_{\eta'}$, the latter given by the Witten-Veneziano formula in Eq.~(\ref{eq:WV-etapr-mass}). One may anticipate that $m_{\bar \eta}$ should be given by a DGMOR-type relation and we will see that will be the case. 

Equating the r.h.s of Eqs.~(\ref{eq:top-suspt1}) and (\ref{eq:top-suspt2}) gives 
\begin{eqnarray}
\label{eq:l-dep-chi}
&&\chi_{\rm QCD}(l^2) = \frac{l^2 - m^2_{\bar{\eta}}}{l^2 - m^2_{\eta'}} \chi_{\rm YM}(l^2)\,,
\end{eqnarray}
and plugging this back into either Eq.(\ref{eq:top-suspt2}) or Eq.(\ref{eq:top-suspt2}) gives 
\begin{eqnarray}
\label{eq:l-dep-b}
&&b^2(l^2) \chi_{\rm YM}(l^2) = - \frac{F^2_{\bar{\eta}}}{2N_f} [m^2_{\eta'} - m^2_{\bar{\eta}}]\,.
\end{eqnarray}
This expression explains the necessity of introducing $b(l^2)$ in the generalized WZW coupling. Since the masses 
are constants, the variation in $b(l^2)$ with $l^2$ must compensate identically for the variation in $\chi_{\rm YM}(l^2)$.

We will now expand the l.h.s and r.h.s of Eqs.~(\ref{eq:l-dep-chi}) and (\ref{eq:l-dep-b}) to quadratic order in $l^2$ and equate expressions at each order to fix the unknowns in these expressions. Expanding Eq.~(\ref{eq:l-dep-chi}) gives 
\begin{eqnarray}
\label{eq:chi+chi-prime}
\chi_{\rm QCD}(0) = \frac{m^2_{\bar{\eta}}}{m^2_{\eta'}} \chi_{\rm YM}(0)\qquad {\rm and}\qquad
\chi'_{\rm QCD}(0) = \frac{m^2_{\bar{\eta}} - m^2_{\eta'}}{m^4_{\eta'}}\chi_{\rm YM}(0) + \frac{m^2_{\bar{\eta}}}{m^2_{\eta'}} \chi'_{\rm YM}(0)\,.
\end{eqnarray}
In the chiral limit where $m_{\bar\eta}\rightarrow 0$ and $m_{\eta'}\rightarrow {\tilde m}_{\eta'}$, we recover  
the results we obtained previously in Paper II, namely, 
\begin{eqnarray}
\label{eq:main-ch-chi}
    \chi_{\rm QCD}|_{m=0}(0) = 0 \qquad {\rm and} \qquad F_{\bar \eta}^2= 2\,N_f\, \chi'_{\rm QCD}|_{m=0}(0) \,,
\end{eqnarray} 
where in the second equality, we employed the Witten-Veneziano formula  in Eq.~(\ref{eq:WV-etapr-mass}). 

Now expanding the l.h.s and r.h.s of Eq.~(\ref{eq:l-dep-b}) to quadratic order and equating the respective coefficients at each order, we get 
\begin{eqnarray}
\label{eq:brl}
b^2(0)\chi_{\rm YM}(0) = - \frac{F^2_{\bar{\eta}}}{2N_f} [m^2_{\eta'} - m^2_{\bar{\eta}}]\qquad {\rm and}\qquad
b(0) \chi'_{\rm YM}(0) + 2 b'(0) \chi_{\rm YM}(0) = 0\,. 
\end{eqnarray}
Again, taking the chiral limit, we see that for consistency with the Witten-Veneziano formula, we will 
need to require that $b(0)=1$. This is satisfying since one recovers precisely the usual WZW vertex when $l^2\rightarrow 0$. From the first equation above, we then obtain
\begin{eqnarray}
\label{eq:WVfm}
m^2_{\eta'} = - \frac{2N_f}{F^2_{\bar{\eta}}} \chi_{\rm YM}(0) + m^2_{\bar{\eta}} 
\end{eqnarray}
which represents a finite mass correction to the Witten-Veneziano formula (\ref{eq:WV-etapr-mass}).

From the second equation (\ref{eq:brl}) we also get
\begin{eqnarray}
\label{eq:bpr-sec}
    b'(0) = -\frac{\chi'_{\rm YM}(0)}{2 \chi_{\rm YM}(0)}\,.
\end{eqnarray}
As we will discuss further shortly, lattice results \cite{Bonanno:2023ple} indicate that this dimensionful coefficient multiplying $l^2$ is very small, indicative of a weak $l^2$ dependence of $b(l^2)$.

\subsubsection{The two-point Green function  $\langle 0|T \phi_5 (x) \Omega(0)|0\rangle$}

This quantity, which also appears in Eq.~(\ref{eq:anomaly-mass-functional}), is obtained by computing $\frac{\delta^2 Z}{\delta \Theta\delta\Pi}(l^2)$. 
\begin{figure}[htb]
 \begin{center}
\includegraphics[width=170mm]{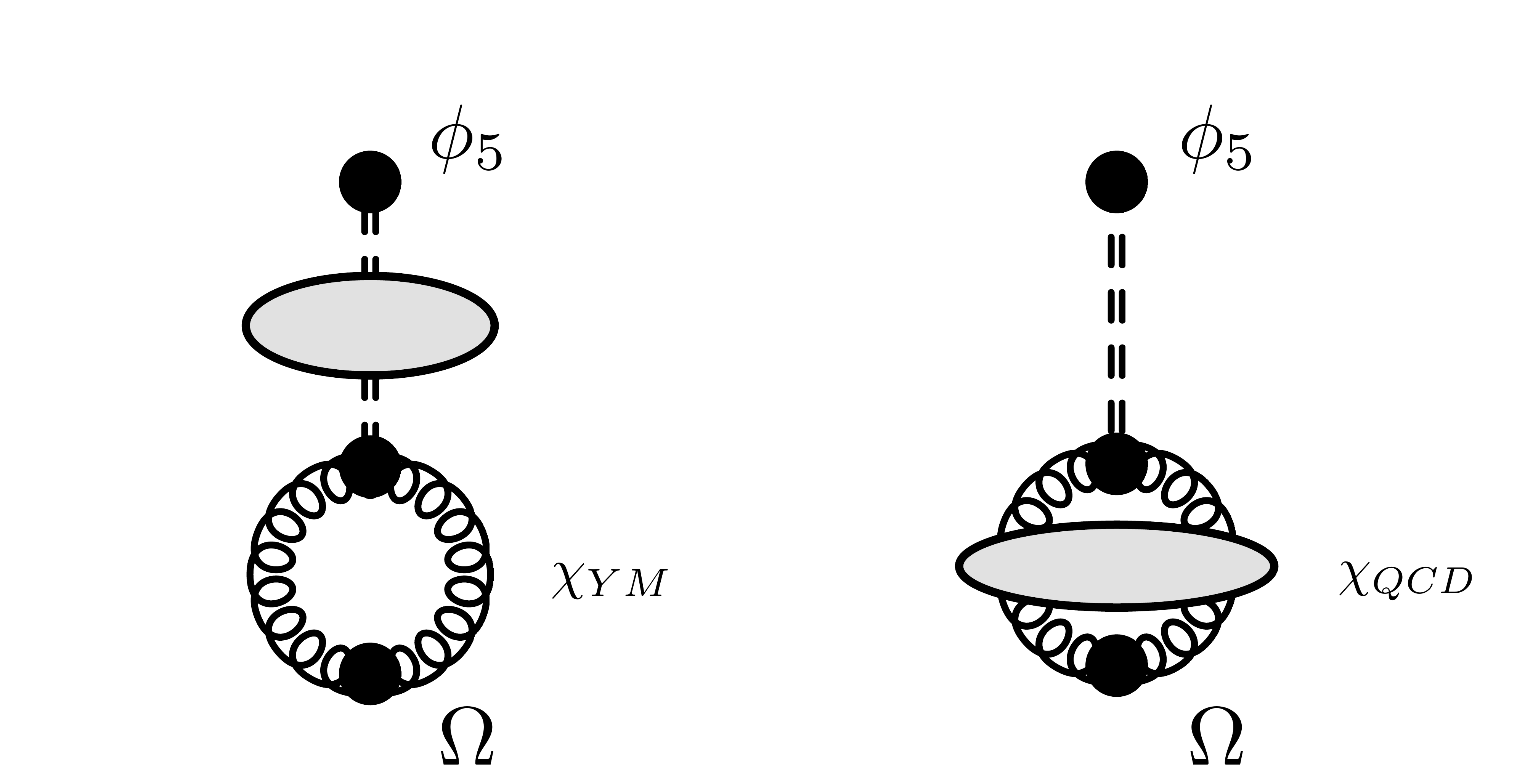}
 \end{center}
\caption{Diagrams representing the two-point Green function $\frac{\delta^2 W}{\delta \Pi \delta\Theta}$ appearing in the chiral Ward identities in Eqs. (\ref{eq:tpf-Om}) and (\ref{eq:tmp-Phy}). Left: The grey blob represents the dressing of the free $\bar \eta$ propagator to generate the physical $\eta^\prime$ propagator. The closed bubble represents the Yang-Mills topological susceptibility. This diagram represents Eq.~(\ref{eq:mixed-1}). Right: The dashed line here corresponds to the propagator with mass $m_{\bar \eta}$ which interacts via the WZW vertex with the full QCD topological susceptibility. The equation corresponding to this diagram is Eq.~(\ref{eq:mixed-2}).}
\label{fig:PTh-gree-func}
 \end{figure}
As previously for the topological susceptibility, we can also write this expression, as shown in Fig.~\ref{fig:PTh-gree-func}, in two equivalent ways. The first, corresponding to the left figure, can be expressed as 
\begin{eqnarray}
\label{eq:mixed-1}
&&\frac{\delta^2 Z}{\delta \Pi \delta\Theta}(l^2) = i \int d^4x\, e^{ilx}\, \langle 0|T \phi_5 (x) \Omega(0)|0\rangle = \frac{1}{l^2 - m^2_{\eta'}} \frac{\sqrt{2N_f}}{F_{\bar{\eta}}} b(l^2)\chi_{\rm YM}(l^2) \,c(l^2)\,,
\end{eqnarray}
while the right figure can be expressed as 
\begin{eqnarray}
\label{eq:mixed-2}
&&\frac{\delta^2 Z}{\delta \Pi \delta\Theta}(l^2) = i \int d^4x\, e^{ilx}\, \langle 0|T \phi_5 (x) \Omega(0)|0\rangle = \frac{1}{l^2- m^2_{\bar{\eta}}} \frac{\sqrt{2N_f}}{F_{\bar{\eta}}} b(l^2)\chi_{\rm QCD}(l^2) \,c(l^2)\,.
\end{eqnarray}
Here $c(l^2)$ is a smoothly behaved interpolating function relating the pseudoscalar quark bilinear $\phi_5$ to the $\eta'$ field. Equating the r.h.s of the above two equations gives precisely Eq.~(\ref{eq:l-dep-chi}), which we would expect for the consistency of our expressions. 

We will now make use of the chiral Ward identities in Eqs.(\ref{eq:tpf-Om}) and (\ref{eq:tmp-Phy}). The first of these can be reexpressed as 
\begin{eqnarray}
&& \frac{\delta^2 Z}{\delta B_\mu \delta\Theta}(l^2) = -i \frac{l^\mu}{l^2}\Big[ 2N_f \chi_{\rm QCD}(l^2) + 2m \frac{\delta^2 Z}{\delta \Pi\delta\Theta}(l^2) \Big]\,.
\end{eqnarray}
Since there cannot be poles in this function because it contributes to the pseudoscalar form factor $G_P$, as shown in Eq.~(\ref{eq:ff-rel2}), the expression in the parenthesis should vanish in the forward limit: 
\begin{eqnarray}
&& 2N_f \chi_{\rm QCD}(0) + 2m \lim_{l^2\to 0}\frac{\delta^2 Z}{\delta \Pi\delta\Theta}(l^2) = 0\,.
\end{eqnarray}
From the r.h.s. of Eq.(\ref{eq:mixed-1}), this gives, 
\begin{eqnarray}
\label{eq:forward-theta-pi}
&&\lim_{l^2\to 0}\frac{\delta^2 Z}{\delta \Pi \delta\Theta}(l^2) = -\frac{1}{m^2_{\eta'}} \frac{\sqrt{2N_f}}{F_{\bar{\eta}}}\chi_{\rm YM}(0) c(0)\,,
\end{eqnarray}
where $b(0)=1$, as shown previously. One thus obtains 
\begin{eqnarray}
&&c(0) = m^2_{\eta'} \frac{2N_f}{2m} \frac{F_{\bar{\eta}}}{\sqrt{2N_f}} \frac{\chi_{\rm QCD}(0)}{\chi_{\rm YM}(0)}
\label{eq:c0-prel}
\end{eqnarray}

Let's now consider the second chiral Ward identity (Eq.~(\ref{eq:tmp-Phy})), which can be expressed as 
\begin{eqnarray}
\label{eq:tpf-BPi}
&& \frac{\delta^2 Z}{\delta B_\mu\delta\Pi}(l^2) = -i \frac{l^\mu}{l^2}\Big[ 2N_f \frac{\delta^2 Z}{\delta \Theta\delta\Pi}(l^2) + 2m \frac{\delta^2 Z}{\delta \Pi\delta\Pi}(l^2) + 2\langle\phi\rangle \Big]\,.
\end{eqnarray}
Again, since the terms inside the parenthesis should vanish in the forward limit, and substituting therein Eq.~(\ref{eq:forward-theta-pi}), and the r.h.s of 
\begin{eqnarray}
\label{eq:vanish-2-chiral}
\frac{\delta^2 Z}{\delta\Pi \delta\Pi}(l^2) = i\int d^4x\, e^{ilx}\, \langle 0|T \phi_5(x) \phi_5(0)|0\rangle = \frac{-1}{l^2- m^2_{\eta'}}c^2(l^2)\,,
\end{eqnarray}
in the forward limit $l^2\rightarrow 0$, we obtain the identity
\begin{eqnarray}
&& - 2N_f \frac{1}{ m^2_{\eta'}} \frac{\sqrt{2N_f}}{F_{\bar{\eta}}} \chi_{\rm YM}(0) c(0) + 2m \frac{1}{m^2_{\eta'}}c^2(0) + 2\langle\phi\rangle = 0\,.
\end{eqnarray}
This simplifies greatly in the chiral limit, allowing us to determine 
\begin{eqnarray}
\label{eq:c0}
&&c(0) = - \frac{2\langle\phi\rangle}{2N_f} \frac{\sqrt{2N_f}}{ F_{\bar{\eta}}}\,.
\end{eqnarray}
Plugging Eq.~(\ref{eq:c0}), as well as the identity 
\begin{eqnarray}
\label{eq:mass-relation}
    m_{\eta^\prime}^2 = m_{\bar \eta}^2 \frac{\chi_{\rm YM}(0)}{\chi_{\rm QCD}(0)}\,,
\end{eqnarray}
obtained from Eq.~(\ref{eq:l-dep-chi}) in the forward limit, back into Eq.~(\ref{eq:c0-prel}), 
we obtain 
\begin{eqnarray}
    m_{\bar \eta}^2 = - \frac{2\,m \langle \phi\rangle}{N_f F_{\bar\eta}^2}\,,
\end{eqnarray}
which is the DGMOR relation we anticipated. 

With the help of the chiral Ward identities, we now have all the ingredients necessary to compute the finite mass 
expression for $\Delta \Sigma$. Combining the rightmost equation in Eq.~(\ref{eq:chi+chi-prime}) with 
Eq.~(\ref{eq:WVfm}), and expanding up to quadratic order in $m_{\bar\eta}$, we obtain 
\begin{eqnarray}
\label{eq:Fchipr-fm}
&&F^2_{\bar{\eta}} = 2N_f \chi'_{\rm QCD}(0) \Big(1 - m^2_{\bar{\eta}} \frac{2  \chi'_{\rm QCD}(0) -  \chi'_{\rm YM}(0)}{\chi_{\rm YM}(0)} \Big) + O(m^4_{\bar{\eta}})\,,
\end{eqnarray}
which defines the finite mass correction to the second identity in Eq. (\ref{eq:main-ch-chi}),  derived in Paper II.

Eq. (\ref{eq:Fchipr-fm}) is one of the principal results of this paper. It allows us to relate $\Delta \Sigma$ to the slope of the topological susceptibility away from the chiral limit. As we will now see, the numerical value of the finite mass correction is quite small. To figure this out, let's further analyze the form of the correction by employing a relation between $\chi'_{\rm YM}(0)$ and $\chi'_{\rm QCD}|_{m=0}(0)$ that we can derive from our previous expressions. Specifically, it can be obtained from Eq. (\ref{eq:tpf-BPi}) using Eqs. (\ref{eq:mixed-1}) and (\ref{eq:vanish-2-chiral}) in the r.h.s of the equation, and writing the l.h.s as\footnote{For some of the details of the derivation, see Appendix B.}
\begin{eqnarray}
\label{eq:tpbmpi-eta-F}
\frac{\delta^2 Z}{\delta B_\mu\delta\Pi}(l^2) = il^\mu \sqrt{2N_f}\,F_{\bar{\eta}}(l^2)\, \frac{-1}{l^2 - m^2_{\eta'}} \, c(l^2)\,.
\end{eqnarray}
Taylor expanding this equation in powers of $l^2$,
we find that while the leading term of the expansion is trivial, at second order it yields the relation
\begin{eqnarray}
\label{eq:bprb}
&&\chi_{\rm YM}(0) b'(0) + \chi'_{\rm YM}(0) = \frac{F^2_{\bar{\eta}}}{N_f}\,.
\end{eqnarray}
Combining this result with Eq. (\ref{eq:bpr-sec}) we obtain\footnote{Note that in our derivation we assume 
$c(l^2)\approx c(0)$ and therefore neglect its slope given by $c'(0)$.}
\begin{eqnarray}
\label{eq:simp-rel}
 \chi'_{\rm YM}(0) = \frac{2}{N_f}F^2_{\bar{\eta}}\,.
\end{eqnarray}
Further, comparing with the second relation in Eq. (\ref{eq:main-ch-chi}), we find 
\begin{eqnarray}
\label{eq:chiprrel}
 \chi'_{\rm YM}(0) = 4 \chi'_{\rm QCD}|_{m=0}(0)\,.
\end{eqnarray}

As a result, taking into account Eqs. (\ref{eq:main-ch-chi}) and (\ref{eq:WVfm}), the finite mass correction in Eq. (\ref{eq:Fchipr-fm}) can be estimated as
\begin{eqnarray}
&&1 - m^2_{\bar{\eta}} \frac{2  \chi'_{\rm QCD}(0) -  \chi'_{\rm YM}(0)}{\chi_{\rm YM}(0)} \approx 1 + 2N_f \frac{m^2_{\bar{\eta}}}{m^2_{\eta'}} \frac{2  \chi'_{\rm QCD}|_{m=0}(0) -  \chi'_{\rm YM}(0)}{F^2_{\bar{\eta}}}
\nonumber\\
&&= 1 - 2N_f \frac{m^2_{\bar{\eta}}}{m^2_{\eta'}} \frac{ 2 \chi'_{\rm QCD}|_{m=0}(0)}{F^2_{\bar{\eta}}} = 1 - 2 \frac{m^2_{\bar{\eta}}}{m^2_{\eta'}}\,.
\end{eqnarray}
In the absence of the anomaly, the Weinberg bound~\cite{PhysRevD.11.3583} would give this mass to be maximally $m_{\bar \eta} = \sqrt{3} \,m_\pi$, corresponding to at most a $12\%$ correction to $F_{\bar\eta}^2$.

We will now finally relate $\Delta \Sigma$ to the slope of the topological susceptibility for the finite mass case. Taking the forward limit of Eq. (\ref{eq:cfuncs-ff}) we obtain the G-T relation 
\begin{eqnarray}
  2M_N \, G_A(0) =   \sqrt{2N_f} \, F_{\bar{\eta}}\, g_{\bar{\eta} NN}\,.
\end{eqnarray}
Here we have employed the fact that the field redefinition from $\phi_5\rightarrow \eta'$ gives the coupling constant $g_{\eta' NN} \equiv -g_{\phi_5 NN}(0)/c(0)$, and further\footnote{The latter assumes that the glueball component of the $\eta^\prime$ is small - for a discussion in our context, see \cite{Shore:2007yn}, and more generally, the review \cite{Bali:2021qem}.} that $g_{\eta'NN}\approx g_{\bar\eta NN} $. 
We can now use the G-T relation to relate $\Delta \Sigma$ to the decay constant $F_{\bar{\eta}}$
\begin{eqnarray}
\Delta \Sigma =   \frac{\sqrt{2N_f}}{2M_N} \, F_{\bar{\eta}}\, g_{\bar{\eta} NN}\,,
\end{eqnarray}
and further utilize Eq. (\ref{eq:Fchipr-fm}) to relate it to the slope of the topological susceptibility
\begin{eqnarray}
\label{eq:Sigma-final-1}
\Delta \Sigma =   \frac{2N_f}{2M_N} \sqrt{\chi'_{\rm QCD}(0)} \Big(1 - m^2_{\bar{\eta}} \frac{2  \chi'_{\rm QCD}(0) -  \chi'_{\rm YM}(0)}{\chi_{\rm YM}(0)} \Big)^{1/2}\, g_{\bar{\eta} NN}\,.
\end{eqnarray}
This is our final result for the finite mass correction to Eq. (\ref{eq:chiral-DeltaSigma}). 

This result for $\Delta \Sigma $ for finite quark mass introduces a correction of only a few percent to the result in Paper II (previously obtained in \cite{Shore:1991dv}) for the chiral limit, 
\begin{eqnarray}
\label{eq:chiral-lim}
\Delta \Sigma|_{m=0}=   \frac{N_f}{M_N} \sqrt{\chi'_{\rm QCD}(0)} \, g_{\bar{\eta} NN}\,.
\end{eqnarray}
Note that the latter, using Eq. (\ref{eq:chiprrel}), can equivalently be written as 
\begin{eqnarray}
\Delta \Sigma|_{m=0} =  \frac{N_f}{2M_N} \sqrt{\chi'_{\rm YM}(0)} \, g_{\bar{\eta} NN}\,.
\end{eqnarray}
We will now discuss briefly the phenomenological consequences of these results. 

\subsection{Large $N_c$ phenomenology}
\label{sec:large-N}
Both $\chi'_{\rm YM}$ and $\chi'_{\rm QCD}$ have been computed in lattice gauge theory with a range of numerical algoritms. Ref.~\cite{Bonanno:2023ple} contains a state-of-the-art computation of the former (and key references) of the status of these computations as well as comparisons to sum rule~\cite{Narison:2021kny} and chiral perturbation theory~\cite{Leutwyler:2000jg} results.  For $SU(3)$ pure Yang-Mills, the recent lattice result is 
$\chi'_{\rm YM}=(17.1\pm 2.1\, {\rm MeV})^2$. For QCD in the chiral limit, there are very few computations of $\chi'_{\rm QCD}$, with significant ambiguities in the computation-see for instance \cite{Koma:2010vx} for a discussion of these. 

As emphasized by Narison, Shore and Veneziano~\cite{Shore:1990zu,Narison:1998aq,Shore:2007yn,Narison:2021kny}, measurements of $\Delta \Sigma$ in polarized DIS can provide key insight into the magnitude of OZI violation in QCD. The OZI rule predicts that the proton's isosinglet axial form factor $G_A$ is equal to $2\sqrt{3}\,G_A^{(8)}$, where $G_A^{(8)}$ is the isooctet form factor; the latter is known from hyperon decays, which gives us~\cite{Narison:1994hv}
\begin{eqnarray}
\label{eq:Sigma-OZI}
    \Delta \Sigma^{\rm OZI} \equiv G_A^{\rm OZI} = 0.579\pm 0.021 \,.
\end{eqnarray}
 From the estimate~\cite{Shore:2007yn} and employing the FLAG central value of $F_\pi=80.3$ MeV~\cite{FlavourLatticeAveragingGroupFLAG:2021npn}, a back of the envelope estimate gives 
\begin{eqnarray}
\label{eq:chi-pYM-OZI}
    \sqrt{\chi'_{\rm QCD,OZI}(0)}= F_{\pi}/\sqrt{6}\approx 32\, {\rm MeV}.
\end{eqnarray}
There exist polarized DIS experimental data from the HERMES~\cite{HERMES:2006jyl} and COMPASS~\cite{COMPASS:2010hwr} collaborations which can be used to estimate the magnitude of OZI violation. Towards this end, with Eq.~(\ref{eq:chiral-lim}), we can write 
\begin{eqnarray}
\label{eq:Sigma-expt}
    \Delta \Sigma^{\rm expt.} = \Delta\Sigma^{\rm OZI}\, \frac{\sqrt{\chi'_{\rm QCD}(0)}}{\sqrt{\chi'_{\rm QCD,OZI}(0)}} \,.
\end{eqnarray}
The HERMES experiment quotes $\Delta\Sigma^{\rm expt.}=0.33\pm 0.011 ({\rm th.})\pm 0.025 ({\rm expt.})\pm 0.028 ({\rm evolv.})$ for $Q^2=5$ GeV$^2$~\cite{PhysRevD.75.012007} while COMPASS quote a $Q^2\rightarrow \infty $ extrapolated value of 
$0.33\pm0.03 ({\rm stat.})\pm 0.05 ({\rm syst.})$~\cite{COMPASS:2006mhr}. For our back of the envelope estimate, we will take the central value $\Delta \Sigma^{\rm expt.}\approx 0.33$. Substituting the numbers in Eqs.~(\ref{eq:Sigma-OZI}) and (\ref{eq:chi-pYM-OZI}) in the above, we find 
\begin{eqnarray}
    \sqrt{\chi'_{\rm QCD}(0)} \approx 18 \,{\rm MeV} \,.
\end{eqnarray}
which, using Eq. (\ref{eq:chiprrel}), gives $\sqrt{\chi'_{\rm YM}(0)} \approx 36 \,{\rm MeV}$, to be compared with the recent lattice result we quoted, with the central value of $17.1$ MeV. 

Note that if we set the slope of the WZW coupling $b'(0)=0$ in Eq. \ref{eq:bprb}, the relation between $\chi'_{\rm YM}(0)$ and $\chi'_{\rm QCD}|_{m=0}(0)$ (\ref{eq:chiprrel}) gets modified to $\chi'_{\rm YM}(0) = 2 \chi'_{\rm QCD}|_{m=0}(0)$, which is consistent with the prediction in \cite{Bonanno:2023ple}. In this case, we would get $\sqrt{\chi'_{\rm YM}(0)} \approx 25 \,{\rm MeV}$. However, as we found (see discussion after Eq. (\ref{eq:l-dep-b})), the dependence of the WZW coupling on $l^2$ is essential for the consistency of the approach, which as we have seen, gives a substantial contribution. 

Nevertheless, for a back of the envelope estimate, the fact that the $\sqrt{\chi'_{\rm YM}(0)}$ estimates are in the same ballpark is very encouraging\footnote{Similar estimates using QCD sum rules, combined with the OZI values, to compute the equivalent of Eq.~(\ref{eq:Sigma-expt}) with $\chi'_{\rm QCD}|_{m=0}(0)$ gives close agreement with the experimental data~\cite{Narison:2021kny}.} and calls for systematic implementation of nonet chiral perturbation theory~\cite{HerreraSiklody:1996pm,Kaiser:2000gs}, as well as further lattice computations of the slope of the topological susceptibility (and its scale evolution) in full QCD, in the polarized DIS framework. Further, since $\Delta \Sigma^{\rm expt.}$ depends on small $x$ extrapolations, data from the future EIC will further help improve the comparison between theory and experiment.

 \section{Summary and outlook}
In Paper I~\cite{Tarasov:2020cwl} and Paper II~\cite{Tarasov:2021yll}, we revisted the role of the chiral anomaly in polarized DIS, a topic of great interest following the ``spin crisis" of the late 1980s and early 1990s~\cite{Anselmino:1994gn,Leader:2001gr}. In addressing this problem, we employed a powerful worldline formalism, which allows one to address perturbative and nonperturbative features of polarized DIS on the same footing. In particular, it provides an elegant heat kernel regularization that allows one to systematically  compute contributions, arising from the phase of the Dirac determinant, that generate the chiral anomaly and Wess-Zumino-Witten terms in the low energy effective action of QCD. 

We computed in Paper I the contribution of the antisymmetric piece of the box diagram that provides the isosinglet contribution to the structure function $g_1(x,Q^2)$. Working in exact kinematics, we demonstrated that 
the box diagram is sensitive to the infrared pole $\frac{l^\mu}{l^2} F {\tilde F}$ of the chiral anomaly in both Bjorken and Regge asymptotics. Since computing $g_1(x,Q^2)$ requires we take the forward limit $l^2
\rightarrow 0$, this raises the question of how this anomaly pole is regularized in physical cross-sections. 

This problem was addressed in Paper II in the chiral limit of zero quark masses. We showed that due to the WZW term coupling the topological charge density to the primordial isosinglet $\bar \eta$ (whose coupling strength is formally suppressed as $1/\sqrt{N_c}$), the anomaly can couple to bubbles of the Yang-Mills topological susceptibility linked by $\bar \eta$ exchanges, with the final $\bar \eta$ coupling to the polarized proton. 
This iteration of Yang-Mills bubbles is nothing but the Witten-Veneziano mechanism whereby the $\bar \eta$ acquires mass, becoming the physical $\eta^\prime$. The consequence for polarized DIS is that the off-forward $\frac{l^\mu}{l^2}$ pole of the anomaly is shifted to $\frac{l^\mu}{l^2-\tilde m_{\eta'}^2}$, where $\tilde m_{\eta'}$ is the $\eta^\prime$ mass in the chiral limit. A further consequence of the anomaly equation, coupled with the Dirac equation, is a Goldberger-Treiman relation between the isosinglet axial vector form factor $G_A$  and the physics  of the $\eta^\prime$ determining the pseudoscalar form factor $G_P$. As a result, we were able to independently recover the prior remarkable result by Shore and Veneziano that $\Delta \Sigma\propto \sqrt{\chi_{\rm QCD}'|_{m=0}(0)}$, where $\chi_{\rm QCD}'$ is the slope of the QCD topological susceptibility, in the chiral limit, evaluated at $l^2=0$. This expression establishes the deep connection between polarized DIS and the topology of the QCD vacuum. 

The worldline formalism developed to address the consequences of the chiral anomaly in polarized DIS applies not just to $\Delta \Sigma (Q^2)$ (the first moment of $g_1(x,Q^2)$) but to the function itself. In particular, we developed in Paper II an effective action for $g_1$ at small $x$ that describes the dynamics of the $\bar \eta$ ``axion" as it now propagates in the excited vacuum comprising highly occupied gluon color charges described by 
the Color Glass Condensate effective field theory~\cite{Gelis:2010nm}. The saturation scale $Q_S(x)$ characterizing this large occupancy can generate ``over the barrier" topological ``sphaleron" transitions if 
$Q_S\sim m_{\eta}'$. As we show in our EFT construction, sphaleron transitions cause a drag-like effect on the propagation of the $\bar \eta$, leading to a rapid quenching of $g_1$ measured by the DIS probe. Thus polarized DIS experiments with protons and light nuclei at the EIC can potentially find empirical evidence for sphaleron transitions in nature.  

All our considerations in Papers I \& II were in the chiral limit. What happens for finite quark masses? This question was examined in this paper. Motivating our work in part was a recent claim in the literature~\cite{Castelli:2024eza} that 
the PVV term that appears in this case in the anomaly equation exactly cancels the pole arising from the AVV terms alone. Indeed, this is a well-known result in QED, and also follows from QED-like perturbative QCD (pQCD) computations of the triangle graphs. To address this issue, we redid our worldline computations for the triangle diagram for finite mass in the general framework where external scalar, pseudoscalar and axial vector external sources are added to the Dirac action. The presence of the pseudoscalar and axial vector sources induces an imaginary part to the worldline effective action. As we discussed previously in Paper II, and reviewed further here, this imaginary term, corresponding to the phase of the Dirac determinant, remarkably can be expressed as a wordline Lagrangian in a form very similar to the worldline Lagrangian corresponding to the real part of the effective action. The AVV triangle is derived by taking the functional derivative of the imaginary part of the worldline effective action with respect to the external axial vector source, and subsequently setting this source to zero. (A simpler version of this AVV derivation was given in Paper I.) The PVV triangle is computed similarly by taking the functional derivative of the imaginary part of the worldline effective action with respect to the external pseudoscalar source and setting it subsequently to zero. 

We first considered the simple QED (or equivalently, pQCD) case. In this case one replaces the 
scalar source term $\Phi$ is simply set equal to the quark mass at the end of the computation. As we will discuss shortly, for QCD in general, $\Phi$ cannot be take as a constant; functional derivatives of the effective action with respect to this quantity are what generate the chiral condensate. In the simpler QED and perturbative QCD case, performing the worldline computation (with the details provided in Appendix A) we showed that indeed the PVV triangle exactly cancels the AVV triangle in the forward limit demonstrating that there is no anomaly pole. 
The result for QED was first shown by Adler and worked out in further detail by Adler and Bardeen. It also provided the essence of the recent claim for pQCD in \cite{Castelli:2024eza}. 

However the dynamics of the anomaly in QCD is far richer and subtler than in QED (or pQCD). This is seen explicitly by computing the chiral Ward identities from the Wess-Zumino effective action. The latter corresponds to now adding to the {\it full} QCD action terms including couplings to the aforementioned external sources, and in addition, the coupling of an external $\theta$ term to the topological charge density. Taking appropriate repeated functional derivations of this action with respect to the sources generates the desired hierarchy of anomalous functional Ward identities. For the problem of interest, this generates for instance the ``one-point" anomaly equation for $J_\mu^5$ but also, as we showed, the two-point Green functions  of $J_\mu^5$ with the topological charge density and the isosinglet pseudoscalar field $\phi_5$. The latter Green functions satisfies a chiral Ward identity that contains a term including the functional derivative with respect to $\Phi$ - in other words, the chiral condensate. Such a contribution does not exist in QED and there would be no way to generate it in pQCD either. But as we showed, it is essential that this chiral Ward identity be satisfied since it is these Green functions, and a Goldberger-Treiman relation, that determines the isosinglet axial charge of the proton, or equivalently, $\Delta \Sigma$. 

A further necessary element in our analysis was polology, which assumes pole dominance of the Green functions, the smooth behavior of couplings, and the Yang-Mills and QCD topological susceptibilities. With the help of these, and large $N_c$ relations, we were able to extract quantitative predictions that can be tested with lattice simulations and polarized DIS experiments. In particular, we computed the finite mass corrections to our result for $\Delta \Sigma$ obtained in the chiral limit. We see that they are very small. However with sufficiently precise experimental data, and lattice QCD computations, one can extract further nontrivial nonperturbative information, and test as well, our polology assumptions. 

These considerations can be extended in several directions. Firstly though, we emphasize that our results indicate that including finite quark masses in the box diagram is unlikely to change the conclusions of Paper II, the most striking of these being the predicted sensitivity of $g_1$ at small $x$ to sphaleron-like transitions. Further, an immediate extension of our work here is to compute the DIS structure function $g_2$ within our worldline formalism. It is well-known that this quantity is sensitive to finite mass effects~\cite{Leader:2001gr} and in particular to the physics of chiral symmetry breaking~\cite{Harindranath:1997qn}. 

Going beyond fully inclusive quantities, further applications include  effects of the anomaly in semi-inclusive and exclusive DIS final states. Of particular interest is the physics of generalized parton distributions-our work in progress indicates that the worldline approach provide a systematic way to go beyond the collinear and high energy approximations often made in the literature that may on occasion, as we demonstrated in this body of work, obscure fundamental nonperturbative physics arising from chiral symmetry breaking and the topology of the QCD vacuum. Many of our considerations can also be extended to discussions of the conformal anomaly, as also discussed recently in \cite{Bhattacharya:2022xxw,Bhattacharya:2023wvy,Bhattacharya:2024geo} and in \cite{Coriano:2024qbr,Coriano:2024wrz}, as well as additional nontrivial anomalous final states~\cite{Vasquez-Mozo,Bilal:2008qx}. The framework developed here to compute the appropriate Wess-Zumino-Witten terms can be extended to address such anomalous contributions to the gravitational form factor of the hadron. Developments in this direction have the potential to provide fundamental insight into nonperturbative features of QCD when confronted with measurements at the Electron-Ion Collider. 
 
 \label{sec:Summary}

\section*{Acknowledgements}
In the course of this body of work, we have benefited greatly from conversations with numerous colleagues on various aspects of this work; we would however like to acknowledge in particular Xabier Feal for his insights into the worldline formalism. This manuscript has also been influenced by discussions with Mike Creutz,  Yoshitaka Hatta, Parmeswaran Nair, Rob Pisarski and Arkady Vainshtein. A.T.'s research is supported by the U.S. Department of Energy, Office of Science, under contract DE-SC0020081. He was also supported by the Center for Frontiers in Nuclear Science at Stony Brook University. R.V's research is supported by the U.S. Department of Energy, Office of Science, under contract DE-SC0012704 and within the framework of the SURGE Topical Theory Collaboration. He was also supported at Stony Brook by the Simons Foundation as a co-PI under Award number 994318 (Simons Collaboration on Confinement and QCD Strings). 

\appendix

\section*{Appendix A: Details of the derivation of worldline results}
\label{appendix:AppendixA}
In this Appendix we will discuss details of the worldline functional integrals in Sec. \ref{sec:Anomaly-quark-mass}. We begin with the general algorithm for such calculations, that can be applied to any worldline diagram, as worked out in \cite{DHoker:1995uyv,Mondragon:1995ab,Schubert:2001he,Tarasov:2019rfp}. To be  concrete, we will use Eq. (\ref{eq:wl-diag-init}) as an example.

We start with calculation of the functional integral over Grassmann variables. The first step in calculating of a worldline functional integral is the separation of zero mode. Since the Grassmann functional integral in the imaginary part of the effective action $W_\mathcal{I}$ has periodic boundary conditions, the corresponding zero mode is not trivial and can be isolated by writing
\begin{eqnarray}
&&\psi_i = \psi + \xi_i\,.
\end{eqnarray}
Now if we take into account that the integral over zero modes
\begin{eqnarray}
\int d^5\psi~ \psi^\mu \psi^\nu \psi^\rho \psi^\sigma \psi^5 = \epsilon^{\mu\nu\rho\sigma}\,,
\label{eq:zmode-gr}
\end{eqnarray}
involves four $\psi$ factors, we find that some terms in Eq.~(\ref{eq:wl-diag-init}) are trivial and we can rewrite the equation as\footnote{Note that in our example only the zero mode $\phi_5$ contribute. In the general case, one has to account for the $\xi^5_\tau$ mode as well.}
\begin{eqnarray}
&&\int d^4x \,e^{ilx} \frac{\delta W_\mathcal{I}}{\delta B_\mu(x)}\Big|^{\rm singlet}_{B_\mu=0;\Pi=0;\Phi=m} = - \frac{\mathcal{E}}{8} \int \frac{d^4k_2}{(2\pi)^4} A_\rho(k_2) \int \frac{d^4k_3}{(2\pi)^4} A_\sigma(k_3)
\label{eq:a-s1}\\
&&\times \int^1_{-1}d\alpha \int^\infty_0 dT \exp\Big[- T \frac{\mathcal{E}d_{000} \alpha^2 m^2}{4} \Big] \int^T_0 d\tau_2 \int^T_0 d\tau_3 ~\mathcal{N} \int_P \mathcal{D}x \exp\Big[-\int^T_0 d\tau \frac{\dot{x}^2}{2\mathcal{E}} \Big] \int \mathcal{D}\xi \exp\Big[-\int^T_0 d\tau \frac{1}{2}\xi\dot{\xi} \Big]
\nonumber\\
&&\times  \int d^5\psi~\psi_5 \Big[ il^\mu \mathcal{E}^2 k_{2\alpha} k_{3\beta} e^{ilx_0} \psi^\rho \psi^\alpha \psi^\sigma \psi^\beta + \frac{4 }{\mathcal{E}} \dot{x}_{0\nu} e^{ilx_0} \Big( i \mathcal{E} k_{3\beta} \dot{x}^\rho_2 \psi^\nu \psi^\mu \psi^\sigma \psi^\beta + i \mathcal{E} k_{2\alpha} \dot{x}^\sigma_3  \psi^\nu \psi^\mu \psi^\rho \psi^\alpha
\nonumber\\
&&  - \mathcal{E}^2 k_{2\alpha} k_{3\beta} (\psi^\nu + \xi^\nu_0)(\psi^\mu + \xi^\mu_0) (\psi^\rho + \xi^\rho_2) (\psi^\alpha + \xi^\alpha_2) (\psi^\sigma + \xi^\sigma_3) (\psi^\beta + \xi^\beta_3) \Big) 
\nonumber\\
&&- 2 d_{000} m^2 \alpha^2 \dot{x}_{0\nu} \int^T_0 d\tau_1 e^{ilx_1} \Big( i \mathcal{E} k_{3\beta} \dot{x}^\rho_2 \psi^\nu \psi^\mu \psi^\sigma \psi^\beta + i \mathcal{E} k_{2\alpha} \dot{x}^\sigma_3 \psi^\nu \psi^\mu \psi^\rho \psi^\alpha 
\nonumber\\
&& - \mathcal{E}^2 k_{2\alpha} k_{3\beta} (\psi^\nu + \xi^\nu_0)(\psi^\mu + \xi^\mu_1) (\psi^\rho + \xi^\rho_2) (\psi^\alpha + \xi^\alpha_2) (\psi^\sigma + \xi^\sigma_3) (\psi^\beta + \xi^\beta_3) \Big)  \Big] e^{ik_2 x_2} e^{ik_3 x_3}
\nonumber
\end{eqnarray}

The integral over zero mode $\psi$ can be calculated using Eq.~(\ref{eq:zmode-gr}), while the remaining integral over the $\xi$ mode should be written as a sum of all possible Wick contractions of the $\xi_\tau$ variables. For example, the last term in Eq. (\ref{eq:a-s1}) can be written as\footnote{Here one has to take into account the anti-commutation property of the Grassmann variables.}
\begin{eqnarray}
&&\int \mathcal{D}\xi \exp\Big[-\int^T_0 d\tau \frac{1}{2}\xi\dot{\xi} \Big] \int d^5\psi ~\psi_5 \Big[ (\psi^\nu + \xi^\nu_0 )(\psi^\mu + \xi^\mu_1) (\psi^\rho + \xi^\rho_2) (\psi^\alpha + \xi^\alpha_2) (\psi^\sigma + \xi^\sigma_3) (\psi^\beta + \xi^\beta_3) \Big]
\nonumber\\
&&= - \Big[ \epsilon^{\nu\mu\rho\alpha} \langle \xi^\sigma_3 \xi^\beta_3\rangle
- \epsilon^{\nu\mu\rho\sigma} \langle\xi^\alpha_2  \xi^\beta_3\rangle
+ \epsilon^{\nu\mu\rho\beta} \langle \xi^\alpha_2 \xi^\sigma_3 \rangle
+ \epsilon^{\nu\mu\alpha\sigma} \langle \xi^\rho_2 \xi^\beta_3 \rangle
\nonumber\\
&&- \epsilon^{\nu\mu\alpha\beta} \langle \xi^\rho_2 \xi^\sigma_3 \rangle
+ \epsilon^{\nu\mu\sigma\beta} \langle \xi^\rho_2 \xi^\alpha_2 \rangle
- \epsilon^{\nu\rho\alpha\sigma} \langle \xi^\mu_1 \xi^\beta_3 \rangle
+ \epsilon^{\nu\rho\alpha\beta} \langle \xi^\mu_1 \xi^\sigma_3 \rangle
- \epsilon^{\nu\rho\sigma\beta} \langle\xi^\mu_1 \xi^\alpha_2 \rangle
+ \epsilon^{\nu \alpha\sigma\beta} \langle\xi^\mu_1 \xi^\rho_2 \rangle
\nonumber\\
&&+ \epsilon^{\mu\rho\alpha\sigma} \langle\xi^\nu_0 \xi^\beta_3 \rangle
- \epsilon^{\mu\rho\alpha\beta} \langle\xi^\nu_0 \xi^\sigma_3 \rangle
+ \epsilon^{\mu\rho\sigma \beta} \langle\xi^\nu_0 \xi^\alpha_2 \rangle
- \epsilon^{\mu\alpha\sigma\beta} \langle\xi^\nu_0 \xi^\rho_2\rangle
+ \epsilon^{\rho \alpha\sigma \beta} \langle\xi^\nu_0 \xi^\mu_1 \rangle\Big]
\end{eqnarray}
where the Wick contraction,
\begin{eqnarray}
&&\langle \xi^\mu_i \xi^\nu_j \rangle = \frac{1}{2}g^{\mu\nu}\Big({\rm sign}(\tau_i - \tau_j) - \frac{2(\tau_i - \tau_j)}{T}\Big) = \frac{1}{2}g^{\mu\nu}\dot{G}_B(\tau_i, \tau_j)\,,
\end{eqnarray}
and the overall minus sign comes from the normalization of the functional integral
\begin{eqnarray}
&&\int \mathcal{D} \xi~ \exp\Big[-\int^T_0 d\tau \frac{1}{2}\xi\dot{\xi} \Big] = -1\,.
\end{eqnarray}

As a result, Eq. (\ref{eq:a-s1}) takes the form
\begin{eqnarray}
&&\int d^4x e^{ilx} \frac{\delta W_\mathcal{I}}{\delta B_\mu(x)}\Big|^{\rm singlet}_{B_\mu=0;\Pi=0;\Phi=m} = \frac{\mathcal{E}}{8} \int \frac{d^4k_2}{(2\pi)^4} A_\rho(k_2) \int \frac{d^4k_3}{(2\pi)^4} A_\sigma(k_3)
\label{eq:a-s2}\\
&&\times \int^1_{-1}d\alpha \int^\infty_0 dT \exp\Big[- T \frac{\mathcal{E}d_{000} \alpha^2 m^2}{4} \Big] \int^T_0 d\tau_2 \int^T_0 d\tau_3 \mathcal{N} \int_P \mathcal{D}x \exp\Big[-\int^T_0 d\tau \frac{\dot{x}^2}{2\mathcal{E}} \Big] \Big[ il^\mu \mathcal{E}^2 k_{2\alpha} k_{3\beta} e^{ilx_0} \epsilon^{\rho\alpha\sigma\beta} 
\nonumber\\
&&+ \frac{4 }{\mathcal{E}} \dot{x}_{0\nu} e^{ilx_0} \Big( i \mathcal{E} k_{3\beta} \dot{x}^\rho_2 \epsilon^{\nu\mu\sigma\beta} + i \mathcal{E} k_{2\alpha} \dot{x}^\sigma_3 \epsilon^{\nu\mu\rho\alpha}
 - \frac{1}{2}\mathcal{E}^2 k_{2\alpha} k_{3\beta} \Big\{
- \epsilon^{\nu\mu\rho\sigma} g^{\alpha\beta}\dot{G}_B(\tau_2, \tau_3)
+ \epsilon^{\nu\mu\rho\beta} g^{\alpha\sigma}\dot{G}_B(\tau_2, \tau_3)
\nonumber\\
&&+ \epsilon^{\nu\mu\alpha\sigma} g^{\rho\beta}\dot{G}_B(\tau_2, \tau_3)
- \epsilon^{\nu\mu\alpha\beta} g^{\rho\sigma}\dot{G}_B(\tau_2, \tau_3)
- \epsilon^{\nu\rho\alpha\sigma} g^{\mu\beta}\dot{G}_B(\tau_0, \tau_3)
+ \epsilon^{\nu\rho\alpha\beta} g^{\mu\sigma}\dot{G}_B(\tau_0, \tau_3)
- \epsilon^{\nu\rho\sigma\beta} g^{\mu\alpha}\dot{G}_B(\tau_0, \tau_2)
\nonumber\\
&&+ \epsilon^{\nu \alpha\sigma\beta} g^{\mu\rho}\dot{G}_B(\tau_0, \tau_2)
+ \epsilon^{\mu\rho\alpha\sigma} g^{\nu\beta}\dot{G}_B(\tau_0, \tau_3)
- \epsilon^{\mu\rho\alpha\beta} g^{\nu\sigma}\dot{G}_B(\tau_0, \tau_3)
+ \epsilon^{\mu\rho\sigma \beta} g^{\nu\alpha}\dot{G}_B(\tau_0, \tau_2)
- \epsilon^{\mu\alpha\sigma\beta} g^{\nu\rho}\dot{G}_B(\tau_0, \tau_2) \Big\} \Big) 
\nonumber\\
&&- 2 d_{000} m^2 \alpha^2 \dot{x}_{0\nu} \int^T_0 d\tau_1 e^{ilx_1} \Big( i \mathcal{E} k_{3\beta} \dot{x}^\rho_2 \epsilon^{\nu\mu\sigma\beta} + i \mathcal{E} k_{2\alpha} \dot{x}^\sigma_3 \epsilon^{\nu\mu\rho\alpha} - \frac{1}{2}\mathcal{E}^2 k_{2\alpha} k_{3\beta} \Big\{
- \epsilon^{\nu\mu\rho\sigma} g^{\alpha\beta}\dot{G}_B(\tau_2, \tau_3)
\nonumber\\
&& + \epsilon^{\nu\mu\rho\beta} g^{\alpha\sigma}\dot{G}_B(\tau_2, \tau_3)
+ \epsilon^{\nu\mu\alpha\sigma} g^{\rho\beta}\dot{G}_B(\tau_2, \tau_3)
- \epsilon^{\nu\mu\alpha\beta} g^{\rho\sigma}\dot{G}_B(\tau_2, \tau_3)
- \epsilon^{\nu\rho\alpha\sigma} g^{\mu\beta}\dot{G}_B(\tau_1, \tau_3)
\nonumber\\
&&+ \epsilon^{\nu\rho\alpha\beta} g^{\mu\sigma}\dot{G}_B(\tau_1, \tau_3)
- \epsilon^{\nu\rho\sigma\beta} g^{\mu\alpha}\dot{G}_B(\tau_1, \tau_2)
+ \epsilon^{\nu \alpha\sigma\beta} g^{\mu\rho}\dot{G}_B(\tau_1, \tau_2)
+ \epsilon^{\mu\rho\alpha\sigma} g^{\nu\beta}\dot{G}_B(\tau_0, \tau_3)
\nonumber\\
&&- \epsilon^{\mu\rho\alpha\beta} g^{\nu\sigma}\dot{G}_B(\tau_0, \tau_3)
+ \epsilon^{\mu\rho\sigma \beta} g^{\nu\alpha}\dot{G}_B(\tau_0, \tau_2)
- \epsilon^{\mu\alpha\sigma\beta} g^{\nu\rho}\dot{G}_B(\tau_0, \tau_2)
+ \epsilon^{\rho \alpha\sigma \beta} g^{\nu\mu}\dot{G}_B(\tau_0, \tau_1)\Big\} \Big)  \Big] e^{ik_2 x_2} e^{ik_3 x_3}\,.
\nonumber
\end{eqnarray}

Following the computation of the functional integral over the Grassmann variables, we will need to calculate the integral over the scalar variables $x$. The calculation is analogous to the calculation of the Grassmann integral. First, we separate the zero mode as 
\begin{eqnarray}
&&x_i = x + y_i\,.
\end{eqnarray}
The integral over the zero mode always yields the momentum conserving $\delta$-function. In $D$ dimensions, we egt
\begin{eqnarray}
&&\int d^Dx ~e^{i(\sum_i p_i) x} = (2\pi)^D\delta^D(\sum_i p_i)\,.
\end{eqnarray}
The remaining functional integral over $y$,  normalized as
\begin{eqnarray}
&&\mathcal{N} \int_P \mathcal{D}y \exp\Big[-\int^T_0 d\tau \frac{\dot{y}^2}{2\mathcal{E}} \Big] = \frac{1}{(2\pi\mathcal{E}T)^{D/2}}\,,
\end{eqnarray}
can be written as the sum of all possible Wick contractions between elementary fields $y_i$ and exponential factors:
\begin{eqnarray}
\langle y^\mu(\tau_i) y^\nu(\tau_j)\rangle = - g^{\mu\nu} G_B(\tau_i, \tau_j)\,,
\end{eqnarray}
and
\begin{eqnarray}
\langle y^\mu(\tau_i) e^{iky(\tau_j)}\rangle = i\langle y^\mu(\tau_i) y^\nu(\tau_j)\rangle k_\nu \,.e^{iky(\tau_j)}\,.
\end{eqnarray}
Once all the elementary fields $y_i$ are eliminated, one has to contract the remaining exponentials, which yields the global factor
\begin{eqnarray}
&&\langle e^{i\sum_i k_i x_i}\rangle = \exp\Big[ -\frac{1}{2}\sum_{i,j} k_{i\mu} \langle y^\mu(\tau_i) y^\nu(\tau_j)\rangle k_{j\nu} \Big]\,.
\end{eqnarray}
As a result, Eq. (\ref{eq:a-s2}) can be written in terms of the worldline propagator $G_B$. For the divergence of Eq. (\ref{eq:a-s2}), we find
\begin{eqnarray}
&&\int d^4x \,e^{ilx} \partial_\mu \frac{\delta W_\mathcal{I}}{\delta B_\mu(x)}\Big|^{\rm singlet}_{B_\mu=0;\Pi=0;\Phi=m} = - \frac{i\mathcal{E}}{8} \int \frac{d^4k_2}{(2\pi)^4} A_\rho(k_2) \int \frac{d^4k_3}{(2\pi)^4} A_\sigma(k_3)
\label{eq:a-s3}\\
&&\times \int^1_{-1}d\alpha \int^\infty_0 dT \exp\Big[- T \frac{\mathcal{E}d_{000} \alpha^2 m^2}{4} \Big] \int^T_0 d\tau_2 \int^T_0 d\tau_3
 \frac{1}{(2\pi\mathcal{E}T)^2} \Big[ \Big( il^2 \mathcal{E}^2 k_{2\alpha} k_{3\beta} \epsilon^{\rho\alpha\sigma\beta}  
\nonumber\\
&& - 4 i k_{3\beta} l_\mu \epsilon^{\nu\mu\sigma\beta} \frac{\partial}{\partial\tau_0}\frac{\partial}{\partial\tau_2}\delta^\rho_\nu G_B(\tau_0, \tau_2)
- 4 i k_{2\alpha} l_\mu \epsilon^{\nu\mu\rho\alpha}  \frac{\partial}{\partial\tau_0}\frac{\partial}{\partial\tau_3}\delta^\sigma_\nu G_B(\tau_0, \tau_3) + 2 i \mathcal{E} l_\mu k_{2\alpha} k_{3\beta} \Big\{
- \epsilon^{\nu\mu\rho\sigma} g^{\alpha\beta}\dot{G}_B(\tau_2, \tau_3)
\nonumber\\
 && 
- \epsilon^{\nu\rho\alpha\sigma} g^{\mu\beta}\dot{G}_B(\tau_0, \tau_3)
- \epsilon^{\nu\rho\sigma\beta} g^{\mu\alpha}\dot{G}_B(\tau_0, \tau_2)
+ \epsilon^{\mu\rho\alpha\sigma} g^{\nu\beta}\dot{G}_B(\tau_0, \tau_3)
+ \epsilon^{\mu\rho\sigma \beta} g^{\nu\alpha}\dot{G}_B(\tau_0, \tau_2) \Big\} 
\nonumber\\
&&\times \Big\{ k_{2\nu} \dot{G}_B(\tau_0, \tau_2) + k_{3\nu} \dot{G}_B(\tau_0, \tau_3)  \Big\} \Big) \exp\Big( l \cdot k_2 G_B(\tau_0, \tau_2) + l\cdot k_3 G_B(\tau_0, \tau_3) + k_2\cdot k_3 G_B(\tau_2, \tau_3) \Big)
\nonumber\\
&&- 2 d_{000} m^2 \alpha^2 \int^T_0 d\tau_1 \Big( - i \mathcal{E} k_{3\beta} l_\mu \epsilon^{\nu\mu\sigma\beta} \frac{\partial}{\partial\tau_0}\frac{\partial}{\partial\tau_2}\delta^\rho_\nu G_B(\tau_0, \tau_2)  
- i \mathcal{E} k_{2\alpha} l_\mu \epsilon^{\nu\mu\rho\alpha} \frac{\partial}{\partial\tau_0}\frac{\partial}{\partial\tau_3}\delta^\sigma_\nu G_B(\tau_0, \tau_3)
\nonumber\\
&& + \frac{i\mathcal{E}^2}{2} k_{2\alpha} k_{3\beta} l_\mu \Big\{
- \epsilon^{\nu\mu\rho\sigma} g^{\alpha\beta}\dot{G}_B(\tau_2, \tau_3)
- \epsilon^{\nu\rho\alpha\sigma} g^{\mu\beta}\dot{G}_B(\tau_1, \tau_3)
- \epsilon^{\nu\rho\sigma\beta} g^{\mu\alpha}\dot{G}_B(\tau_1, \tau_2)
\nonumber\\
&&+ \epsilon^{\mu\rho\alpha\sigma} g^{\nu\beta}\dot{G}_B(\tau_0, \tau_3)
+ \epsilon^{\mu\rho\sigma \beta} g^{\nu\alpha}\dot{G}_B(\tau_0, \tau_2)
+ \epsilon^{\rho \alpha\sigma \beta} g^{\nu\mu}\dot{G}_B(\tau_0, \tau_1)\Big\}  \Big\{ l_{\nu} \dot{G}_B(\tau_0, \tau_1) + k_{2\nu} \dot{G}_B(\tau_0, \tau_2)
\nonumber\\
&& + k_{3\nu} \dot{G}_B(\tau_0, \tau_3)  \Big\}  \Big) \exp\Big( l \cdot k_2 G_B(\tau_1, \tau_2) + l\cdot k_3 G_B(\tau_1, \tau_3) + k_2\cdot k_3 G_B(\tau_2, \tau_3) \Big) \Big] (2\pi)^4\delta^4(l + k_2 + k_3)\,.
\nonumber
\end{eqnarray}

Our task now is to simplify this result of the calculation of the worldline functional integrals. First, we need to integrate by parts with respect to the proper time variables terms with the second derivative of the worldline propagator. Assuming that the background gluons are on the mass-shell, after some lengthy algebra we arrive at\footnote{Here we use the identity $g^{\mu\nu}\epsilon^{\alpha\beta\rho\sigma} +g^{\mu\alpha}\epsilon^{\beta\rho\sigma\nu}+g^{\mu\beta}\epsilon^{\rho\sigma\nu\alpha} +g^{\mu\rho}\epsilon^{\sigma\nu\alpha\beta}+g^{\mu\sigma}\epsilon^{\nu\alpha\beta\rho} = 0$.}
\begin{eqnarray}
&&\int d^4x e^{ilx} \partial_\mu \frac{\delta W_\mathcal{I}}{\delta B_\mu(x)}\Big|^{\rm singlet}_{B_\mu=0;\Pi=0;\Phi=m} = - \frac{i}{4} \int \frac{d^4k_2}{(2\pi)^4} A_\rho(k_2) \int \frac{d^4k_3}{(2\pi)^4} A_\sigma(k_3) \epsilon^{\rho\alpha\sigma\beta} k_{2\alpha} k_{3\beta}
\label{eq:a-s4}\\
&&\times \int^1_{-1}d\alpha \int^\infty_0 dT \exp\Big[- T \alpha^2 m^2 \Big] \int^T_0 d\tau_2 \int^T_0 d\tau_3
 \frac{1}{(4\pi T)^2}  \Big[ 4 i k_2 \cdot k_3 \Big(  1 - \dot{G}^2_B(\tau_0, \tau_3) + 1 - \dot{G}^2_B(\tau_0, \tau_2)
\nonumber\\
&&  + \dot{G}^2_B(\tau_2, \tau_3) - ( \dot{G}_B(\tau_0, \tau_2) + \dot{G}_B(\tau_2, \tau_3)  + \dot{G}_B(\tau_3, \tau_0)  )^2 \Big) \exp\Big( - k_2 \cdot k_3 ( G_B(\tau_0, \tau_2) + G_B(\tau_0, \tau_3) - G_B(\tau_2, \tau_3) )\Big)
\nonumber\\
&&- 8 i m^2 \alpha^2 k_2 \cdot k_3 \int^T_0 d\tau_1 \Big( 
- \dot{G}_B(\tau_3, \tau_1)  \dot{G}_B(\tau_1, \tau_0) -  \dot{G}_B(\tau_2, \tau_1) \dot{G}_B(\tau_1, \tau_0) 
\nonumber\\
&&+ ( \dot{G}_B(\tau_1, \tau_0) +  \dot{G}_B(\tau_0, \tau_3)  + \dot{G}_B(\tau_3, \tau_1) )^2
+ (  \dot{G}_B(\tau_1, \tau_0) + \dot{G}_B(\tau_0, \tau_2) + \dot{G}_B(\tau_2, \tau_1) )^2
\nonumber\\
 &&-  ( \dot{G}_B(\tau_2, \tau_0) +  \dot{G}_B(\tau_0, \tau_3) +  \dot{G}_B(\tau_3, \tau_2) )^2
-\dot{G}^2_B(\tau_1, \tau_2) - \dot{G}^2_B(\tau_1, \tau_3) + \dot{G}^2_B(\tau_2, \tau_3)\Big)
\nonumber\\
&&\times \exp\Big( - k_2\cdot k_3 ( G_B(\tau_1, \tau_2)+ G_B(\tau_1, \tau_3) - G_B(\tau_2, \tau_3) ) \Big) \Big] (2\pi)^4\delta^4(l + k_2 + k_3)\,,
\end{eqnarray}
where we also fixed $\mathcal{E}=2$ and $d_{000}=2$.

To further simplify the equation, we apply the identities
\begin{eqnarray}
 \dot{G}_B(\tau_1, \tau_2) + \dot{G}_B(\tau_2, \tau_4) + \dot{G}_B(\tau_4, \tau_1) = -G_F(\tau_1, \tau_2)G_F(\tau_2, \tau_4)G_F(\tau_4, \tau_1)\,,
\end{eqnarray}
\begin{eqnarray}
1 - \dot{G}^2_B(\tau_i, \tau_j) = \frac{4}{T} \,G_B(u_i, u_j)\,,
\label{b2dot}
\end{eqnarray}
and
\begin{eqnarray}
&&\int^T_0 d\tau_2 \int^T_0 d\tau_3 \int^T_0 d\tau_1 \exp\Big[ -k_2 \cdot k_3 (G_B(\tau_1, \tau_2) + G_B(\tau_1, \tau_3) - G_B(\tau_2, \tau_3) )\Big]
\nonumber\\
&&\times \Big( - \dot{G}_B(\tau_3, \tau_1) \dot{G}_B(\tau_1, \tau_0) - \dot{G}_B(\tau_2, \tau_1) \dot{G}_B(\tau_1, \tau_0) \Big) = 0\,,
\end{eqnarray}
which, after taking into account the rotational invariance of the proper time integrals, leads to
\begin{eqnarray}
&&\int d^4x \,e^{ilx} \partial_\mu \frac{\delta W_\mathcal{I}}{\delta B_\mu(x)}\Big|^{\rm singlet}_{B_\mu=0;\Pi=0;\Phi=m} = 4 \int \frac{d^4k_2}{(2\pi)^4} A_\rho(k_2) \int \frac{d^4k_3}{(2\pi)^4} A_\sigma(k_3) \epsilon^{\rho\alpha\sigma\beta} k_{2\alpha} k_{3\beta} \int^1_{-1}d\alpha \int^\infty_0  \frac{dT}{T}
\nonumber\\
&&\times \exp\Big[- T \alpha^2 m^2 \Big] \int^T_0 d\tau_2 \int^T_0 d\tau_3 \frac{1}{(4\pi T)^2} (  1 -  2 m^2 \alpha^2 T ) k_2 \cdot k_3 \Big( G_B(\tau_0, \tau_2) + G_B(\tau_0, \tau_3)  - G_B(\tau_2, \tau_3) \Big)
\nonumber\\
&&\times \exp\Big[ - k_2\cdot k_3 ( G_B(\tau_0, \tau_2)+ G_B(\tau_0, \tau_3) - G_B(\tau_2, \tau_3) ) \Big] (2\pi)^4\delta^4(l + k_2 + k_3)\,,
\label{eq:a-s5}
\end{eqnarray}
which greatly simplifies Eq. (\ref{eq:a-s3}).

Rescaling the proper time variables as $\tau_i = T u_i$, and integrating over $T$, we obtain
\begin{eqnarray}
&&\int d^4x e^{ilx} \partial_\mu \frac{\delta W_\mathcal{I}}{\delta B_\mu(x)}\Big|^{\rm singlet}_{B_\mu=0;\Pi=0;\Phi=m} = \frac{1}{4\pi^{2}}   \int \frac{d^4k_2}{(2\pi)^4} A_\rho(k_2) \int \frac{d^4k_3}{(2\pi)^4} A_\sigma(k_3) \epsilon^{\rho\alpha\sigma\beta} k_{2\alpha} k_{3\beta} (2\pi)^4\delta^4(l + k_2 + k_3)
\nonumber\\
&&\times \int^1_0 du_2 \int^1_0 du_3 \int^1_{-1}d\alpha \Big( 1 - \alpha^2 m^2 \frac{ 3 \mathcal{G}(u_0, u_2, u_3) + \alpha^2 m^2 }{ ( \mathcal{G}(u_0, u_2, u_3) + \alpha^2 m^2 )^2} \Big)\,,
\label{eq:a-s6}
\end{eqnarray}
where we introduced the notation
\begin{eqnarray}
&&\mathcal{G}(u_0, u_2, u_3) \equiv k_2 \cdot k_3 \Big( G_B(u_0, u_2) + G_B(u_0, u_3) -  G_B(u_2, u_3)\Big)\,.
\end{eqnarray}
Integrating over $\alpha$, we get 
\begin{eqnarray}
&&\int d^4x e^{ilx} \partial_\mu \frac{\delta W_\mathcal{I}}{\delta B_\mu(x)}\Big|^{\rm singlet}_{B_\mu=0;\Pi=0;\Phi=m} = \frac{1}{4\pi^{2}} \int \frac{d^4k_2}{(2\pi)^4} A_\rho(k_2) \int \frac{d^4k_3}{(2\pi)^4} A_\sigma(k_3) \epsilon^{\rho\alpha\sigma\beta} k_{2\alpha} k_{3\beta} (2\pi)^4\delta^4(l + k_2 + k_3)
\nonumber\\
&&\times \int^1_0 du_2 \int^1_0 du_3 \Big( 2 - \frac{2 m^2}{\mathcal{G}(u_0, u_2, u_3) + m^2} \Big) \,.
\label{eq:a-s7}
\end{eqnarray}

Finally, integrating over proper time variables in the first term we get
\begin{eqnarray}
&&\int d^4x e^{ilx} \partial_\mu \frac{\delta W_\mathcal{I}}{\delta B_\mu(x)}\Big|^{\rm singlet}_{B_\mu=0;\Pi=0;\Phi=m} = \frac{1}{2\pi^{2}} \int \frac{d^4k_2}{(2\pi)^4} A_\rho(k_2) \int \frac{d^4k_3}{(2\pi)^4} A_\sigma(k_3) \epsilon^{\rho\alpha\sigma\beta} k_{2\alpha} k_{3\beta} (2\pi)^4\delta^4(l + k_2 + k_3)
\nonumber\\
&&\times \Big( 1   
- m^2 \int^1_0 du_2 \int^1_0 du_3 \frac{1}{\mathcal{G}(u_0, u_2, u_3) + m^2} \Big)\,.
\label{eq:a-s8}
\end{eqnarray}

\section*{Appendix B: Derivation of Eq.~(\ref{eq:tpbmpi-eta-F})}
\label{appendix:AppendixB}
Let us consider the calculation of the $T$-product
\begin{eqnarray}
&&\langle 0|TJ_{5\mu}(x)\phi_5(y)|0\rangle = \theta(x^0 - y^0)\langle 0|J_{5\mu}(x)\phi_5(y)|0\rangle + \theta(y^0 - x^0)\langle 0|\phi_5(y)J_{5\mu}(x)|0\rangle\,.
\end{eqnarray}
Assuming that this two-point function is saturated by $\bar{\eta}$ exchange, we insert the complete set
\begin{eqnarray}
&&\int \frac{d^3\vec{k}}{(2\pi)^3} \frac{1}{2E_k}|\bar{\eta}(k)\rangle \langle \bar{\eta}(k)| = 1\,,
\end{eqnarray}
where $E_k = + \sqrt{|\vec{k}|^2 + m^2_{\eta'}}$. This gives,
\begin{eqnarray}
&&\langle 0|TJ_{5\mu}(x)\phi_5(y)|0\rangle = \theta(x^0 - y^0)\int \frac{d^3\vec{k}}{(2\pi)^3} \frac{1}{2E_k}\langle 0|J_{5\mu}(x)|\bar{\eta}(k)\rangle \langle \bar{\eta}(k)| \phi_5(y)|0\rangle
\nonumber\\
&&+ \theta(y^0 - x^0) \int \frac{d^3\vec{k}}{(2\pi)^3} \frac{1}{2E_k} \langle 0|\phi_5(y)|\bar{\eta}(k)\rangle \langle \bar{\eta}(k)| J_{5\mu}(x)|0\rangle\,.
\end{eqnarray}

Now using the definition of the $\bar \eta$ decay constant $\langle 0| J_{5\mu}(x)|\bar{\eta}(k)\rangle = i k_\mu \sqrt{2N_f} F_{\bar{\eta}}(k^2)\, e^{ikx}$, and the interpolating function $c(p^2)$ defined as
\begin{eqnarray}
&&\langle 0| \phi_5(y)|\bar{\eta}(k)\rangle = \int \frac{d^4p}{(2\pi)^4}\, e^{ipy}\, c(p^2)\langle 0| \bar{\eta}(p)|\bar{\eta}(k)\rangle = c(k^2)\, e^{iky}\,,
\end{eqnarray}
we obtain
\begin{eqnarray}
&&\langle 0|TJ_{5\mu}(x)\phi_5(y)|0\rangle 
\nonumber\\
&&=  \Big( \theta(x^0 - y^0)\int \frac{d^3\vec{k}}{(2\pi)^3} \frac{1}{2E_k} e^{ik(x-y)} + \theta(y^0 - x^0) \int \frac{d^3\vec{k}}{(2\pi)^3} \frac{1}{2E_k} e^{-ik(x-y)} \Big) i k_\mu \sqrt{2N_f} F_{\bar{\eta}}(k^2) c(k^2)
\end{eqnarray}

The equation can be rewritten as
\begin{eqnarray}
&&\langle 0|TJ_{5\mu}(x)\phi_5(y)|0\rangle 
= \int\frac{d^4l}{(2\pi)^4} \frac{i}{l^2 - m^2_{\eta'} + i\epsilon}  e^{-il(x-y)} il_\mu \sqrt{2N_f} F_{\bar{\eta}}(l^2)\, c(l^2)
\end{eqnarray}
which finally yields us the result in Eq.~(\ref{eq:tpbmpi-eta-F}):
\begin{eqnarray}
&&\frac{\delta^2 Z}{\delta B_\mu\delta\Pi}(l^2) = i\int d^4x \,e^{il x} \langle 0|TJ^\mu_5(x)\phi_5(0)|0\rangle 
= i l^\mu \sqrt{2N_f} F_{\bar{\eta}}(l^2) \frac{-1}{l^2 - m^2_{\eta'}} \,c(l^2)\,.
\end{eqnarray}

\bibliography{wlines}
\end{document}